\documentclass[prb,aps,twocolumn,showpacs,floatfix]{revtex4}
\usepackage{graphicx}
\begin{document} 
\title
{Structure prediction for cobalt-rich decagonal AlCoNi from pair potentials}

\author{Nan Gu, M. Mihalkovi\v c~\protect\cite{MM-addr}, and C. L. Henley} \affiliation{Dept. of Physics, Cornell University, Ithaca NY 14853-2501}

\def\FUTURE#1{}
\def\LATER#1{}
\def\TODO#1{}
\def\remark#1{{}}

 \def\MM#1{}
 \def\CLH#1{}
 \def\NAN#1{}
\def\OMITb#1{{}}

\def\rr{{\bf r}} 
\def\zbar{{\bar z}}
\def\Zbar{{\bar Z}}
\def\fiveand{`5 \& 5' }

\def\Decagon{13\AA{}~decagon}
\def\Decagons{13\AA{}~decagons}
\def\Dec{13\AA{}~D}
\def\Decs{13\AA{}~Ds}
\def\Star{Star cluster}
\def\Stars{Star clusters}
\def\PB{PB}
\def\PBs{PBs}

\def\HHor{{{\cal H}_{\rm or}}}
\def\Jor{{{\cal J}^{\rm or}}}
\def\EAF{E_{\rm AF}}
\def\EFM{E_{\rm FM}}
\def\Eav{E_{\rm av}}
\def\Ediff{E_{\rm diff}}
\def\Nat{N_{\rm at}}
\def\Zbar{{\bar Z}}
\def\puck{{\mu}}
\def\Nch{{N^{ch}}}
\def\Jpuck#1{{{\cal J}^{\rm puck}_{#1}}}
\def\HHpuck{{{\cal H}_{\rm puck}}}

\begin{abstract}
A systematic, decoration-based technique to discover the atomic structure 
of a decagonal quasicrystal, given pair potentials and experimentally 
measured lattice constants, is applied to the ``basic'' cobalt-rich decagonal 
Al-Co-Ni quasicrystal.  First lattice-gas Monte Carlo simulations are 
performed, assuming the atomic sites are vertices of a rhombus tiling with 
edge 2.45\AA{}.  This phase is found to be dominated by 13\AA{} diameter 
decagon-shaped clusters, each with a pentagon of Co atoms at the center.
These, and another smaller cluster, decorate vertices of a ``binary tiling''
with rhombus edge 10.4\AA{}.  Further simulations with a restricted site 
list show that Al arrangements on the borders of the 13\AA{} decagon 
cluster form Hexagon, Boat, and Star tiles with edge 2.45\AA{}; they 
indicate specific sites for Co versus Ni atoms, and how the structure 
adapts to small composition changes. In the second half of the paper, 
relaxation (augmented by molecular dynamics annealing) is used to obtain 
realistic structures. The dominant new feature is a set of linear 
``channels'' attractive to Al atoms and running transverse to the layers. 
Each is typically occupied by three atoms in four layers, implying puckering 
and a spontaneous period doubling to $c\approx 8$\AA.  Puckering favors 
pentagonal long range order of the cluster orientations.  Our simulation 
captures most features of the related $W$-AlCoNi crystal, except for its 
pentagonal bipyramid motif.
\end{abstract}

%% 61.16Bg

\maketitle

\section{Introduction}
\label{sec:intro}

\CLH{CLHNEW: Intro was reordered, paragraph on method moved here from Sec II}

This paper recounts the results of a project to simulate the structure
of decagonal quasicrystal Aluminum-Nickel-Cobalt $d$(AlCoNi) in the
``basic Co'' (cobalt rich) phase purely from energy minimization principles.  
Of the equilibrium decagonal quasicrystals, $d$(AlCoNi) has (in some of its
modifications) the highest structural quality and has received the most
study.  Studies of the phase diagram indicate that, e.g., decagonal
$d$(Al$_{72.5}$Co$_{18}$Ni$_{9.5}$) is stable (at higher temperatures
only), whereas $d$(AlCo) is metastable only~\cite{grushko-alco}.  

Recently, a computational approach was proposed for discovering the
atomic structure of any decagonal quasicrystal, given no information
except a set of pair potentials, the quasilattice constant, and the
periodic lattice constant; it was applied first to $d$(AlCoNi) in the
``basic Ni'' phase;~\cite{alnico01,alnico02}.  
In the study described here (and briefly reported
elsewhere~\cite{Gu-letter,Gu06-ICQ9}), the same approach is applied to
``basic Co'' for the first time, and shown to work.  As in the ``basic
Ni'' case, our final structure description is in terms of a
supertiling with a large quasilattice constant, but here different
clusters and tiles enter than in the ``basic Ni'' case.

Since the sensitivity of the structure to the precise composition
is one of the issues in this paper (e.g. in Sec.~\ref{sec:bNi-compare}),
and since known structures of crystalline ``approximant'' phases
greatly illuminated our understanding of the related quasicrystals
in the past, we shall pause to review what is known in the Al-Co-Ni
phase diagram.

The decagonal portion of the Al-Co-Ni phase diagram is fragmented 
into several modifications occupying small domains.~\cite{Ri96b,Ri98}.  
Of these, those showing the simplest diffraction
patterns are the so-called ``basic Nickel'' phase near the Ni-rich
composition Al$_{70}$Co$_{10}$Ni$_{20}$ and so-called ``basic
Cobalt'' near the Co-rich composition Al$_{70}$Co$_{20}$Ni$_{10}$.
Several high-resolution X-ray structure determinations were carried
out for the ``basic Ni'' phase~\cite{xray}, but studies of the ``basic
Co'' phase have lagged.  
An interesting aspect of the Co-rich portion
of the phase diagram is the fivefold 
(rather than tenfold) symmetric decagonal phase,~\cite{li-5fold}.
%% originally discovered in 
%% the composition $d$(Al$_{70}$Co$_{15}$Ni$_{10}$Tb$_{5}$
in particular $d$(Al$_{72.5}$Co$_{20}$Ni$_{7.5})$ and
$d$(Al$_{72.5}$Co$_{19}$Ni$_{8.5})$~\cite{ritsch-5fold}, 
also $d$(Al$_{71.5}$Co$_{25.5}$Ni$_3$)~\cite{Ri96a}.
This and other Co-rich modifications show superstructure diffraction peaks, 
indicating modulations of the ``basic'' structure:
$d$(Al$_{72.5}$Co$_{17.5}$Ni$_{10}$), 
similar to the ``fivefold'' modification~\cite{Ri98},
and $d$(Al$_{71}$Co$_{20}$Ni$_{9}$), which has a 
period of 61\AA{} in one direction and thus was called 
the ``one-dimensional quasicrystal''~\cite{Ri00}.
Throughout the phase diagram, the {\it quasilattice constant} 
$a_0$ is close to $2.45$\AA{}.

\CLH{CLHNEW: to MM, is that satisfactory def'n of $a_0$?}
\MM{to CLHnew, sure!}
There is a solved periodic crystal approximant of ``basic Co'',
$W$(AlCoNi) structure~\cite{Su02}. There are
further modifications near to the ``basic Ni'' composition
as well as near $d$(Al$_{70}$Co$_{15}$Ni$_{15}$, and
another (partially solved) approximant~\cite{steurer}, with 
unit cell $37.5$\AA{} $\times 39.5$\AA{} $\times 8$\AA, 
and composition Al$_{71}$Co$_{14.5}$Ni$_{14.5}$.
The ``basic Co'' phase has a $4.08$\AA{} stacking period, like ``basic
Ni'', but it shows much stronger diffuse scattering than ``basic Ni'',
\cite{Fr00a,Fr00b} in such a way as to indicate a local doubling of
the $c$ periodicity (to $8.16$\AA{});  the known large approximants
also have $c\approx 8$\AA{}.
[Later in this paper (Sec.~\ref{sec:doubled}), 
we shall address the stabilizing effects of this period doubling.]

Our general technique is the same as those used in the previous work
on ``basic Ni''.~\cite{alnico01,alnico02,alnico04}.  We begin with a
small-scale rhombus tiling and discover general motifs and patterns.
These patterns usually have a geometry consistent with an inflated
Penrose tiling: we define a new model using that tiling, and the
patterns observed at the small scale are promoted to be fundamental
objects on the inflated tiling. By restricting configurations and
increasing the size of fundamental objects, we can run simulations on
larger and larger unit cells without excess degrees of freedom,
speeding up the MMC process considerably.

The outline of the paper is as follows.  After reviewing the technique
and the information needed in its set-up (Sec.~\ref{sec:set-up}), we
present initial results in Sec.~\ref{sec:fixed-site} from Monte Carlo
lattice-gas simulations using a discrete site-list, both with the
initial edge-$a_0$ rhombi and also with $\tau a_0$-edge bilayer
rhombi; in particular, the whole structure is built from two cluster
modifs -- the \Decagon~ and the \Star.  Next, Sec.~\ref{sec:ideal-deco} 
codifies this by describing an ideal decoration, which requires
specification of the \Decagon~ orientations as well as the optimum
placement of a subset of easily moved Al atoms.

In Sec.~\ref{sec:relax}, we pass on to molecular dynamics and
relaxation studies that break free of the discrete-site lists; these
reveal troughs (which we call ``channels'') in the potential energies
felt by Al atoms which lead to local disruption of the layering of the
atoms, and a breaking of the two-layer peridiocity assumed in previous
sections.  Here and in Secs.~\ref {sec:decLRO} and
\ref{sec:puckerLRO}, we take up the correlations in the atoms'
displacements, and also how this determines the an ordering of the
orientations of \Decagon~ clusters which reduces the system's symmetry
to pentagonal.  We conclude with an application to W(AlCoNi), the
best-known approximant of Co-rich quasicrystal $d$(AlCoNi), in
Sec.~\ref{sec:W}, and a discussion (Sec.~\ref {sec:discussion})
of the key results and the limits on their validity.

\section{Methods and input information}
\label{sec:set-up}

In this section, we lay out the procedures of the simulation, as well
as the assumptions and facts that all our results depend on.

\subsection{Methods}
\label{sec:methods}

Our methods are a combination of Metropolis Monte Carlo (MMC),
relaxation, and molecular dynamics (MD). We first perform MMC on a set
of fixed sites. We create this set by make use of a {\it tiling} of
Penrose rhombi on each layer and by placing atomic sites on each of
the rhombi using a {\it decoration}. Fig.~\ref{fig:tileflips} shows
Penrose rhombi and two decorations that we use. [See
Appendix~\ref{app:code} for a more detailed description of the
decoration and tiling.]

Penrose rhombus tilings (even random ones) have a natural inflation
rule whereby the same space can be retiled with rhombi whose edges are
a power of the golden ratio $\tau\equiv(\sqrt 5 +1)/2 \approx 1.618$
multiplied by the orginal edge length.  In this paper, we will make
use of rhombi with edges $a_0 = 2.45$\AA{}, $\tau a_0
\approx$3.96\AA{}, and $\tau^3 a_0\approx$10.4\AA{}; we shall also
mention a similar tiling with edges $\tau^2 a_0$ that applies to the
Ni-rich $d$(AlNiCo) phase, a structure closely related to the one we
are investigating.

A unit cell can be tiled in many different ways with the same number
of Penrose rhombi; this is physically important since the different
tilings correspond to many different configurations of atomic sites
that are consistent with the same physical cell and the assumptions
based on the lattice constants.  To explore this degree of freedom, we
perform MMC on the rhombic tiles by using rearrangements of three
Penrose rhombi (and the atoms on them) that preserve their collective
hexagon outline. The two rhombus configurations for which this is
possible are shown in Fig.~\ref {fig:tileflips}.

\begin{figure}
\includegraphics[width=1.7in,angle=0]{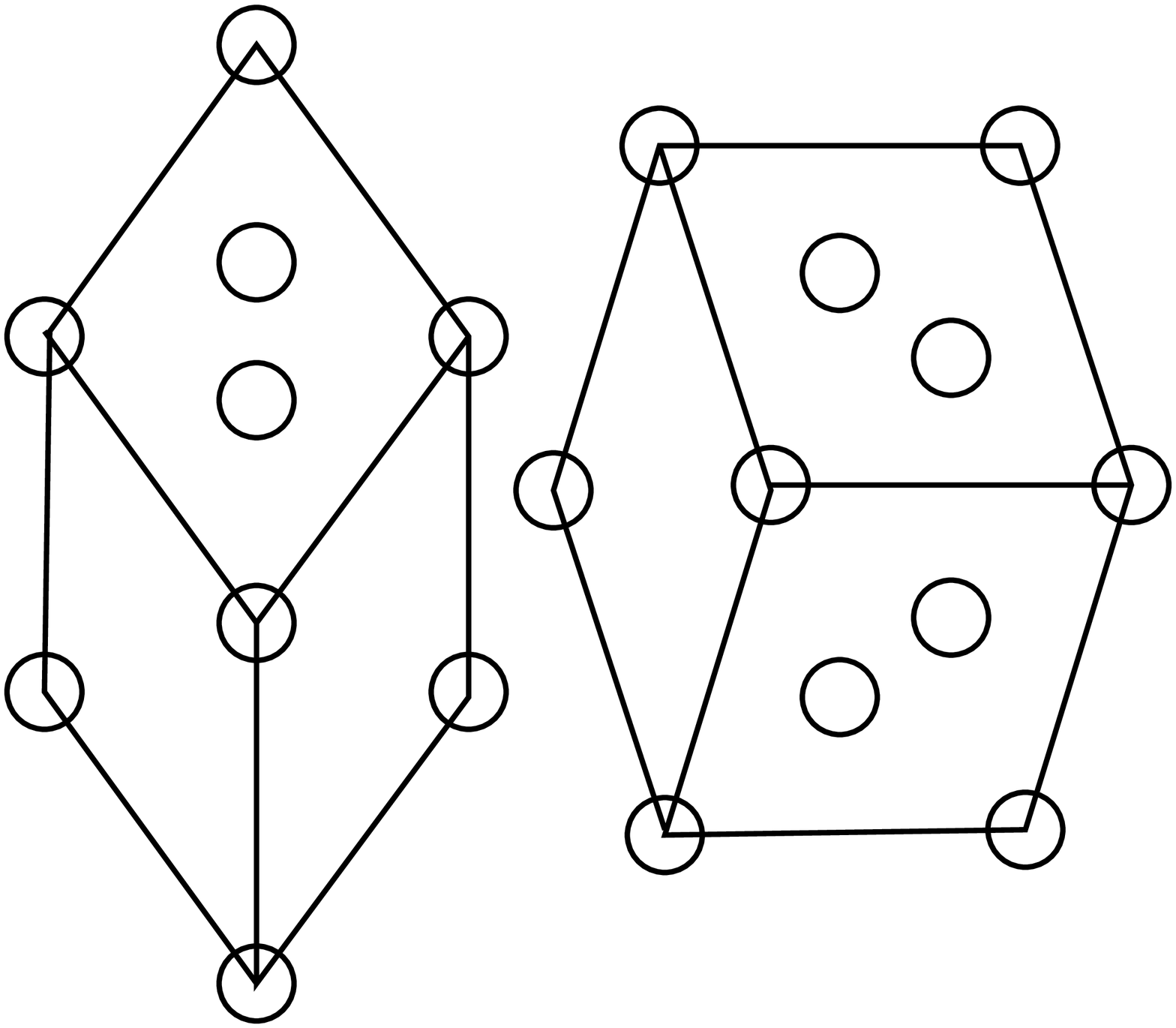}
\caption{These two configurations of Penrose rhombi, with edge
$a_0\approx 2.46 $\AA{}, can be flipped with respect to their
asymmetric axes as a way to move sites around.
\LATER{CLH suggested somewhere to show the 4A tiles' site list as well, 
and Marek suggested to show the inflation relationship.}  }
\label{fig:tileflips}
\end{figure}

The MMC is performed on a temperature schedule specified by the
beginning and ending inverse temperatures $\beta = {{1}/{k_{B}T}}$
along with an inverse temperature step $\Delta \beta$. At each
temperature step, a set number of MMC operations per site are
performed. After we find a configuration with this ``fixed-site''
method, we can remove the site list restriction, and use relaxation
and MD to find a structure that is more energetically favorable by our
potentials.

\CLH{CLHNEW: A paragraph on our general recipe was moved into the
introduction}

Why is our method to start with tilings, decorations, and discrete sites,
and iterate this (as outlined in the Introduction), rather than 
immediately perform MD?
The reason is that the energy surface of $d$(AlCoNi) in configuration
space contains many local energy minima. A pure MD program would be
almost certain to be trapped in a glassy configuration. [Even with a
small number of atoms and a simpler set of potentials, extremely long
MD coolings were necessary in order to produce recognizable (but still
quite defective) quasicrystals by brute force.~\cite{roth97}]

\subsection{Input information}
\label{sec:methods-input}

The only experimental inputs into the procedure are 
lattice constants, composition, and and density;
the only theoretical input is the pair potentials.
For the initial trials, one must also make a discrete choice
of which size of rhombus to use -- the quasilattice constant 
of a decagonal tiling is defined only modulo factors of $\tau$
and one must decide how many atomic layers are to be simulated.
\MM{NEW I would replace by the following statement:
``and one must decide how many atomic layers are bound to a layer
of the tiling.''}

\subsubsection{Pair potentials}
\label{sec:pairpot}

%%%%%%%%%%%%%%%%%%
\begin{table} 
\begin{tabular}{|lrccr|}
\hline
A-B  & $R_0^{\rm 0.1eV}$ (\AA{}) & ($i$)& $R_i$ (\AA{}) &  
$V_{A-B} (R_i)$ (eV) \\
\hline \hline
Al-Al &2.62 & (0)  & 2.49 & +0.351 (hc)\\
\hline
Al-Co &2.00 & (0)  & 2.30 & -0.285 (hc)\\
     &&  (1)  & 2.38 & -0.302 \\
     &&  (2)  & 4.44 & -0.035 \\
\hline
Al-Ni & 2.02 & (0)  & 2.25    & -0.152 (hc) \\
     &&  (1)  & 2.38 & -0.192 \\
     &&  (2)  & 4.37 & -0.030  \\
\hline
Co-Co & 2.48 & (0)  & 2.73  & +0.045 (hc) \\
     &&  (1)  & 2.68  & +0.040  \\
     &&  (2)  & 4.49 & -0.091 \\
     &&  (3)  & 6.44 & -0.033 \\
\hline
Co-Ni &2.48 & (0)  & 2.62 & +0.050 (hc) \\
     &&  (1)  & 2.67 & +0.044 \\
     &&  (2)  & 4.42 & -0.081 \\
     &&  (3)  & 6.39 & -0.029 \\
\hline
Ni-Ni &2.46 & (0)  & 2.63   & +0.051 (hc) \\
     && (1)  & 2.64 & +0.051 \\
     &&  (2)  & 4.34 & -0.075 \\
     &&  (3)  & 6.30 & -0.027 \\
\hline
\end{tabular}
\caption{ \footnotesize 
  Pair potential minima $R_i$ for Al-Co-Ni. 
  The ``(0)'' well is the hard core radius, defined as the minimum 
radius actually found in a relaxation (after molecular dynamics
annealing) of an example configuration; $R_0^{\rm 0.1eV}$
is defined by $V_{\rm A-B}(R_0^{\rm 0.1eV}) = +0.1$eV.
Minima are listed only for $|V_{\rm A-B}(R_i)| >$0.025~eV.}
\label{tab:potentials}
\end{table}
\remark{MM, 9/05: The correction  TM-TM terms were 
shown to have a simple effective analytic form 
${\rm const} (2.55\hbox{\AA}/r)^b$, where $16 < b < 22$.}

\remark{MM 6/05: 
For our sample xyz.rel.1 (ideal 8A MD+relaxed), 
the minimal Al-TM distances after relaxation were
2.25\AA{} (for Al-Ni) and 2.33A{} (for Al-Co).}

The six pair potentials (for the combinations of Al, Co, and Ni) were
generated using Moriarty's ``Generalized Pseudopotential
Theory''~\cite{Mo-GPT}, as modified using results from {\it ab initio}
calculations to add a repulsion correcting the forces between
TM-TM nearest neighbors~\cite{Al-GPT}, attributed to many-body
terms beyond the pair terms given by GPT.
A standard cutoff radius of 7\AA{} was normally used.
Even modified, the potentials are
imperfect in their unreliable handling of TM-TM nearest neighbors and
their neglect of three-body interactions~\cite{marek-GPT}.  

%% The quantitative phase diagram is somewhat problematic since 
The  same potentials may be used over the interesting composition range,
even though they implicitly depend on electron density, 
because the lattice constants fortunately compensate so as to keep the electron
density nearly constant (over this range).  A major post hoc justification 
for the pair potentials is the correct prediction of binary and ternary phase
diagrams~\cite{widom-GPT-phased}.
In particular, the (corrected) ternary GPT potentials predict the
correct Co-Ni chemical ordering in the approximant 
$X$(Al$_{9}$[Co,Ni]$_4$)~\cite{Kat06}
and it seems in $W$(AlCoNi)~\cite{Mi06-ICQ9}.

\remark{MM 9/05. At present, the EAM potentials of Gaehler et al
are not set up to reproduce ternary phase diagrams, or chemical ordering Co-Ni;
they are adapted to a high-T condition for a specific Ni-rich
composition.}

\LATER{It should be noted that the pair-potential total energy
omit a contribution related to the overall
density of the electron  sea; thus, our
pair potentials are not suitable to decide
which density is realistic:  they should only
be used to compare configurations of the same density.
[Marek's note (9/05) in the context of short bonds.]}

Radii at which these potentials have minima are given in
Table~\ref{tab:potentials}, as well as a ``hardcore'' radius.  [Plots
of the same potentials are in Fig.~1 of Ref.~\onlinecite{alnico01}.]
As was noted previously~\cite{Wid96,alnico01} the salient
features of such potentials are (i) a very strong Al-TM
nearest-neighbor well, which is 1.5 times as strong for Al-Co as for
Al-Ni; (ii) a rather strong TM-TM {\it second} neighbor well [TM-TM
first neighbors are unfavorable because they would deprive TM of Al
neighbors] (iii) no Al-Al interaction to speak of except the hard
core.

A cartoon recipe for an optimum structure is (i) satisfy the TM-TM
interactions by a relatively uniform spacing of TM atoms (ii) place as
many Al as possible in the Al-TM first wells, limited by the Al-Al
hardcore.  In principle, the Al-TM optimization might constrain the
TM-TM lattice, but in fact the considerable freedom in placing Al's
allows these tasks to be separated.  
(The main operation of the Al-Al
constraints is presumably to select a subset of TM arrangements, which
would be practically degenerate if only the TM-TM potentials were
taken into account.)

\subsubsection{Cell, lattice constant, density and composition}
\label{sec:setup-contents}

Decagonal quasicrystals are quasiperiodic (at least on average) in just two
dimensions. In this decagonal plane, we assume the atomic configuration can be
described reasonably well by a tiling of Penrose rhombi with edge length 
$a_0=2.45 $\AA{} quasilattice constant, which is experimentally determined.
In the dimension normal to the
quasiperiodic plane, the $c$-axis, the quasicrystal repeats after a number of
two-dimensional quasiperiodic layers are stacked on top of each other 
with a uniform separation $c/2=2.04$\AA{} taken from experiment.

Periodic boundary conditions are always adopted in all three
dimensions: the constraint that this be consistent with a rhombus
tiling permits only a discrete family of simulation cells. The cell
sizes we chose are especially favorable since they permit a tiling
which is close to having five-fold symmetry [in the frequency of the
various orientations of rhombi or other objects in the tiling We label
our unit cells by their dimensions, $a\times b \times c$, where the
stacking period (almost always 4.08\AA{}, and often omitted) comes
last (see Table~\ref{tab:cells}).  The largest part of our studies
were done on the ``32 $\times$23'' cell, which conveniently has
dimensions large enough to accomodate a variety of (dis)ordered
arrangements, but small enough to be tractable.  We too rarely used
the ``20$\times$ 38'' cell, which has exactly the same area, but a
more elongated shape.  The 20$\times$ 23 cell has an area smaller by
$\tau^{-1}$ than the standard 32 $\times$23; we call it ``half-W'' as
we used it less often than the ``W-cell''. That was so called since it
has the same dimensions as the approximant $W$-AlCoNi; we employed the
``W-cell'' even when not trying to predict the $W$-phase structure,
for it too has a convenient size.  We made the least use of the
``20$\times$20 centered'', which is quite small 
(half the 32$\times$23 cell).  
For a special purpose we once used the 12$\times$14 cell, 
which is shorter by a factor $\tau^{-1}$ in each direction
than the ``half-W'' cell; we call it the 
``Al$_{13}$Co$_4$'' cell,  as it is the same size
as the orthorhombic variant of that crystal.

\begin{table} 
\begin{tabular}{|llrrr|}
\hline
Name & symmetry & a (\AA{}) & b (\AA{}) & $\gamma$ \\
\hline \hline
32$\times$23 ``standard'' & rectangular &  32.01 &   23.25  &  
($90^\circ$) \\ 
20$\times$20 centered     & oblique    & 19.78  &   19.78  &  
$72^\circ$   \\
20$\times$38 ``elongated''& rectangular &  19.78 &   37.62  &  
($90^\circ$) \\ 
20$\times$23 ``half-W''   & rectangular & 19.78  &   23.25  &  
($90^\circ$) \\ 
40$\times$23  ``W-cell''  & rectangular & 39.56  &   23.25  &  
($90^\circ$) \\ 
12$\times$14 ``Al$_{13}$Co$_4$'' & rectangular & 12.22  &   14.37  &  
($90^\circ$) \\ 
\hline
\end{tabular}
\caption{ \footnotesize Unit cells used in this work. 
Note the 20$\times$20 is the primitive
cell of the 32$\times$23 {\it centered} rectangular lattice, but
in an oblique lattice setting so as to give 
primitive vectors $a$ and $b$ correctly (with angle
$\gamma$  between them).}
\label{tab:cells}
\end{table}

The ``basic Co'' phase of $d$(AlNiCo) is experimentally known to have
a period $c'=2c=8.16 $\AA{}, but -- up till the relaxation studies of
Sec.~\ref{sec:relax} -- we always simulated a cell with a period $c$.
In other words, our philosophy (as in Ref.~\onlinecite{alnico01}) was
to discover as many features as possible in the simplest (4.08\AA{}
period) framework, and only later to investigate deviations from this.
A partial justification is that an {\it approximate} 4.08\AA{}
periodicity is expected, and found: {\it many} of the atoms
do repeat with that period, modulo small offsets.  Ideally, though,
one should only take the layer spacing from experiment, and investigate
cells with different numbers of layers, so as to let the simulation
reveal any additional modulations that may increase the period.

\LATER{JUSTIFY what is the reasonable range of compositions and densities.}
%%%%%%%%%%%%%%%%%
Most of our simulations used a standard density~\cite{FN-bracketdensity}
of 0.068 atoms/\AA{}$^3$ and a composition Al$_{70}$Co$_{20}$Ni$_{10}$.
Variations of a few percent were tried for special purposes; in
particular, our $W$(AlCoNi) simulation (Sec.~\ref{sec:W})
used density $\sim$0.071 atoms/\AA{}$^3$ and composition
Al$_{72}$Co$_{21}$Ni$_{7}$.

\remark{Nan mentioned: We also investigate changes in occupation on 
the 4.0\AA{} tiling when the density is changed.  [I left out tests on the 
composition shifts (Ni-richer/Co-richer) because these were performed before 
I knew how to automate simulations. There were about 3 simulations of these at
each composition.]}

\LATER{Nan: I also ran tests on the W cell, which had a density range of 68-78.}

In simulations specifically exploring the effect of atom density, 
we varied it over a range of roughly 
$0.066$\AA{}$^{-3}$ to $0.074$\AA{}$^{-3}$ ; this is unphysically
loose at one extreme, and unphysically overpacked at the other.
A range of roughly $0.066$\AA{}$^{-3}$ to $0.072$\AA{}$^{-3}$ 
is internally ``legitimate''; our diagnostic for this
is that the run-to-run variance of the energy should
not be too large.  If we took into account competition with 
other structures in the Al-Ni-Co phase diagram, or if we used
the densities appearing in actual approximant phases, presumably
the density range would be much smaller.
The actual W phase~\cite{Su02} has a reported density 0.0708\AA{}$^{-3}$,
or 0.0703\AA{}$^{-3}$ when fractional occupancies are
best resolved~\cite{Mi06-ICQ9}.
The atomic density in some decagonal approximants is 
0.0724 \AA{}$^{-3}$ for Al$_5$Co,
0.0695 \AA{}$^{-3}$ for Al$_{13}$Co$_4$
(in the mC32 structure variant using the standard
nomenclature \cite{Pearson}),
or 0.0687 \AA{}$^{-3}$ for  Al$_3$Ni.
\remark{Densities for the Al-Co approximants
from MM 9/05; he doesn' use Al13Co4.oP102
or mC102 as there are some fractional occupancies.}

\remark{MM 9/05: fractional occupancies
are resolved in such a way that the actual density is 0.0703 (with
two sites per unit cell vacant).}

%%%%%%%%%%%%%%%%%%
\begin{table}
\begin{tabular}{|lrrrl|}
\hline
Model   &    Al (\%) & Co (\%) & Ni (\%) & density (\AA{}$^{-3}$) \\
\hline \hline
Standard initial   &  70   &  20  & 10   &  0.068x \\
$W$(AlCoNi)        &  71.7 &20.8  &  7.5 &  0.070x \\
``basic Ni'' ideal &  70.0 & 9.3  & 20.7 & 0.0706 \\
idealized W-cell   &  70.1 & 22.4 &  7.5 &  0.071x \\
\hline
\end{tabular}
\caption{ \footnotesize
Composition and density comparison for various
structure models.  Experimental densities are
surprisingly hard to measure accurately, and
composition of the ideal crystal structure is
rarely identical to actual compositions of the
samples.  (Sources: ``Basic Ni'', Ref.~\onlinecite{alnico01}, 
Sec.~III B;
$W$-AlCoNi, Ref.~\onlinecite{Su02,Mi06-ICQ9}.)
\remark{MM 9/05: W composition here uses values from the best VASP
model of MM and MW; this should be more accurate than
the diffraction-data refined values (with mixed occupancies.)}
\LATER{Does this table need any other column, or row?}
}
\label{tab:deco-count}
\end{table}

\section {Fixed-site simulations}
\label{sec:fixed-site}

\begin{figure}
\includegraphics[width=2.9in,angle=0]{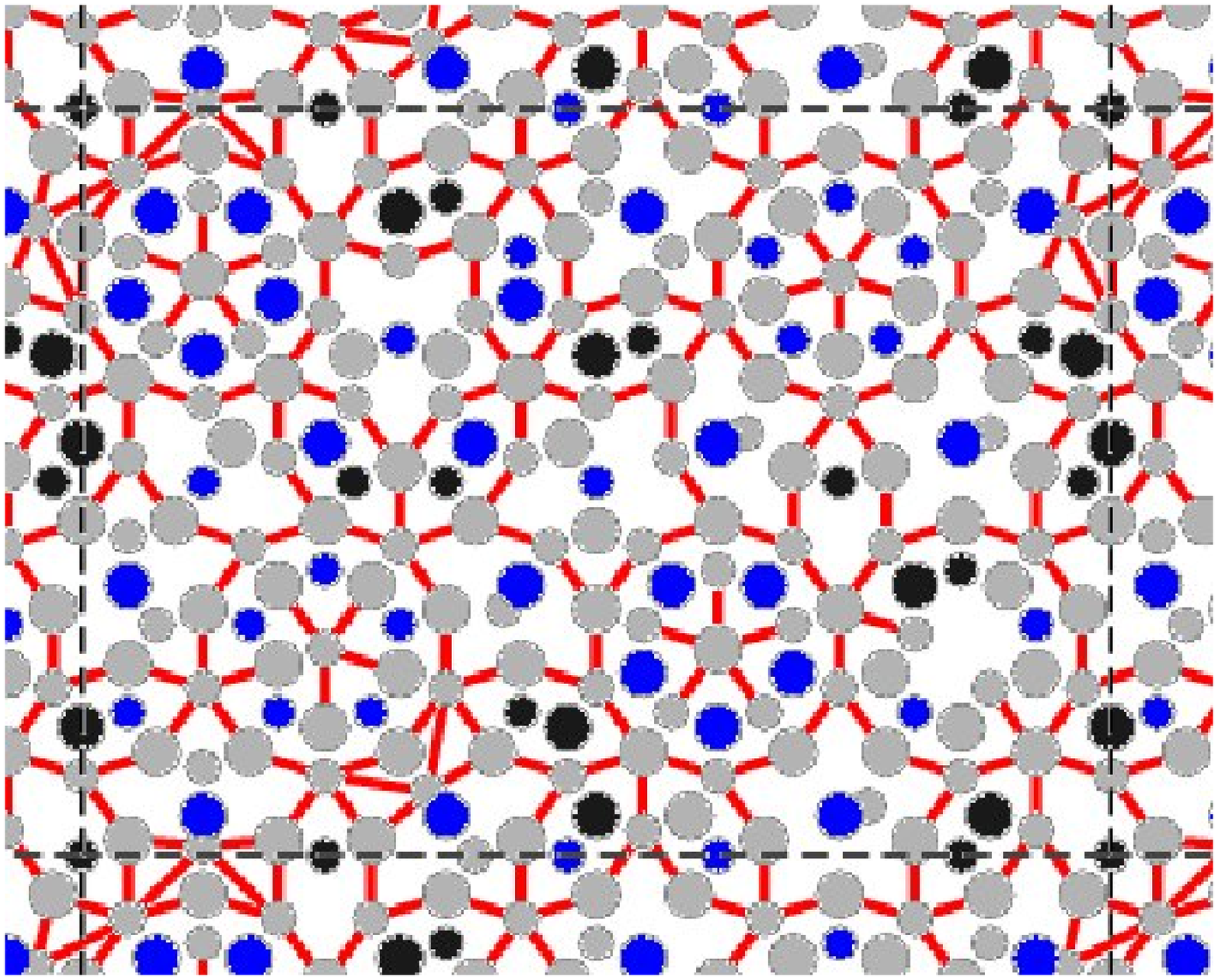}
\caption{[Color] Typical result of Monte Carlo simulation using
2.45\AA{} edge tiles, on the 32$\times$23 tiling.
\LATER{Convert to black-and-white, and revise color code.}
The composition is Al$_{0.700}$Co$_{0.198}$Ni$_{0.101}$,
with  207 atoms in the cell,
and the unrelaxed energy  is -0.4419 eV/atom.
Black circles are Ni, blue are Co, and gray are Al.
The [red] lines are a visual aid to mark Al-Al neighbors
in different layers, separated by 2.45\AA{} in projection.
These mostly form a hexagon-boat-star-decagon tiling as
described in Sec.~\ref{sec:ideal-deco}.
\remark{In this example, the decagon-center clusters 
have antiparallel orientations. }
Outer rings of \Decagons~ can be made out, as well
as \Stars, but have many imperfections, e.g.
%%% Figure file for 2.45A is ~marek/uniconfig.ps (12/05)
``short'' (2.25\AA{}) Al-Co bonds 
(see App.~\ref{app:short-bonds}.)
\LATER{Instead show,
``The lines are a visual aid automatically  drawn to connect pairs of
TM atoms 4.43\AA{}-4.49\AA{} apart''?}
}
\label{fig:smalltiles}
\end{figure}

In this section, we describe two stages of Metropolis Monte Carlo
simulations using discrete site lists, and the key structure motifs
that emerged from them.  It is important to note that in this kind of
run, we are {\it not} averaging quantities over the ensemble, {\it
nor} are we analyzing the final configuration after the
lowest-temperature anneal.  Instead, we pick out the lowest-energy
configuration which has appeared during the entire run.  This
procedure, since it singles out the low-energy fluctuations, may give
meaningful results even when performed at surprisingly high
temperatures.~\cite{FN-highT}

\subsection {Exploratory simulation using small  tiles}
\label{sec:MC-a0}

A series of annealing simulations and relaxations at the level were
performed using the edge $a_0$ rhombi.  Most of these runs were done
on the 32$\times$23$\times$4 unit cell with our standard composition
of Al$_{70} $Co$_{20}$Ni$_{10}$ and our standard atomic density of
0.0682 atoms/\AA{}$^3$.  That unit cell was small enough that it was
not computationally prohibitive to simulate, yet large enough that
motifs could form without strong constraints by the periodic boundary
conditions.~\cite{FN-cellsize}
Our aim at this stage is to allow the configurations as much freedom
as possible to discover the correct local patterns appropriate to this
composition. With the atoms restricted to tile-based sites and 
the composition and density fixed at our 
standard values~\cite{FN-bracketdensity},
there still seemed to be sufficient freedom to find good local order, 
as there had been in the ``basic Ni'' simulation~\cite {alnico01} (and in
much earlier lattice-gas simulations~\cite{widom-cockayne}).

\CLH{12/05: The following edited from some remarks of MM email
9/05, in response to CLH's query about runs in the special
20x38 tiling  to test decagon-decagon placement. But
it referred to the standard run, so moved here.  Then was melded
with Nan's existing description, which had
different numbers -- I kept Nan's mostly, but expressed them
in sweeps (which is suggested from the way Marek had
expressed it).}

\remark{\MM (2/06) notes: 
At each temperature, we were doing:
10 * ( 10 short swaps/short-bond + 10 long-swaps/long-bond +
       + 2 tile-flips/vertex ).
So we did 100 sweeps at each
temperature, and a sweep has also 0.2 attempted tile flips/vertex.
}

\CLH{to MM CLHNEW: are the numbers right, now?}
To define a Monte Carlo sweep for the 2.45\AA{} edge tiling, 
we must delve into some technicalities.
As we just noted, there are two basic kinds of updates,
swaps of atoms between sites and tile-flips.  
A ``sweep'' is taken to have one attempted site-swap for each
``short bond'', defined as any pair of (candidate) sites separated 
by less than 3\AA{}.
In addition, every sweep contained $\sim 1$ attempted swap
per ``long bond'', defined to have a separation 
of 3 -- 5 \AA{}.  
Our standard 2.45\AA{} tile simulation 
on the 32$\times$23 cell had 680 candidate
sites (occupied by about 210 atoms), with about 9.5
short bonds per site and about 67 long bonds per site.
Finally , each sweep also had 0.2 attempted tile flips per 
tile vertex.
\MM{to CLHnew, corrected 0.02 to 0.2 (there were 
10*2 tile flips/vertex at each temperature, not 2!)}
%%%%%%%
\remark{
The standard settings Marek sent in 9/05 differ
in having 1 swap per long bond, and also somewhat
more sweeps per temperature step.  Nan, you said 
there were 1000-1500 short range steps per site:
you meant ``candidate site'', not ``atom'', right?
Also Marek said $\beta$ initial = 2, increments 1 or 2.}

A annealing temperature schedule 
typically began at $\beta=4$ and decreased
to $\beta=20$ in increments of $\Delta \beta = 1$ or $0.5$,
where $\beta$ is measured in (eV)$^{-1}$. 
At each temperature, 100-200 sweeps were performed.
\remark{I get 100-160 from converting Nan's number,
while Marek wrote 200.}
The {\it lowest} energy encountered during each annealing was saved.
(A similar search method was used in Ref.~\onlinecite{jeong94}.)
The annealing cycle was repeated 20 times;
the whole set of annealings took about 5 hr on an AMD Athlon 
2.1GHz processor.
Simulations were run with different temperature schedules, but the
results were not noticeably different.  

%%%%%%%%%
\remark{Each complete ``inner-most'' loop was cycled 20 times
at each T.  It contained a sequence of short swap stage
(10 swaps per short bond), long swap stage 
(10 swaps per long bond), and tile flip (2 attempts per vertex).}
%%%%%%%%%
%%%%%%%%%
\remark{(CLH added). And WITHOUT new initial conditions.
The intent was that the high temperatures would have the
effect of re-randomizing the configuration.
However, the TM-TM network has a higher
``melting temperature'' and our procedure does not
really erase all memories.}
%%%%%%%%

It should be remarked that the lowest temperatures would
have been more appropriate for a deterministic  decoration
forcing a good atomic structure, so that quite small energy
differences are being explored.  In these exploratory 
2.45\AA{} tile simulations, even $\beta=2$ eV$^{-1}$ 
(about three times the melting temperature)
can give good structures (keep in mind that the best
configuration is saved from each annealing.)
Note that the tile-flip degrees of freedom freeze out  while the 
temperature is still high.  Which atom configurations 
are available at lower temperatures depends sensitively on the
site-list.

No configuration found by 
MC annealing (even on the 4\AA{} tiling, see Subsec.~\ref{sec:MC-4A})
had energy as low as the idealized tiling in Fig.~\ref{fig:RulesWideal}.
We believe this is an artifact of the 
very limited sitelist. The TM arrangement freezes at
medium temperatures and  becomes frozen at low temperatures, 
as the only MC moves with a small energy difference
are Al hops to a vacant site (with -- perhaps --
Co/Ni swaps at somewhat higher temperatures).
A site which is good for a TM is generally not good for Al, and vice versa,
so there are no low-energy Al/TM swaps; 
a rearrangement of more than two atoms is needed 
to accomplish such a change.
\remark{c 6/05, Nan said ``Al/vacant is usually the move
with the least energy differential. I can't make the same
statement about Ni/Co.''}

\remark{MM suspects this is why simulation does
not find the low E states.  CLH points out a good
site list would not remedy this, but e.g.
three-atom move in the MC.
That would be a ``cluster update''
-- one usage of the word ``cluster'' which was not
mentioned in an extensive discussion 
at the conference ICQ9~\cite{discussion-ICQ9}.
Indeed, a good way to think
of the ``tiles'' or ``clusters'' that emerge from our
simulations  is as the smallest groupings of atoms that 
are favorable to update together. MM suggests that another
remedy in  the MC toolbox is ``tempering''. }

\remark{An idea of CLH (spring 05):
To understand the competition between different low-energy
configurations in an annealing simulation, let us assign
each excited state to the ``nearest'' low-energy configuration.
The number of such excited states defines an entropy. 
If a ``wrong'' (not lowest energy) configuration has a higher
entropy, then at  medium temperatures it may
have a lower free energy, and so the system tends to freeze
into this configuration rather than the ``right'' one.}

A typical result is shown in Fig.~\ref{fig:smalltiles};
this has total content Al$_{145}$Co$_{41}$Ni$_{21}$ corresponding
to our standard conditions.  The most striking feature was
that the TM atoms organized into a well-patterned sublattice,
reminiscent (in $z$-projection) to the packing of pentagons, stars,
and partial stars which is one of the canonical representations of
Penrose's tiling.~\cite{penrose-pentagons} 
The TM atoms configured
themselves to be $\sim 4.5$\AA{} apart, inviting a speculation that
the longer range patterns are enforced by the TM-TM interactions,
while the Al atoms flow around like hard spheres and fill in the gaps.
Indeed, there were different ``freezing temperatures'' for the TM-TM 
quasilattice and the Al-TM interactions: that is, the TM-TM lattice 
is well established at a temperature much higher than that necessary 
to rearrange the Al atoms.

Similar TM patterns are seen in all Al-TM decagonals (with many variations
having to do with the placement of the two TM species
and the larger-scale arrangement of these large pentagons).  
So as to best highlight this tiling-like network (and other 
medium-range structural features) to the eye, our graphics processing 
was automatically set to show a line connecting every pair of TM atoms 
in different layers and separated by $\tau a_0 \approx 4.0$\AA{}
in-layer.  
\LATER{Such TM-TM lines are seen in Fig.~\ref{fig:dds} (a).
It is not true for the present version.}

A striking effect at this stage is how the Al atoms in the two layers
organize themselves into a {\it one-layer} network (with edge 
2.45\AA{}: see Fig.\ref{fig:smalltiles}). 
The even vertices are all in one layer and odd vertices
within the other layer, so this represents a kind of symmetry-breaking
and long-range order that has propagated through the entire simulation
cell.  In fact, we can already recognize the 2.45\AA{}-edge 
Decagon-Hexagon-Boat-Star (DHBS) tiling, to be elaborated in
Sec.~\ref{sec:small-DHBS}.
Along with this order, and probably driving it, the TM atoms
also obey this alternation, except they go in the opposite
layer to the layer Al would have gone into.  Among other things,
that produces large numbers of TM-TM pairs in different layers, 
separated by $\tau (2.45) \equiv 4.0$\AA{} in-plane  and hence
by 4.5\AA{}, as described in the previous two paragraphs.

\LATER{The above 3 paragraphs perhaps need reordering}

\subsection{Fundamental cluster motifs}
\label{sec:clusters}

\subsubsection{\Decagon~cluster}

\LATER{CLHNEW In a couple places in the paper, CLH used the notation 
"Co(3)", which is very convenient, without ever defining it.}

It became apparent that at the Al$_{70}$Co$_{20}$Ni$_{10}$
composition, our pair potentials favor the creation of many
Al$_5$TM$_5$ rings surrounded by two more concentric rings with
approximate fivefold and screw decagonal symmetry.  The object as a
whole will be called the ``\Decagon ''(\Dec) for its diameter
($2\tau^2 a_0 \approx 12.8$\AA{}).

\begin{figure}
\includegraphics[width=3.1in,angle=0]{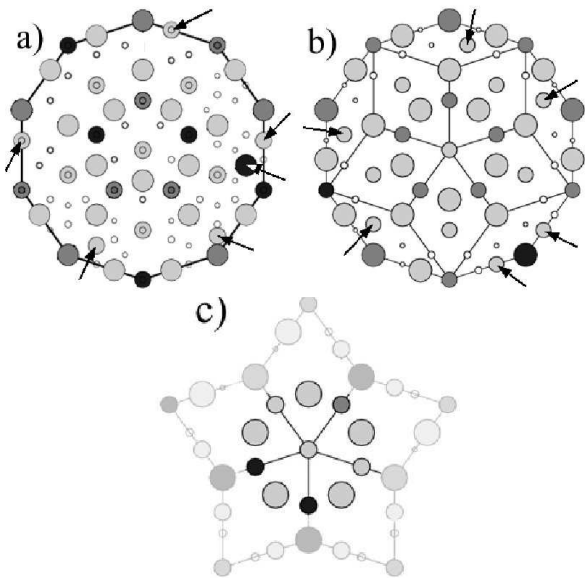}
\caption{
(a)\Decagon~created on the 2.45\AA{} random rhombus
tiling at $0.068 {\text{\AA{}}}^{-3}$ point density
Al$_{70}$Co$_{20}$Ni$_{10}$ composition.
Small empty circles denote unoccupied sites. 
Large and small circles are atoms located in the upper
and lower layers, respectively. (Double circles are overlapping 
circles in two layers.)
Black color denotes Ni, dark gray Co, and light gray Al.
%%%%%%%%%%%%%%%%%%%%
(b)A \Decagon~ formed on the 4.0\AA{} tiling and divided into 4.0\AA{}
scale Penrose tiles. 
Here and in (a), arrows point to atom sites 
(all in ring 2.5 or ring 3) that violate the cluster's $5m$ symmetry.  
(c) A \Star~created on the 4.0\AA{} tiling. Only the darker atoms are 
considered part of the \Star; the lightened atoms belong to 
adjacent \Stars~or \Decs.
\LATER{This figure should be made bigger, by placing a, b across the
  column and then c underneath} }
\label{fig:dds}
\end{figure}

%%%%%%%%%%%%%%%%%%%%%%
These tiles are decorated by a site list which
favors (but does not absolutely force) a \Decagon~ to appear when 
Metropolis Monte Carlo
is performed. Notice the significantly decreased site list and
enchanced ordering, as compared with the version of \Dec~ in (a).

\begin{itemize}
\item [1.]
At the center there is a single Al atom. 

\item [2.]  Ring 1 is ten atoms (Al$_5$TM$_5$) that we call the
\fiveand cluster. In projection, they form a decagon, but they
alternate in layer, so the symmetry element of the column is
${10_5}/m$.
The sites in the same layer as the central Al are (almost always) TM
sites; the other five sites are normally Al.
\remark{CLH worries this is an abuse of notation:
This is not really a point group or a space group, it is the
symmetry of a 1-dimensional object (a stack) in 3-space.
MM 9/'05 wrote ``${10_5}/m$ is correct I believe.''}

\item [3.]
Ring 2 consists of another ten Al atoms; in $z$ 
projection, each atom is  
along a ray through an 
atom of Ring 1 (but in the other layer). 

\item [4.]  
Ring 3 is on the outer border of the decagon which has
edges of length $4.0$\AA{}. In projection, there is a TM atom on each
corner alternating in layer (so the actual TM-TM separation is
$4.46$\AA{}).  These TM atoms sit in the same layer as their Al
neighbor in Ring 2.  In addition, almost every 13\AA{} decagon edge
has an Al atom dividing it in the ratio $\tau^{-1}:\tau^{-2}$, sitting
in the opposite layer from the TM atom on the nearer corner.  These Al
atoms are usually (but not always) closer to the corner TM's that are
in the same layer as the ring 1 TM atoms (see Fig.~\ref{fig:dds}).

\item [5.]  
Between rings 2 and 3 are candidate sites which are
occupied irregularly by Al, which we will call collectively ring
`2.5'.  [The rules for placement of the ring 2.5 and ring 3 Al will be
explored much later (Sec.~\ref{sec:small-DHBS}).]

\end{itemize}

At the stage of the $a_0=2.45$\AA{} tile simulation, virtually every \Dec~ has
imperfections, and there are variations between Co/Ni or Al/vacant in
even the best examples; in a typical \Dec~ only 80\% of the atoms
conform to the above consensus structure, but that is already
sufficient to settle what the ideal pattern is.

\subsubsection{\Star}
\label{sec:star-cluster}

Filling the spaces between the \Decagons, we identify another 11-atom
motif similar to the Al$_6$Co$_5$ center of the \Dec: a pentagonal
antiprism, in which one layer is all Al atoms,
and the other layer is centered by an Al atom.
The difference is the five atoms in the second pentagon are
only ``candidate'' TM sites; they have mixed occupation, with 
roughly 60\% TM (usually Ni) and 40\% Al.  We shall call this 
small motif the ``\Star~'', associating it with the
star-shaped tile that fills the space in a ring of five adjoining
\Decagons.  (The atoms along the edges, however,
are not formally counted as part of the \Star:
they normally belong either to the outer edge of a 
\Decagon~ cluster, or else to another 11-atom \Star.)

Such centers were evident in the 2.45\AA{}-edge
simulations, but they appear more clearly as repeated patterns in the
4\AA{}-edge simulations. (That tiling must fill the space between
\Decs~ by 4\AA{}-edge Hexagon, Boat, and Star tiles; the internal
vertex of each of those tiles gets a \Star~ centered on it.)

\LATER{CLH: Is the alternation a consequence of the grand alternation 
along 2.5\AA{} tile edges?}

The local symmetry around the center of a \Star~ is
fivefold (unlike the tenfold local symmetry around the \Dec).
Adjoining \Stars~actually overlap [if we represent them 
by the star of five rhombi as in Fig.~\ref{fig:dds}(c)]
and necessarily have opposite orientations; furthermore, 
the respective central Al atoms (and surrounding candidate-TM sites) 
are in alternate layers.
Thus, all \Stars~ can be labeled ``even'' or ``odd'' according 
to their orientation.

\FUTURE{One (clipped) reason says ``Since it is
impossible to tile all of two-dimensional space with \Decagons,
the remainder of space must be filled with \Stars.''
From that viewpoint, the \Star~ interior structure is wholly
determined by the relative positioning with respect to \Decs~and
other \Stars.  Is that true?}

\CLH{to MM: did we adopt the 4.0 A tiling already knowing that it 
would model Star as well as Decagon clusters?}
\MM{to CLH: The general setup for 4A tiling was inspired by TM atoms
network. The toy-Hamiltonian model designed for refining Decagon-based
models - also based on 4A tiling - I knew it will lead to 4.5A DHBS
tiling, that was the intention, to have the site-list under control.}

\subsection{Relationships of neighboring decagons}
\label{sec:Dec-relation}

\FUTURE{MM 9/05. 
I think this issue may be related
to a ``fun'' simulation mentioned later, in which we would take 4.5A
DHBS tiling, and replace decoration of the tau-smaller 2.45 objects.
this would create D-D contacts for example.}

The next step in our general method, after a cluster motif is identified,
is to discover what geomeric rules govern the network of cluster centers.
Those rules are usually expressed as a list of allowed inter-center 
vectors, which become the ``linkages''~\cite{Hen91ART} of our
model geometry.  Often, a mild further idealization of this network
converts it into random tiling. At that point, one is ready to 
proceed with the next stage of simulations, based on decorations of
this tiling.

\LATER{Foreshadow here that it will be Binary Tiling...}

So in the present case, 
once the \Decagon~ is identified as the basic cluster of our structure,
the question is how two neighboring ones should be positioned.
(The relative orientations of their pentagonal centers will be
left to Sec.~\ref{sec:decLRO}).  
As always, the fact that a cluster appears frequently suggests it is 
favorable energetically, and that one of the geometric rules should 
be rule to maximize its density.  Yet the more closely we place 
clusters, e.g. overlapping, the more imperfection (deviation from the 
ideal, fivefold symmetric arrangement) must be tolerated in 
each cluster; when the clusters are too close, this cost 
negates the favorable energy. (Note that even if the 
clusters do not appear to overlap, it is conceivable that
a further concentric shell should have been included in the
definition of the ideal cluster. In this case, the clusters 
-- properly defined -- are still classhing.)

\begin{figure}
\includegraphics[width=3.1in,angle=0]{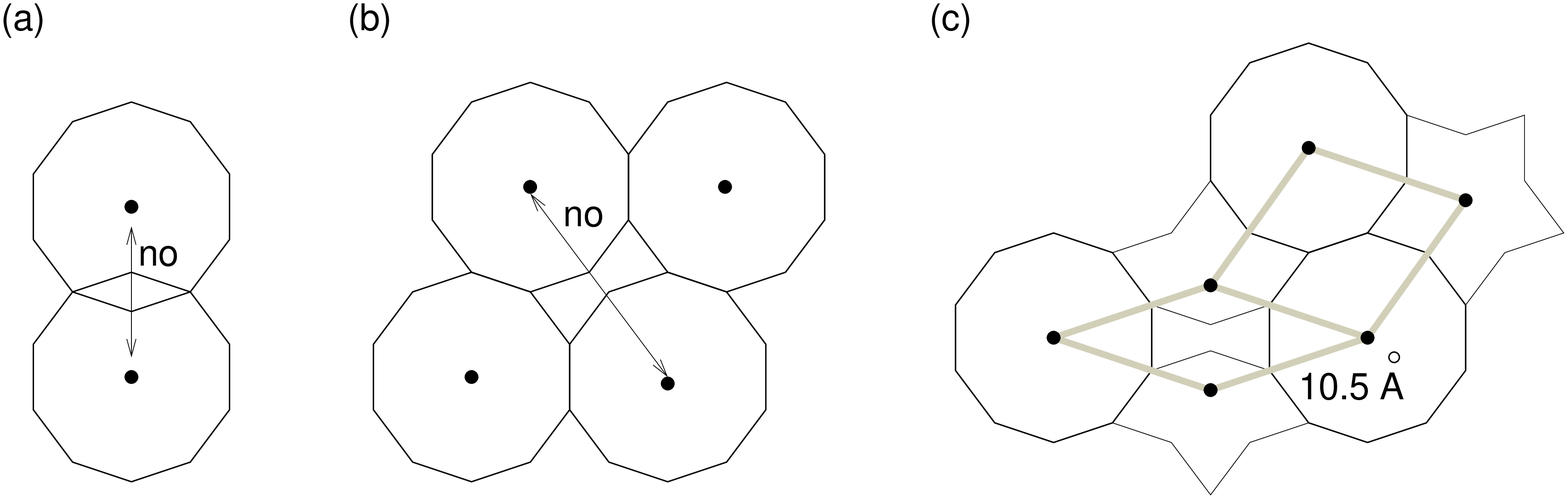}
\caption{Possible ways  
for \Decagons~ to adjoin.
(a), (b) Unfavorable ways  (c) Actual pattern, forming
the  
``Binary tiling'' (edges shown in gray).
\LATER{Should we add (d) and (e) showing  Decagon packings
for the half-W and the 32x23 unit cells?}
}
\label{fig:d-adjoin}
\end{figure}

We considered the four candidate linkages 
in Fig.~\ref{fig:d-adjoin} (a,b,c), but concluded
that only the linkages of
Fig.~\ref{fig:d-adjoin} (c) were valid.
Of course, the frequency by which such patterns appeared 
spontaneously in our simulations was one clue:
edge-sharing is the commonest relationship between \Decs.
[Cluster relations like Fig.~\ref
{fig:d-adjoin}(a,b)  did appear on occasion in the 2.45\AA{}-tile
simulations, particularly when the simulation cell lattice parameters
did not permit a network using only the favorable separations.]
Beyond that, we addressed the question more quantitatively 
by ad-hoc tests in which we  arranged that a configuration
would (or could) include one of the  rarer linkage types, 
and then compared its energy with a configuration
having the common linkages, or checked which of two locations
was likelier for another cluster to form.
These tests are detailed in Appendix~\ref{app:dec-rel-test}.

\LATER{Meld this with above.
A natural guess would be that these clusters, since
they appear repeatedly among the minimum-energy configurations from
the MC runs, surely are well adapted to the potentials, therefore we
ought to maximize their packing density.  But at first, of course, it
was not evident just where to draw the cluster's boundaries: how much
of the ``motif'' is mandatory (the part to be repeated), and how much
is adaptable to overlaps etc.?  }

In the Fig.~\ref{fig:d-adjoin}(a)  linkage,
cluster centers  are separated $\tau^3 a_0$, 
and the clusters overlap by a thin Penrose rhombus
with edge 4\AA{}.  In two places
a ring 2 (Al) site of one cluster coincides with a
ring 3 (Co) site of the other one, so 
modifications are mandatory for a couple of atom occupancies.
This linkage is motivated by the possibility
that the small decagon (bounded by the ring 2  Al atoms)
is the key cluster. (Indeed, that decagon, of edge 2.45\AA{},
is one unit of an alternative structural description
we shall introduce in Sec.~\ref{sec:ideal-deco}.)

Let us forbid overlaps of \Decagons~ henceforth, and assume
that the shortest linkage is edge sharing, a length 
$1.176 \tau^3 a_0 \approx 12.2$\AA{} (here $1.176 = 2 \sin 72^\circ$) 
The densest packing of \Decs~ would then~\cite{Hen86}
be the vertices of 4\AA{}-edge Hexagons, Boats, and Fat
72$^\circ$ rhombi; that would include many separations by
$\sqrt{5} \tau^2 a_0 \approx 14.4$\AA{}, like the one across a
fat rhombus's short diagonal as in Fig.~\ref{fig:d-adjoin}(b).  
\remark{In the 4.0\AA{} tiling,  a ``bowtie'' is left 
over in the middle of the decagons.}
This linkage also turns out to be disfavorable 
(Appendix~\ref{app:dec-rel-test}).
The reason appears less obvious than in the overlapping cases,
where there were atom conflicts.  One viewpoint (adopting
the analysis of Sec.~\ref{sec:ideal-deco}, below) is that this
relationship would not allow the space between \Decagon~ centers
to be filled with 2.45\AA{}~ Hexagons, Boats, and Stars.
A more direct reason is that at the closest approach, the TM
atoms on the respective Decagons's corners (in different layers)
are separated by just $a_0$ in layer, or a total distance
of 3.19\AA{}, which (see Table~\ref{tab:potentials}) is
disfavorable.

\CLH{CLHNEW to MM: Last sentence, do you agree?}
\MM{to CLHnew: yes}

We are left, then, with a network in which the angles
are multiples $(2\pi/10) n$, where $n\ge 3$.  We believe
that the densest possible packing under these
constraints is when the \Decs~ sit on the Large sites of
a Binary tiling of rhombi 
with edge $\tau^2 2.45$\AA{}$ \approx 12.2$\AA{},
as in Fig.~\ref{fig:d-adjoin}(c).
(In this tiling, Large circles sit always on vertices of 
the short diagonal of the Fat hexagon and of the long
diagonal of the Thin hexagon, and Small circles he
other way around: this defines an edge matching
rule that still allows random-tiling freedom~\cite{binary}.)
The second closest possible separation of cluster centers  
is $1.176 \tau^4 a_0$ (the long diagonal of the Thin Penrose rhombus).  
The \Star~ clusters go on the Small sites of the Binary tiling.

\LATER{The best {\it a posteriori} justification for the Binary tiling network 
(at the compositions we are exploring) 
is that it perfectly describes the \Dec~ arrangement in the
$W$-AlCoNi approximant phase (see Sec.~\ref{sec:W}).}

\remark{Nan had formulated the Decagon-linkage story in
  terms of small ring-2 (2.45A-edge) decagons + 2.45A-stars, 
  as "competing" -- in the sense that the system almost  wants
  to phase separate between these.
  ``What is actually being maximized is a weighted sum of the number
of small decagans plus CoAl$_9$ motifs.''
  (From that viewpoint, the 13A Decagons are a sort-of accidental phenomenon:
  one might easily expect to  find an occasional
  small decagon surrounded by an incomplete 13A Decagon.)
   I rewrote that  story to
  say the small decagons and 2.45A-stars are "complementary" parts
  of how the energy gets minimized -- i.e. they are attracted, 
  rather than a phase-separation which is frustrated by packing
   geometry.
  (That implies the "real" cluster is actually a 13A Decagon 
EXTENDED to include those 2.45A-edge Stars.)
  CLH don't think this was fully resolved.}

\MM{(9/05) suggests that, in the 4A fixed-site simulations,
we're lucky that CoAl$_9$ clusters formed robustly at all:
apparently that was a consequence of the medium-low density,
that prefers HS over BB pairs (2.45A).
In 4A simulation, MM would say the
question what's the real story is a subtle one.}
\CLH{to MM: What alternative are you referring to that
you think competes with CoAl$_9$ clusters?  What is in
the BB pairs, don't they also have CoAl$_9$ clusters???}

It should be noted that,
in a small or moderate-sized system, the choice of periodic
boundary conditions practically determines the network of \Decagons~
(assuming the number of them is maximized). Consequently, in some
unit cells the placement of \Dec~linkages is frustrated, while in others
it is satisfied.  Those cells mislead us, obviously, regarding
the proper linkage geometry; worse, they may mislead us at the prior
stage of identifying cluster motifs.

Thus, is conceivable that certain system shapes
would favor or disfavor the appearance of \Dec~ clusters, opposite to
the infinite-system behavior at the same composition and density.
If (as is likely) a significant bit of the cluster stabilization
energy is in the linkages, and if there is a competing phase based on 
other motifs, then the frustration of \Dec~ linkages 
in a particular cell might tip the balance toward the phase based on
the other geometry.
\remark{Thus the apparent phase boundary would have
shifted by the boundary conditions, a species of finite size effect.}

These considerations show why it was important, even in the earliest
stages of our exploration, to explore moderately large system sizes 
(a too small system would not even contain a motif as large
as the \Dec); and also why it must be verified that
results are robust against changes in the system shape (i.e.
periodic boundary effects).
To address this issue, we ran additional simulations on the
20$\times$38$\times$4 cell (Table~\ref{tab:cells})
%%% 37.62\AA{}$\times$19.78\AA{} $\times$4.08\AA{} 
with the same volume and atom content 
as our standard 32$\times$23$\times$4 simulation.
\remark{When the atom
content is exactly the same, it is valid to compare energies per atom
between the two tilings, without the need to construct tie-lines etc.}
%%%%%%
The lowest energy configurations on this tiling also 
maximized occurrences of non-overlapping \Decs, although
extensive annealing was needed (see App.~\ref{app:overlap-linkage}).

\LATER{CLHNEW:  MM (2/06) pointed out it will be hard for the
reader to visualize these networks.
This should be coordinated with the planned improvements
to Fig.~\ref{fig:d-adjoin}.}

In sections \ref{sec:ideal-deco} and \ref{sec:decLRO}, we shall
consider decoration rules that divide either the \Decagon~ network,
or the \Star~ network, into even and odd clusters.  It should be 
recognized that the Binary tilings that fit in the cells 
in Table~\ref{tab:cells} are non-generic from this viewpoint.
The \Dec~ network has no odd-numbered rings (if it did, that
would frustrate any perfectly alternating arrangement of 
cluster orientations).  A corollary is that 
the \Stars~always appear in (even/odd) pairs:
there is never an isolated Star cluster, or any odd grouping.

\LATER{CLHNEW: Is this redundant with -- 
should it be combined with -- Subsec.~\ref{sec:fixed-site-20A}?}

As detailed in Appendix~\ref{app:20A}, 
recent structural studies~\cite{oley06,abe06-ICQ9}
and simulations~\cite{Hi06} suggested a
cluster geometry based on even larger clusters, with
linkages $\tau$ longer than the edges in our tiling.
Our initial simulation cells, although much
larger than those used previously for the ``basic Ni'' 
phase~\cite{alnico01}, were too small to discover such
a cluster. 
Apart from the possibility of the \PB~ cluster motif 
(Sec.~\ref{sec:W-PB}), the atomic structure of the large cluster models
is very similar to ours; in particular, the 20\AA{} cluster
is just a \Decagon~ with two additional rings
Consequently the large-cluster model
must be quite close in energy to ours (and to a family of similar
structures), so it is amazing that a clear pattern 
(as we found in Appendix~\ref{app:20A})
can ever emerge at the 2.45\AA{} stage, even with annealing.
We cannot decide at present which 
structure is optimal for our potentials.

\subsection{Relating structure to potentials}
\label{sec:struct-pot}

\LATER{Rewrite these to refer to ``$V_{\rm AlCo}$'', etc.''}

In this subsection, we pause to rationalize the stability of the motifs 
and structural features identified so far,
in terms of the ``salient features''  of the
pair potential interactions 
(noted in Sec.~\ref{sec:pairpot}).  
Notice that, at this stage, {\it no} assumptions need
be made that atoms in these clusters have stronger energetic
binding compared to the other atoms.~\cite{ICQ9-disc}
To use clusters as a building block
for subsequent structural modeling, it suffices that they are the
most frequent large pattern appearing in the structure.
(It is convenient if the clusters have a high local symmetry, too.)

We start out explaining some general features .
First, the TM (mostly Co) atoms are
positioned $\sim$4.5\AA{} apart, right at the minimum of the second
(and deepest) well of the potential $V_{\rm Co-Co}(r)$.  Second, every
Co atom has as many Al neighbors as possible -- nine or ten
-- and as many of those at a distance $R_1 \approx$2.45\AA{}, 
close to the particularly strong minimum of $V_{\rm Al-Co}(r)$
(see Table \ref{tab:potentials}).
Such coordination shells are, roughly, solutions of the
problem of packing as many Al atoms as possible 
on a sphere of radius $R_1$,
subject to the constraint of a minimum Al-Al distance (hardcore radius)
of $R_0=2.6$\AA{} to 2.8\AA{}, which is an fair idealization
of the potential $V_{\rm Al-Al}(r)$.
Furthermore, since every TM atom is maximizing its coordination by Al atoms,
TM-TM  neighbor pairs are as rare as possible (and they usually
involve Ni, since the Al-Co well is deeper than the Al-Ni well).
These features are
also true of the ``basic Ni'' phase and other Al-TM compounds.

\remark{This packing problem is equivalent to packing disks of angular radius
$\sin^{-1}(R_0/2R_1)= \pi/5$ on the surface of a sphere. 
To demonstrate that the maximum number of disks is nine,
let us divide the unit
sphere's surface into (disjoint!) Voronoi domains.  The minimum area
of each disk's Voronoi domain is $4\pi/9.466$ (attained when it is a
spherical pentagon of inradius $\pi/5$.)  The the sphere's area is
$4\pi$, so no more than 9.466 disks can be packed.
In the Al$_9$Co coordination shell, the two Al atoms above and below 
the pentagon's plane 
have separations $R_{\rm Al-Co}= R_{\rm Al-Al} = 2.57$\AA{}, 
so our assumptions do not hold strictly.
But in the relaxed configurations (see Sec.~\ref{sec:relax}), 
the shell after the
Al pentagon atoms pucker is close to a valid hard-disk packing.}

\LATER{Check references about Co-Ni chemical order.}

Based on an electron channeling technique called ALCHEMI,
it was claimed~\cite{sai00} that for 
a d(Al$_{72}$Co$_{8}$Ni$_{20}$) alloy, the Ni and Co atoms are
almost randomly mixed on the TM sites.
Modeling studies (whether of that Ni-rich composition~\cite{alnico01},
or the results in this paper for the Co-rich case) suggest that,
on the contrary, substantial energies favor specific sites for Ni and Co
so the structure is genuinely ternary, not pseudobinary.
%%%%%%%%%%%%%%%
\remark{Our studies, of course, are at $T=0$.  Maybe they could be 
mixed at higher $T$, though I doubt it: Al-TM is the strongest 
potential well, and Al-Co is stronger by a factor 1.5 than Al-Ni.
The issue will be settled by neutron diffraction, which Steurer
hopes to do at ISIS (see email 7/05 after my visit to him).}

To rationalize the $d$-AlCoNi structure in a more detailed way, we
must recognize it is locally inhomogeneous in a sense:
it is built from two kinds of small motif -- small decagons
plus Al$_9$Co clusters -- which have quite different 
composition and bonding. (A third small motif that is neglected by
our approach is the $W$-AlCoNi pentagon, a kind of pentagonal bipyramid
cluster, which will be discussed in Sec.~\ref{sec:W}.)
To explain these smaller motifs, we must anticipate part off
the descriptive framework of Sec.~\ref{sec:ideal-deco},
in which a 2.45\AA{}-edge ``HBS'' tiling will be introduced.

\remark{Presumably,  if there was a good way to
make packings with one kind of motif, the system would phase separate.
It does not, and in instead the disparate motifs pack together to form 
larger motifs (such as the \Dec): why?
One hypothesis is that packings of just one kind of motif are frustrated
geometrically; an alternate theory would be an attractive interaction
between the two types of motif, which is a necessary component of
the energy stabilizing the mix.}

\FUTURE{One cannot, {\it a priori}, rule out the possibility of a phase  
separation into a Decagon-rich and Star-rich structures.}

\subsubsection{Small decagon}

\remark{In principle, since $R_{Co-Co}\approx 4.5$\AA{} is very nearly 
$\tau$ times  
a typical atom separation, it should be possible to 
place Co atoms almost with  
uniform spacing $R_{\rm Co-Co}$  and 
without any Co-Co contacts, up to Co content  
$\tau^{-3}\approx$
23.6\%.   Possibly, though, one only accomodates $\sim$ 13\% Co
if one is using the CoAl$_9$ shell (below, Sec.~\ref{sec:CoAl9}).
In any case, it is somewhat surprising to find the pentagon
of Co neighbors.}

The heart of the \Dec~ is a smaller decagon (edge $a_0=2.45$\AA{}),
bounded (in projection) by Al atoms (ring 2).  This motif seems to 
be particularly characteristic of Co-rich structures. 
Despite the strong tendency to avoid  Co-TM pairs, as 
mentioned just above, this cluster has a ring of {\it five} Co neighbors.

Our best explanation is that a conjunction of several Al's is required
in order to compress the Al-Al bonds as short as 2.57\AA{}, but that
is advantageous since it allows the Al-Co bonds to be correspondingly
shortened to 2.45\AA{}, the bottom of the deep Al-Co potential.  The
site-energy map (see Fig.~\ref{fig:siteE-ideal}) shows that the
interactions of the ring-1 Co
atoms are not very well satisfied, compared to ring-3 Co atoms. 
On the other hand, the ring-1 Al and (especially) the central Al 
are well satisfied.

\subsubsection{The CoAl$_9$ coordination shell}
\label{sec:CoAl9}

This motif consists of a pentagon of
five Al atoms centered by Co in one layer, capped by two more 
Al atoms in the layer above and two in the layer
below, so the Co atom has coordination 9 by Al.  
(A complete pentagon of this sort is centered in each of the
2.45\AA{}-edge Star tiles visible in Fig.~\ref{fig:RulesWideal}.)
Actually, this motif is almost always surrounded 
(in projection) by a larger pentagon of five TM, lined up
with the Al pentagon, but we shall not treat these
Co as part of the motif. They are (often) centers
of neighboring CoAl$_9$-type clusters, as described
in the next paragraph.

Two CoAl$_9$ motifs might be packed by joining the pentagons
with a shared edge (two shared Al), but that would create an energetically
unfavorable Co-Co distance ($R_{\rm Co-Co}=3.96$\AA{}).
If instead two pentagons shared a corner (one Al at the 
midpoint of the Co-Co line), then 
$R_{\rm Co-Co}=4.9$\AA{} which is also disfavored.  
The only way to achieve a
favorable $R_{\rm Co-Co}\approx 4.5$\AA{}
is to place the two Co in different layers,
with some Al atoms from the pentagon around one Co
capping the pentagon around the other Co, 
and vice versa.  
That is, more or less, the arrangement found around the perimeter of 
every \Decagon~ cluster: 
Finally, if CoAl$_9$ motifs on the perimeters of two \Decs~ are
shared, it corresponds to an edge-sharing linkage,
and the centers will be $12.2$\AA{} apart, 
consistent withthe 10.4\AA{}-edge Binary tiling (Subsec.~\ref{sec:struct-pot}).

The CoAl$_9$ motif was equally important in 
the ``basic Ni'' phase~\cite{alnico01}.

\LATER{merge preceding observation
with the justifications of linkages by potentials???}

\subsubsection{Site energies map}
\label{sec:siteE}

\remark{MM 9/05. We did not use  the site energy here as a 
``diagnostic'', but in retrospect this kind of figure 
is very informative, for example when comparing 4A and 8A results,
as the worse 4A model atoms turn into reasonably happy atoms after
8A relaxation. (So we were somehow lucky that these bad sites were
tolerated in 4A simulation).}

A diagnostic which was useful in prior investigations using pair
potentials~\cite{Mi96b} is the ``site energy'' for site $i$,
   \begin{equation}
        E_i = \frac{1}{2} \sum _j V_ 
{ij}(R_{ij})
   \end{equation}
   where $V_{ij}(R)$ is the proper potential for the species occupying
   sites $i$ and $j$, and $R_{ij}$ is their separation.

\begin{figure}
\includegraphics[width=3.1in,angle=0]{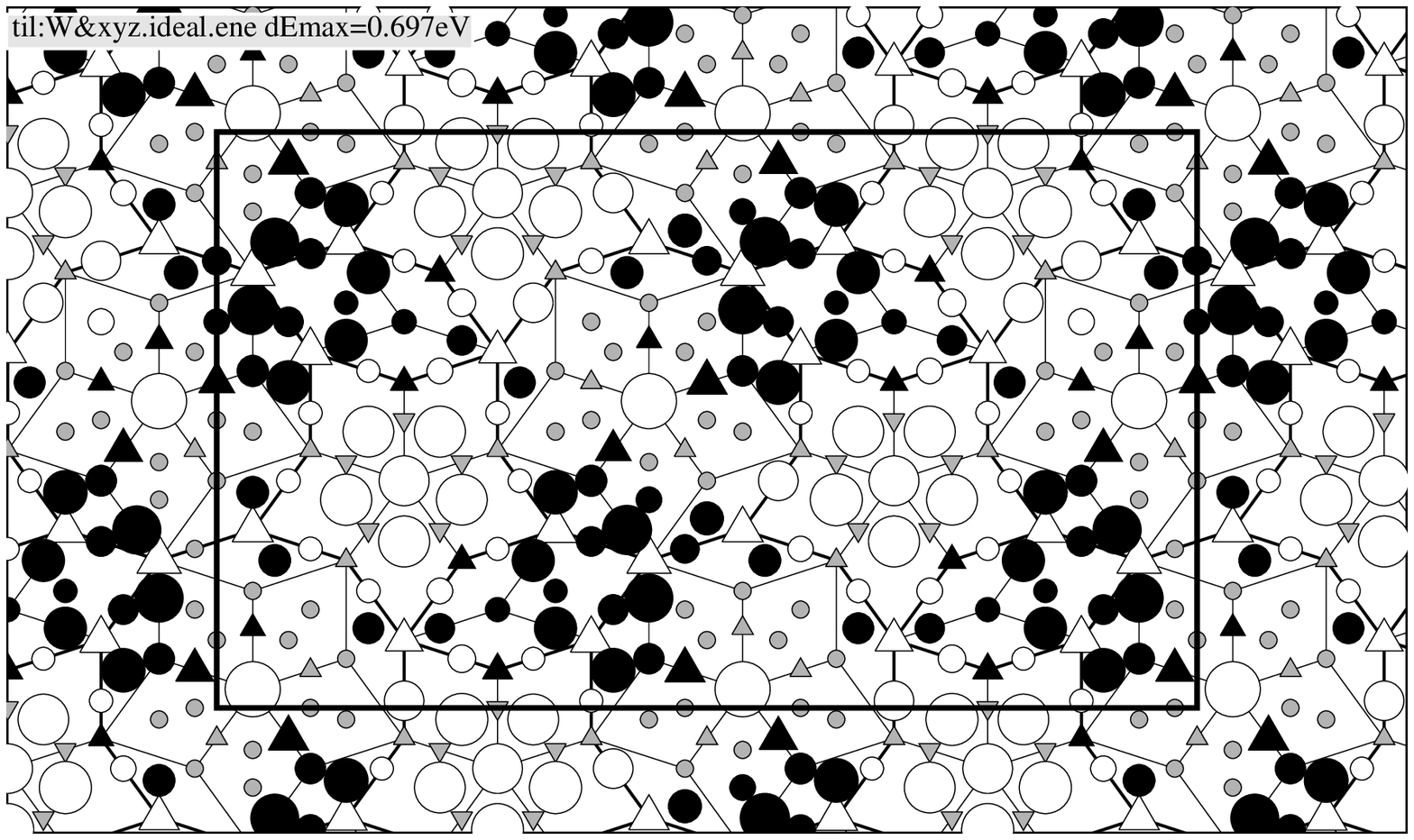}
\caption{Site energies for ideal sites in the configuration of
  Fig.~\ref{fig:RulesWideal}. 
Here black (white) filled symbols indicate a disfavorable (favorable)
deviation, compared to the mean for that species.
Circles are Al, up-pointing triangles are Co, and down-pointing
triangles are Ni.
%%%%%%%%%%%%%
\remark{The convention here is black = positive (high) E.}
%%%%%%%%%%%%%
\LATER{MM 9/05 (responding to a query about this figure 
referring ahead).  MM can easily create a new figure,
that would be relevant for this stage. 
MM would say such a figure is more relevant for simulation created samples
(as here)  than for the idealized model.} 
}
\label{fig:siteE-ideal}
\end{figure}

   It is revealing to plot $E_i$ graphically
   (Fig.~\ref{fig:siteE-ideal}).  
The symbols represent each atom's site energy minus
the average (over the cell) of the site energies for that species,
which is our crude surrogate for the chemical potential.  
The energies are strikingly non-uniform between different
places in the structure.
\LATER{``(The atomic configuration is an idealized one
from Sec.~\ref{sec:ideal-deco}, which has   total energy 
than any other fixed-site structure we found.'' CHECK IS THAT TRUE?}  
An extremely good site energy is obtained for the Al atoms
in the even \Star.  The Co atoms on the \Dec~perimeter, as expected, 
are much more satisfied than those in the interior.  
The variable Al atoms in the \Dec~ are the least satisfied,
also as expected. The overall picture was not
very different when this diagnostic is applied to
configurations that, after MD and relaxation,
developed puckering with the variable Al entering ``channels'' 
(see Sec.~\ref{sec:relax})

The configuration shown is taken from the idealized structure model
of Sec.~\ref{sec:ideal-deco}.
There is a strong contrast between good and bad Al sites;
(this is reduced but not eliminated by
relaxation and molecular dynamics as in Sec.~\ref{sec:relax}).
Bad energies are often seen in neighborhoods which are
somewhat ``overpacked'' by Al atoms;
when MD is performed (Sec.~\ref{sec:relax}), Al atoms are
observed to run from these sites to other places
which are missing Al atoms.
(For example, 
in the 2.45\AA{}-edge DHBS tiling of Sec.~\ref{sec:ideal-deco}, 
three adjacent 2.45\AA{} Boats is overpacked; if two of them are
converted to the combination Hexagon + Star, energy could 
be lowed by puckering as in Sec.~\ref{sec:relax}.)
\FUTURE{Mention the Al-potential-map as a 
good way to locate places which could take another Al?}

\subsection{4\AA{} rhombus simulations}
\label{sec:MC-4A}

\LATER{We flip back and forth between saying $a_0$/$\tau a_0$
  and 2.45\AA{} / 4.0 \AA{}.  Let's use $a_0$, and remind the reader
  oftener.}

The $a_0=2.45$\AA{}-tile simulations are inadequate, 
to resolve further details of the atomic structure, such as
the exact occupation of ring 2.5/3, or the interactions between \Decs,
as these are decided by small energy differences  that get
overwhelmed by the frequent incorrect occupancies at this level.
A new simulation is needed using larger tiles and with a site
list reduced as guided by the 2.45\AA{}-tile results
It might have been appropriate to try an edge $a_0$ hexagon-boat-star
tiling (as done in Ref.~\onlinecite{alnico01}). However, we chose to go
directly to an inflated rhombus tiling with edge $\tau a_0 \approx
4$\AA{}, which is convenient for decomposing the \Decagons (as they
have the same edge).

The starting point  is that space is tiled with large
${\tau}^3 a_0$ edge length rhombi in a binary tiling.~\cite{binary}
The Large disk vertices (which have a local ten-fold symmetry) are 
then the centers of the \Dec, as argued in Subsec.~\ref{sec:Dec-relation}.
On the actual
simulation scale ($\tau a_0\approx 4.0$\AA{}), each \Dec~is represented
by five fat rhombi arranged in a star, with five thin rhombi surrounding 
them to form a decagon with five-fold symmetric contents. A decagon in the
4.0\AA{} scale rhombus decomposition is shown in
Fig.~\ref{fig:dds}(b).

\LATER {MM: Cite Fig.~\ref{fig:tileflips}, once it 
is augmented to show the 4A tiles with their sitelist?
MM 9/05 later remarked, it doesn't seem essential to show such figure.}

As compared to the $2.45$\AA{} site list, (i) instead of having
independent tilings in the two layers, we now have just one; (ii) the
alternation in layers between the sites separated by a $2.45$\AA{}
edge is now built in; (iii) there are not many places where the site
list allows even a possibility of close distances; (iv) a large
fraction of the candidate sites get occupied -- the only question is
which species.  Thus, the $4.0$\AA{} site list is partway to being a
deterministic rule.
It should be emphasized that this 4.0\AA{} decoration is {\it not}
well defined on an arbitrary rhombus tiling since the inflated
(${\tau}^3 a_0 $) tiling must follow a binary tiling scheme. The
decoratable tilings are a sub-ensemble of the rhombus random tilings.

The rhombi outside the \Decs~ are grouped into  4\AA{}-edge Hexagon, 
Boat, and Star tiles (a \Star~ is centered on the interior rhombus 
vertex of each of these tiles).  For example, towards the left
side of Fig.~\ref{fig:d-adjoin}(c), two 4\AA{} Stars are seen
with an overlap (shaped like a ``bowtie'') that is resolved
by converting either one to a Boat.  This decoration of the 
4\AA{} rhombi produces a slightly different sitelist,
depending on which way such overlaps are resolved, 
but this did not seem to make a difference for the sites
which are actually occupied.~\cite{FN-sitelist}
We shall occasionally refer to this version of the
4.0 \AA{} rhombus tiling as the ``4.0\AA{} DHBS'' tiling.

\remark{MM 2/06: Originally, the site-list for the sparse-sitelist
4A decoration for DHBS was NOT symmetric with respect to rhombuses,
as it was intened to use minimal number of sites needed, as concluded
from the low-T 2.45A configurations. But later on (App. on 20A decagons)
Nan symmetrized the site list, with the surprising result.}

The $\tau a_0$ scale tiling and site list can
be naturally deflated back to $a_0$-edge Penrose tiles,  and
these in turn can always grouped into a $a_0=2.45$\AA{}-edge 
Decagon-Hexagon-Boat-Star (DHBS) tiling 
%%% that approximately obeys the five-fold and screw ten-fold symmetries. 
which is used in Sec.~\ref{sec:ideal-deco} and later as a 
basis of description.
\remark{In Sec. IV to pin down the fixed-site behavior of ring 2.5/3,
and later as a way of considering the effects of relaxing the fixed site list.}

This stage of Metropolis simulation uses only atom swaps,
tile flips being disallowed (they would almost always be rejected).
We enforce a reduced site list, but do not fix any occupation:
any atom (or note) may occupy any sit.
%%%%
The initial inverse temperature was typically $\beta = 10$ or $\beta = 20$, 
and increments were $\Delta \beta = 1$ or $0.5$ in the 4.0\AA{}-tile
annealing runs.~\cite{FN-4Aanneal} 
(Higher temperatures are not needed since in
the $4.0$\AA{} simulations, it is very easy for TM atoms to find their
`ideal' sites.) The reduced temperature makes the Al occupancy less
random than before.

\subsubsection{Use of toy Hamiltonian to generate tilings}
\label{sec:MC-4A-tileH}

\remark{In this discussion, we may write ``\Dec'' as an abuse of
language, denoting the ``star-decagon'' pattern of five Fat and five
Thin rhombi with five-fold symmetry.  But these always did in fact 
get occupied by \Decagon clusters -- at least, in all systems small enough
that atomistic lattice-gas simulations could be carried out on the
4\AA{} tiling.}

To generate appropriate tilings of 4.0\AA{} 
as a basis for these second-stage lattice-gas simulations, 
we performed pure tile-flip MC simulations using an
artificial ``tile Hamiltonian'' as a trick.
The main term in the Hamiltonian 
was $-N_{\rm star}$: here $N_{\rm star}$ is the count of 
star-decagons of 4\AA{} rhombi that 
are bound to ``level 0'' ($\nu=0$) sites, using the nomenclature of
App.~\ref{app:code}. (The level 0 condition ensures that 
such decagons cannot overlap, but only share edges.)

In effect, then, we are maximizing
the density of  non-overlapping \Decagons, with the constraint
that the spaces between \Decs~ are always tiled with 
4\AA{}-edge HBS tiles.   Every resulting tiling
(even in very large cells) was always a Binary tiling 
with edge $\tau^3 a_0 = 10.4$\AA{}
as described in Sec.~\ref{sec:Dec-relation}, with a star-decagon
on every Large vertex and a star of five fat rhombi on
every Small vertex.  We conjecture that maximizing the
frequency of non-overlapping star-decagons {\it rigorously} 
forces a Binary supertiling;
many other examples are known in which maximization of
a local pattern leads to a (random) supertiling, decorated with
smaller tiles.~\cite{jeong94,Hen98}

\remark{By the way, our requirement that Penrose rhombi fill space -- in
particular, the area between \Decs~ (strictly speaking, between
star-decagons of rhombi) -- handily excludes the energetically
unfavorable \Dec~ relationship in Fig.~\ref{fig:d-adjoin}(b), where
the interstice has the shape of a bow-tie.}

There is a large ensemble of degenerate ground states of 
this Hamiltonian, which differ (i) in the Binary tiling
network, and (ii) the detailed filling of the 4\AA{} HBS
tiles between the star-decagons.
Additional terms were used to remove the second kind of 
degeneracy so that every Binary tiling was still degenerate, 
but there was a unique (or nearly unique) decomposition of
every Binary tiling configuration into 4\AA{} rhombi.

In the Binary tilings\cite{binary}, the Small vertices may occur
isolated, but most often form chains.  In the ``half-W'' (or W) unit
cell (Fig.~\ref{fig:d-adjoin}(d), the chains are unbounded (extending
in the $y$ direction), whereas in the 32$\times$23 cell and also the
20$\times$38 cell the chains are just two vertices long.
\remark{The surrounding \Decs~ form a fat hexagon in that case.}

\FUTURE{MM 6/05 did a new simulation favouring Decagon+Star.
``Originally, I thought of this as 2.45 edge length. But it also makes
sense with 4A edge length.''  Need to get report on the results.}

\subsubsection {Results of $4.0$\AA{} edge simulations}
\label{sec:4A-results}

%%% Added from a message of MM c 12/18/05}
The post-hoc justification of the 4.0\AA{} tile decoration is
is that its configurations have an energy typically
about 0.006 eV/atom lower than a 2.45\AA{} result such as
Fig.~\ref{fig:smalltiles}, even though it has a {\it reduced} site list.
(These lower energy configurations were found in less time and 
at a lower temperature, too, than on the 2.45\AA{} tiling.)
On the $20\times 38$ tiling, the actual low energy configurations
found after a 2.45\AA{}-level run of long duration are similar to the
those in Fig.~1 of Ref.~\onlinecite{Gu-letter}, which
was created from $4.0$\AA{} simulations.  

\CLH{to Nan: The preceding sentence was inherited from an 
earlier draft -- but why say the 20x38 tiling is like that in
Letter fig 1? (that is, in what way -- what feature -- are
they similar?)}

This suggests to us that this limited ensemble includes
all of the lowest-energy states of the original ensemble;
the removal of some sites simply keeps the MC from getting
stuck in local wells of somewhat higher energy.
The most problematic issue of local environments
excluded by the site-list reduction was the ``short''
Al-Co bonds, discussed in Appendix \ref{app:short-bonds}.

We found the \Decagon~ to be  robust, forming 
in our usual 32$\times$23 cell
over a range
of compositions Al$_{0.7}$Co$_{0.3-x}$Ni$_{x}$ for
$x=0.05$ to $0.15$ (with the standard density), and
also over a range of atom densities 0.066 to 0.076.  \AA{}$^{-3}$
(at the  standard composition Al$_{0.7}$Co$_{0.2}$Ni$_{0.1}$). 
(These were later checked by simulations with the same atom content
on the 4.0\AA{} scale tiling of Subsec.~\ref{sec:MC-4A}.)
In the ``$W$(AlCoNi)'' unit cell, \Decs~ were checked to apppear at 
densities 0.069 to  0.072 with composition Al$_{0.718}$Co$_{0.211}$Ni$_{0.071}$.
Additionally, we confirmed \Dec~ formation when the potentials were cut off
at radius 10\AA{} as well as the standard 7\AA{},
or with standard conditions in every unit cell from Table~\ref{tab:cells}.

\remark{Nan also noted Temperature variations [in K]: (2900), 580 
and presumably all intermediate values.}

\remark{Nan put the exact energies from the original density and composition 
tests in $\sim$ngu/Summer04/Comptest/results and
  $\sim$ngu/Summer04/PDtest/results.}

We can now go beyond the idealized description of idealized clusters, to
note some tendencies for variations (especially the TM placement).
Although these may be expressed in the language of a rule,
they are at this point only statistical biases
(primarily based on our 23$\times$32$\times$4 unit
cell with our standard composition and density, 
and mostly using simulations on the 4\AA{}-tiling site
list of Sec.~\ref{sec:MC-4A}.)
% as described in Sec.~\ref{sec:MC-a0}.  
Only in Sec.~\ref {sec:ideal-deco}
will these observations be turned into actual rules.

\remark{Nan comments that the reduced site list simplifies the
task of recognizing recurring patterns:
  on the 2.45\AA{}, several configurations may try to
  represent the same physical configuration.
On the other hand, MM criticizes any ``explanation''
based on specific bond lengths available in the sparse
site list, since competing configurations have already
been hidden.}

Although we presented rings 2, 2.5, and 3 as having 10-fold
symmetry, that is an oversimplification and many site occupations
get modulated according to the orientation of the core TM pentagon;
(Thus it will not be surprising that a long-range order of
the orientations develops, as detailed in Sec.~\ref{sec:decLRO}.)
In particular, the ring 3 Al atoms along the decagon's edges usually are 
placed in a layer different from that of the ring 1 TM atoms, which 
means that  (in projection) these Al are alternately displaced
clockwise and counterclockwise from the bond center.
However, whenever Ni occupies a ring-3 TM site, both the adjacent
ring-3 Al atoms tend to adopt the sites in the opposite layer, at a
distance of 2.54\AA{} from the Ni, regardless of the
core orientation. (Note the adjacent ring-3 TM sites are
very likely Co, and this displacement puts the Al-Co distance 
to 2.45\AA{},  nearly the bottom of the Al-Co well which
beats the the Al-Ni attraction.)
Finally, if we draw a line from the center of a \Dec~through an Al atom in
ring 1 and extend it through the vertex of the \Dec, the site immediately
outside of the decagon along this line (in projection) 
has a preference for TM with very strong tendencies towards Ni.  (If not 
occupied by Ni or Co,  such sites are most often Al rather than vacant.) 
This induces a relationship between the core orientation
and the placement of the \Stars~ that are richer in Ni.

Changes in the net Al density -- forced, in our simulations,
when we changed the overall density while keeping stoichiometry
constant -- are accommodated by the 2.5 ring.
(The \Star~ is less flexible: it has a fixed number of atoms.)
To anticipate Sec.~\ref{sec:ideal-deco}, the ring 2.5/3 Al's
can be alternatively described as the vertices of a 
Decagon-Hexagon-Boat-Star tiling with edge 2.45\AA{}, 
and the Al count can be increased by replacing Hexagons and Stars
by Boats.

\subsection{Effects of TM composition changes}
\label{sec:Co-Ni}

\CLH{Question: is this section too short?  Are sections E-F-G
divided  and ordered properly?}

The TM sites in the \Decagon (found in ring 1 and ring 3) are
normally Co (and otherwise are always Ni)
This was checked by a special series of lattice-gas Monte Carlo runs
in which only Ni/Co swaps were enabled; this confirmed a Co
preference in ring 1.
However, when there is an excess of Ni atoms -- because either the
Ni fraction or the overall density has been increased --
Ni atoms start to appear in ring 2.5 of the \Dec~
(in which case the nearby Al atoms behave somewhat differently
from their regular patterns). Excess Ni atoms even
enter some ring 1 TM sites, in which case the
neighboring ring 2.5 (Al) sites are less likely to be occupied
(as expected, in light of the powerful Al-Co potential).

When Ni atoms are added at the expense of Co, they typically substitute
first for Co(3) on the boundary of a \Dec, on sites adjacent to Ni of a \Star.
This presumably disrupts the puckering units that would otherwise
be centered on (some of) those Co's.  

\begin{figure}
\includegraphics[width=1.85in,angle=0]{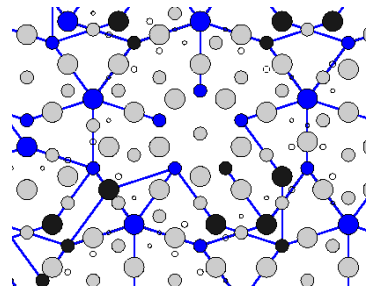}
\caption{[color] ``Arrow'' configuration at high Ni density.
This pattern is seen at lower right, on the edge of
a \Decagon~in which one Co(1) from the inner ring has
been converted to Ni.
The overall atom density (0.074 \AA$^{-3}$) is somewhat above
the physical range.  At a more realistic density, it appears
the same ``arrow'' configurations and Ni substitutions occur, 
but much less frequently.
The lines in this figure connect pairs of TM atoms in different
layers and separated by 4\AA{} in the $xy$ plane.
The color conventions for species  are the same as 
Fig.~\ref{fig:smalltiles}.
%%%%%%%%%%%%%%%%%%
\remark{Nan advised. ``For good examples see
/afs/msc.cornell.edu/home/henley/ngu/
Spring05/orientations/ferro/Run5
Just choose any of the \#.ps files in there.''
This is file 31.ps, cropped using ``gimp''.}
%%%%%%%%%%%%%%%%%%
}
\label{fig:Ni-arrow}
\end{figure}

We observed how Ni atoms are incorporated {\it without} decreasing Co,
when the atom density was varied while
the same lattice constant and the standard composition
Al$_{70}$Co$_{20}$Ni$_{10}$ were maintained.
%%%%%%%%%%%%%%%%%%
\remark{Since the TM superlattice arrangement
seemed to be largely governed by TM-TM interactions, CLH speculates
the results would be similar if we increased Co and Ni at the 
expense of Al, or if we simply added Ni atoms. }
%%%%%%%%%%%%%%%%%%
In this case, Ni atoms typically enter 
ring 2.5 in the \Decagon, creating a local pattern of TM occupations that
we call the ``arrow.''  This is convenient to describe in the language
of the 4\AA{} rhombus tiling.  Say that a \Dec~ corner site is lined 
up with a Co(1) [ring 1] site and occupied by Ni,
and also has am Ni nearest neighbor in an adjacent \Star: call
these sites Ni(3) and Ni($s$), respectively.  Then additional ``Ni(2.5)''
sites appear inside the \Dec, in the same layer as the Ni(3).
The head of the ``arrow'' is the 72$^\circ$ angle 
that Ni($s$) makes with the two Ni(2.5), as in Fig.~\ref{fig:Ni-arrow}.

\LATER{The Star cluster shown in the figure would be an Odd
Star, in view of the Decagon orientation.  Recall the Co(1) 
atoms are destined to be mirror (flat) layers
in Sec.~\ref{sec:relax}), when atoms are allowed to pucker
out of the ideal sites.  I need to scan the figure files 
to check whether the layer/orientation behavior in the
Figure is truly typical... The added Ni form a pentagon
all in the puckering layer, with two Al(1) and Ni(3).
See email CLH to Nan 1/10/06.}

The five TM's [Ni(3) + 2 Ni(2.5) + 2 Co(1)] form a regular pentagon, 
centered on the Al(2) of the same layer.  This Al(2) is also surrounded
by Al(1) + 2Al(2)  + 2 Al(3) in the other layer,
so the combination is an Al$_6$(TM)$_5$ just like the core of a
\Decagon, except that only two of the TM's are Co, and also the
pentagon of five Al's is quite distorted in this case.  

%%%%%%%%%%%%%%
%%%%%%%%%%%%%%%%%% 
%%%%%%%%%%%%%%%%%%%%%%%%%%%%%%%%%%%%%%%%%%%%%%%%%%

The density threshold, above which ``arrows'' appear, 
was 0.068\AA{}$^{-3}$  for the 32$\times$23 tiling, and
0.071\AA{}$^{-3}$ for  for the W-cell tiling.
The difference might be due to our 
enforcing the standard stoichiometry  in both cells, 
although the ideal Ni:Co ratio must differ since 
the \Star:\Dec~ ratio for these cells is, respectively
1:1 and 2:1.
\remark{The \Star/\Dec~ ratio is 2:$\tau$ in the 10-fold tiling.}

Ni atoms are very often found in \Stars.
When two \Stars~ adjoin, it makes a pair of candidate-TM sites
from the respective \Stars, and these often form a Ni-Ni pair.
However, these candidate-TM sites have a large number of Al 
neighbors, hence one of these is viable site for Co occupation
(in which case the other becomes Al).
In general, $\sim$2.5\AA{} TM-TM bonds, wherever they are found,
will usually be Ni-Ni since this maximizes the number of 
$\sim$2.5\AA{} Al-Co contacts (recall the Al-Co well is
deepest, Al-Ni being only the second deepest well)
\LATER{to MELD:
\Stars~ often come in contact with
one another; then a possible-TM site from one \Star~ is $2.54$\AA{}
away from such a site of the adjacent \Star.  This atom pair is found
usually to be Ni-Ni, but Al-Co bond occurs sometimes.}

\subsection{Comparison to Ni-rich decoration}
\label{sec:bNi-compare}

\FUTURE{MM asks: does Co-rich have more Al-TM contacts than Ni-rich?
  CLH notes that Sec.~\ref{sec:CoAl9} would imply the number of 
contacts depends on the ratio $R_{Al-Al}/R_{Al-TM}$, 
but this ratio isn't different between Co and Ni.  
MM rejoins, that's the point: 
small differences can make big and trasparent consequences 
at medium range scale!
The question is, why does Co want to either have
no close TM neighbor at all (like in Ni-rich phase);
or when it must have TM neighbors as at higher Co content,
instead of forming pairs with one close TM-TM contact, 
(like Ni in Ni-rich), they instead form
pentagons with 2 TM-TM close contacts? 
It's hard to see a {\it simple} reason for that 
arising from pair potentials.}

In this subsection, we compare our present results to previous work 
on the ``basic Ni''.  phase~\cite{alnico01,alnico02,alnico04}.

Our path at this point is actually somewhat different from that taken
for the basic Ni structure~\cite{alnico01}. In the case of ``basic
Ni'', a hexagon-boat-star (HBS) tiling with a
$2.45$\AA{} edge length was used in the analog of our second stage
simulations. This was followed by a third stage using an inflated HBS tiling 
with edges $\tau^2 a_0\approx 6.5$\AA{} (with a deterministic decoration).
That description was simple, because (to a good approximation) 
the decoration was context-independent, i.e. has the
same approximate energy independent of which tiles were adjacent.
Specifically, all edges were decorated in the same way, and there was
no strong constraint relating the Al atoms in the tile interiors to
the surrounding tiles.
This rule was checked by a simulation at the ideal composition, and the
resulting configurations  were
identical to the ideal decoration, apart from one or two
defects per simulation cell.  

That template cannot be completely transferred for our
Co-rich phase. In this case, it is harder to neglect instances of
Co/Ni substitution.  In particular, though the TM atoms on the
boundary of the decagon object should be idealized as Co, there are
special environments in which they clearly are converted to Ni, which
introduces a context-dependence into the decoration.  Also, there are
complicated rules for Al atoms around the outer border of the decagon
(i.e. in rings 2.5 and 3), as well as for the occupation of TM atoms
in the five candidate sites of the \Star. These degrees of freedom
interact with the tiling geometry, as well as each other.  

Our choice for ``basic Co'' second stage was to go directly to the
4.0\AA{} DHBS tiling of Sec.~\ref{sec:MC-4A} (which is essentially a
10\AA{} edge Binary tiling), thus building in an assumption of the
frequency and low energy of the \Decagon~ cluster.  We were not really
able to reach a third stage simulation, which properly would have
required a complete understanding of puckering and its interactions.
Indeed, in Sec.~\ref{sec:ideal-deco} we will present a deterministic
decoration rule, for a particular composition, taking into account the
tendencies noted in Sec.~\ref{sec:MC-4A} and Sec.~\ref{sec:Co-Ni}.
But this rule is more speculative than the ``basic Ni'' rule of
Ref.~\onlinecite{alnico01}, in particular no MC simulation 
reproduced its energy (they were higher, by at least a small energy

\MM{(2/06) I would say MANY came near to it, but none was
quite as low as this one.}

\LATER{(Move to Discussion?) MM 2/06
I'm now quite sure that the reason was
the sparseness of the site list, such that the energy
landscape had weird ups and downs (barriers). }

\FUTURE{Presumably, after we understand the simulations in
  enough detail, we should be able to resolve these complications and
  formulate a deterministic decoration rule, using tiles larger than
  the $4.0$\AA{} edge rhombi.}

\subsubsection{Competition of basic-Ni and decagon based structures}

We now turn to the physical question of the competition 
between the Basic-Ni and Basic-Co structure variants
in the Al-Co-Ni phase diagram.
The ``basic Ni'' phasse is defined by frequent
NiNi nearest-neighbor pairs (forming zigzag chains along
the $z$ directions), and Co at centers of a HBS tiling
with edge $a_0\approx 2.45$\AA{}, without any 5-fold
symmetric motif; whereas ``basic Co'' is defined by the two
types of 11-atom pentagonal clusters that form the
centers of \Decagons~ and \Stars.
Now, in Subsec.~\ref{sec:Co-Ni}, it is described how added Ni atoms
appear inside the \Dec~ as Ni(2.5), adjacent to Co(3).
If we also replace this Co(3)$\to$Ni(3), we get 
a Ni-Ni pair (TM in a pair always  tends to be Ni to  free up
Co to have more Al neighbors, since Al-Co has a stronger bond
than Al-Ni as we have repeatedly remarked.)
It is indistinguishable from 
the characteristic Ni-Ni pair in the ``basic Ni'' phase 
of $d$-AlNiCo~\cite{alnico01}. 
In other words, the motifs of that phase
are appearing continuously as the composition gets richer in Ni.
[In the language of the 2.45\AA{}-edge-DHBS small tiling introduced in
Sec.~\ref{sec:ideal-deco},
the small tile around that TM(3) must become
a Hexagon, like the tile in the ``basic Ni'' decoration~\cite{alnico01}.]

We incompletely explored this competition 
by some variations in the site list, in the unit cell
size/shape, or in composition.
It appears there is a barrier between the basic-Ni
and basic-Co structures in our simulations, perhaps
a thermodynamic barrier or perhaps merely a kinetic one due to 
our handling of the degrees of freedom.
Thus, there is no assurance that simple brute-force
simulation will reach the best state.  The only
reliable criterion is to anneal each competing
phase to a minimum-enegy state,  and compare the
respective energy values. 

\remark{MM email 6/14/05: In the  2001 b-Ni work, we DID check explicitly 
the possibility of formation of 20\AA{} 
(YES, 20A) D-clusters with 5-gonal center, 
by including some extra sites such that the decagon center could get 
5-fold symmetric decoration even within 6.5A-HBS framework.
(MM thinks it amounted to associating some extra sites with Boat,
these should be then effective when 2H+B form 20A Decagon.)
The Decagons did not form. But now MM suspects they WOULD have
formed here and there, but could not because the pattern
formation is related to an interplay within a larger radius,
and the HBS framework simply did not allow those correlations
to develop.  One possible reason is that the ``ring 2.5'' sites 
were not  all available in the site list.  Now that we know the 
importance of the [13A Decagon] [and know what its atomic
structure is], we would have certainly set that up more carefully.}

\remark{More on basic-Ni/basic-Co competition from MM 6/22/05}

\remark{MM notes: The 10.4A is a tiling of Hexagons packed in alternating
fashion like in o-Al13Co4; parallel hexagons would be the m-Al13Co4)} '

\remark{MM: I found CLOSE competition (energy difference nearly zero)
  of the two decorations throughout the Ni-Co composition range, with
  subtle preferences for [basic Ni] type decoration on Ni-rich side
  and Al6Co5 type on Co-rich side.  And these were sensitive to Al
  content.}

\LATER{CLH to MM: Which way did the sensitivity to Al go?
CLH needs to check the MM's old report, from 4/04)}

We used the 12$\times$14 simulation cell for a
direct study of the competition of the ``basic Ni'' and ``basic Co''
kind of structure; they were found to be practically degenerate 
in energy throughout the Ni-Co composition range.
But in a similar simulation in the standard 32$\times$23 cell, the
preference for the Al$_6$Co$_5$ rings was much stronger.  Our
interpretation is that the Al$_6$Co$_5$ cluster is not robustly stable
by itself, but only when surrounded by the other rings of the
\Decagon.  Since the 12$\times$14 unit cell is too small to 
allow a proper ring 3, the full benefit of the \Dec~ arrangement is lost 
and the balance is tilted towards the ``basic Ni'' type of structure, 
which is built of smaller (2.45\AA{}-edge HBS) tiles and has no 
frustration in a cell this size.

\LATER{The next para. refers to statements, somewhere, about
the bNi/decagon competition being perhaps a matter of TM content
more than Ni/Co content.  
`The mentioned ``tilt'' in the phase line is consistent with
observations we made about the influence of composition.'
Find where those remarks are.
Find, too, where I incorporated Marek's notes from his 2004 report.}

It is interesting to note here that the theoretical phase boundary found 
by Ref.~\onlinecite{Hi06} in the Al-Co-Ni composition space, running
roughly from Al$_{76}$Co$_{24}$ to Al$_{70}$Ni$_{30}$, 
corresponds fairly well to the domain~\cite{Gru04} in which 
decagonal Al-Co-Ni is thermodynamically stable. In other words, 
$d$(AlCoNi) occurs at all only when the two competing structure
types are close in energy.

A useful diagnostic for the phase competition was used by
Hiramatsu and Ishii~\cite{Hi06}, which might be called
the weighted differenced pair distribution function.
One takes the difference of the pair distribution function 
(as a function of radius) 
between two competing phases, and multiplies it by the pair potentials.
The large positive and negative peaks then reveal which potential
wells favor which kind of structure.  The dominant contributions turned out
to be Al-TM nearest-neighbor wells favoring the decagon-based structure,  
and Al-Al nearest-neighbor repulsion favoring the basic-Ni structure.

\LATER{That is consistent with the phase diagram line as found
by Ref.~\onlinecite{Hi06} and also with the trends found by
us [FIND THEM].}

\subsubsection{20 \AA{}  decagons?}
\label{sec:fixed-site-20A}

We have just observed 
%%% (Sec.~\ref{sec:Al6Co5-question}) 
that using the wrong size of unit cell might spuriously
exclude the optimal type of tile or cluster. 
Thus we may well worry whether even our standard unit cells 
are large enough to obtain the most correct structure.

Unfortunately, it is not feasible to simulate larger cells using
the 2.45\AA{} random-tiling lattice-gas. 
It would be necessary instead to devise a new decoration,
which is more constrained than the 2.45\AA{} sitelist
of Sec.~\ref{sec:MC-a0}
but less constrained than the 4\AA{} rhombus
decoration of Sec.~\ref{sec:MC-4A}.  Alternatively, as 
some conjectured atomic structures are available based on
20 \AA{} decagons~(see Appendix~\ref{app:20A}),
one might design a decoration which can represent
structures built of either 13 \AA{} decagons or 20\AA{} decagons.

The same caveat (about the unit cell size)  applies to earlier work 
by some of us on the ``basic Ni'' modification.\cite{alnico01}
In that case, too, electron microscopy studies had
suggested structure models having 20\AA{} diameter clusters
with pentagonal symmetry~\cite{abe06-ICQ9}.

\remark{The TEM studies really only show the TM sites, which scatter stronger. }

\section{Idealized decoration}
\label{sec:ideal-deco}

In this section, we present an explicit model structure, derived by
idealizing the simulation results of Sec.~\ref{sec:fixed-site}, as a
decoration of a 10.5\AA{}-edge Binary tiling.  
%%%% As in the case of ``basic Ni'', 
Such idealizations are necessarily speculative -- 
they go beyond the simulation observations that inspire them;
nevertheless, they are important for several reasons.
First, they make available an explicit model for decoration or
diffraction.  
It is trivial to construct a quasiperiodic Binary tiling;
\remark{It is a well-known subset of the Penrose tiling vertices.}
decoration of this specifies a quasiperiodic atomic structure,
which may be expressed as a cut through a five-dimensional structure, 
and compared to other models 
formulated that way.~\cite{xray,Yama-AlCoNi-5D,del06}.
(It should not be forgotten that the rules also allow the decoration 
of {\it random} tilings, which among other things can be used 
to simulate diffuse scattering.)

Second, we hope that a well-defined rule for chemical occupancy
corresponds to an energy minimum, in that all the good sites for a
particular species are used, and no more.  For this reason, it is
quite natural that an idealized model has a somewhat different stoichiometry
and/or density than the simulations it was abstracted from.  
Once we have an ideal model, 
the effect of small density or composition variations 
may be described by reference to it.
The ultimate validation of an idealized model is that it provides
a lower energy than any simulations with the same atom
content (and lower than other idealized models we may try).

The main issue in passing to a complete rule is to systematize 
the Al arrangements in rings 2.5 and 3 of the \Decagon
(which are apparently irregular, and surely not fivefold symmetric), 
and secondarily the TM arrangements in the \Star.  
This will impel the introduction (Subsec.~\ref{sec:small-DHBS}
of yet another tiling, the 2.45\AA{}-edge decagon-hexagon-boat-star 
(DHBS) tiling.

It should be recognized that the details of variable Al around the edge of
the \Decagon~ are crucially modified by relaxation, as will be
reported in Secs.~\ref{sec:relax} and \ref{sec:puckerLRO}.
Nevertheless, we first describe the structure as it emerges within the
fixed-site list because (i) this is the path that our method
necessarily leads us along; (ii) most of the structure ideas of the
fixed-site list have echoes in the more realistic relaxed
arrangements. In particular, the ring 2.5 and ring 3 patterns
(including short bonds) become the ``channels'' for Al atoms of
Subsec.~\ref{sec:channels}; the 2.45\AA{} HBS tiles in
Subsec.~\ref{sec:ideal-deco} and the puckering units
of Subsec.~\ref{sec:pucker-units} are centered on the same Co chains; 
and finally, the fixed-site
explanation of the ``ferromagnetic'' order of \Decagon~ orientations is
closely related to the puckering explanation (Sec.~\ref{sec:decLRO}).

\subsection{Inputs for the decoration rules}

Next we give the starting assumptions (based on Sec.~\ref{sec:fixed-site})
which consist of (i) guidelines for the best local environments, given
the (fairly artificial)  assumption of the fixed-site; (ii) the underlying
tile geometry which is to be decorated.

\subsubsection{Guidelines for atom  placement}
\label{sec:deco-guidelines}

The description inferred from MC runs left undecided (i) the choice of Co 
versus Ni on sites
designated ``TM'' in the \Dec; (ii) the choice of Ni, Co, or Al on the
sites designated ``candidate TM'' in the \Star; (iii) the location of
Al sites in rings 2.5 and 3 of the \Dec.  
We seek the minimum energy choices, 
guided by the salient features of the pair potentials 
in Table \ref{tab:potentials} and by the typical configurations resulting 
from simulations on the 4.0\AA{} tiles (Sec.~\ref {sec:4A-results}).
%%% Subsections~\ref{sec:struct-pot} 
To resolve details, we also used spot tests (in which selected
atoms were flipped by hand) and the site energy function 
(Sec.~\ref{sec:siteE}).  

Guideline 1, the strongest one, is the TM-TM superlattice,
with separations $\sim 4.5$\AA.
Note that though Al-TM potentials are stronger than TM-TM, 
the negligible Al-Al potential seems to allow the TM-TM 
interaction to dominate the TM placement.
This spacing should be enforced particularly for Co-Co, since
that potential is somewhat stronger than Co-Ni or Ni-Ni.

\remark{Guideline 1 suggests the \Decagon~ vertices should all
  be Co.}

Guideline 2 is  to maximize number (and optimize the distance)
of nearest-neighbor Al-Co contacts, since this  
potential well is very favorable.  A corollary 
is that TM-TM nearest neighbor pairs tend to 
be Ni-Co or Ni-Ni (with the glaring exception of
five Co in the \Dec's core),
so as to increase Al-Co at the expense of Al-Ni bonds.
(This last fact is 
more important in a Ni-rich composition~\cite{alnico01}.)

Guideline 3  is that in the central ring of the
\Star, the favorable location for Ni (occasionally Co)
is on the line 
joining its center to that of an adjacent \Decagon, whenever that line
  passes over an Al (rather than a TM) atom in ring 1 of the \Dec.  
(That line is an edge of a 10.5\AA{} binary tiling rhombus).

%%%%%%%%%%%%%%%%%%%%%%%%%%%%%%%%%%%%%%%%%%%%%%%%%%%%%% 
%%%%%%%%%%%%%%
\begin{figure}
\includegraphics[width=3.3in,angle=0]{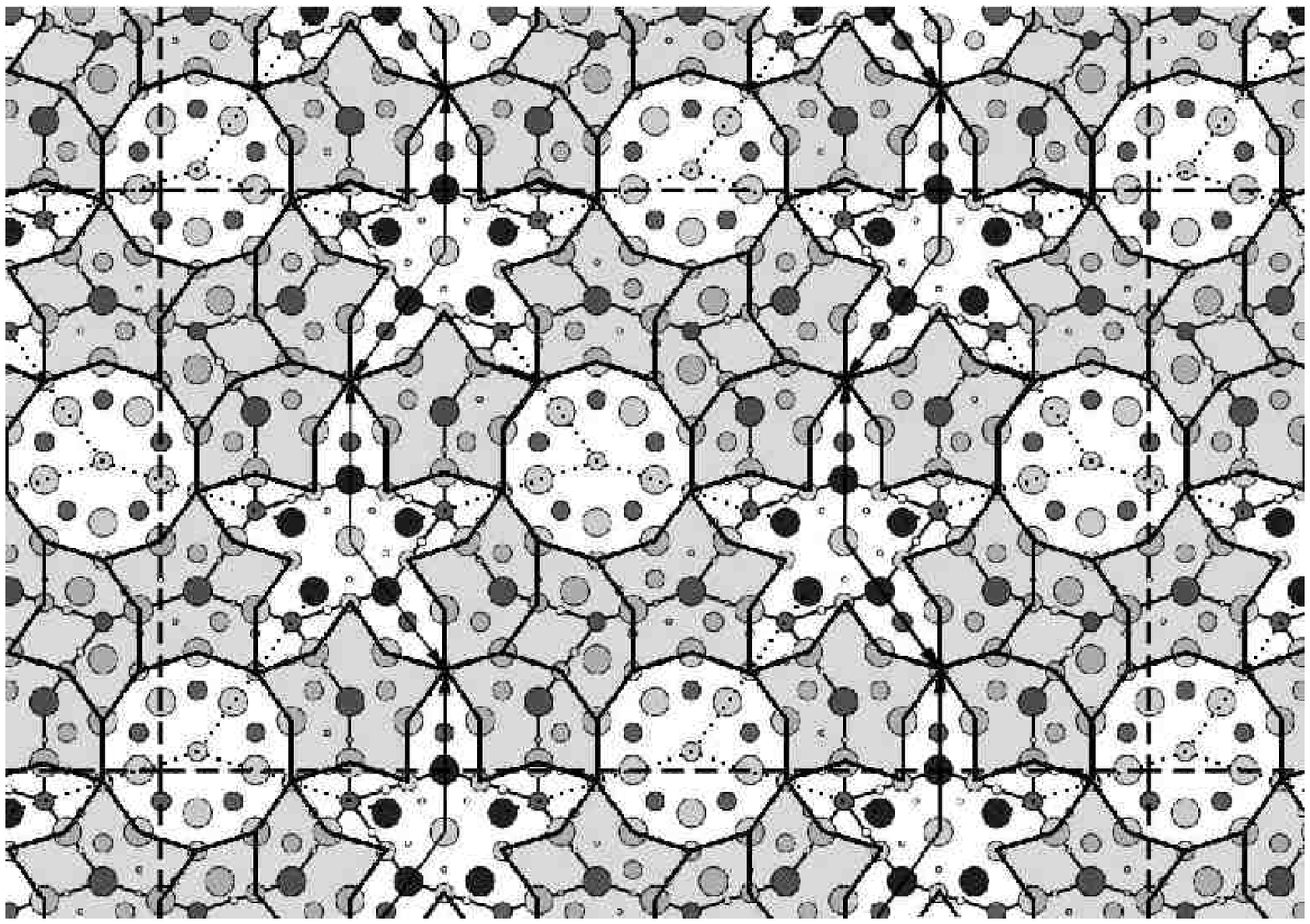}
\caption{Idealized atom decoration for a bilayer ($c\approx 4$\AA{})
  structure on the 40 $\times$ 23 tiling, given an arrangement of
  \Decagons~ (shown by light lines).
%%%%%%%%%%%%%%%%%
  \remark{The orientation is assumed to be the same in every \Decagon, which
  breaks the global 10-fold point symmetry down to 5-fold.
The particular composition here is slightly Ni poor (7.5\% vs.
    10\% for standard and ~8.5\% for experimental W.}
%%%%%%%%%%%%%%%%%
  Atom species and layer are identified by same convention as in earlier
  figures.  The \Decagon~ edges are mostly covered
  by 2.45\AA{} Stars and Boats, shaded gray,
which specify the placement of
  Al atoms in the Decagon's ring 2.5 and ring 3.  
%%%%%%%%%%%%
\CLH{To Nan: I have hacked at your caption, because much of
its story is told in the text -- and in earlier figures.
If we had one figure and one paragraph to say everything,
it would do a pretty good job of that!}
%%%%%%%%%%%%
The sites along the lines connecting adjacent \Stars~ are TM 
(Ni on the even glue cluster and Co on the odd one), as marked by arrows.  
The even \Stars~ (normally) have Ni in the directions towards the nearest 
Decagons, as pointed out by dotted lines; these also mark edges of
the 10.5\AA{} Binary HBS tiles.}
\label{fig:RulesWideal}
\end{figure}
%%%%%%%%%%%%%%%%%% 
%%%%%%%%%%%%%%%%%%%%%%%%%%%%%%%%%%%%%%%%%%%%%%%%%%

\subsubsection{Binary Hexagon-Boat-Star tiling}
\label{sec:binary-HBS}

Following Subsecs.~\ref{sec:clusters} and~\ref{sec:Dec-relation}, 
our decoration is based on a packing of \Decagon~ clusters and \Stars~ 
on the 10.5\AA{}-edge binary tiling.
We anticipate the results of Sec.~\ref{sec:decLRO} by orienting
the \Decagons~ all the same way.  This has strong implications for the
\Stars.  The latter sit on ``small''
vertices of the Binary tiling, which (as is well known)
divide bipartitely into ``even'' and ``odd''sublattices: 
every 10.5\AA{}~ rhombus has one vertex of either kind. 
Because of the \Decagon's fivefold symmetric core, the
adjacent even \Stars~ are not related to it the
same as adjacent odd \Stars.
When the \Dec~ cores are all oriented the same, 
then the \Stars~ of one sublattice -- we shall call it Even --
have every candidate TM site aligned with ring-1 Al of the
adjacent \Dec~ which (by Guideline 3) is favorable for 
TM occupancy. On the other hand, in the Odd \Stars~ 
the only sites favorable for TM are the ones adjoining 
a TM-filled site in the adjacent even \Star; the rest of
the sites are favorable for Al.

The strong even/odd distinction, and the lack of a 
prominent pattern on the Odd \Stars, inspires a slightly
different way of representing the 10.5\AA{} tile geometry.
If one erases the vertices that center the Odd \Stars, and the
binary-tiling edges that connect to them, the remainding vertices and
edges form a hexagon-boat-star tiling with 10.5\AA{} edges.  
This defines a random tiling model called the ``Binary HBS tiling''.
(Ref.~\onlinecite{Mi06-ICQ9} introduced this term, for a different Al-Co-Ni 
decoration using 4\AA{}-edge tiles, but it has implicitly
appeared in some prior decagonal models.)
This is not equivalent to the ordinary random HBS tiling, 
since it is still constrained by additional colorings of the vertices as 
``large'' or ``small'', carried over from the Binary tiling.
However, it {\it is} essentially equivalent to the random 
Binary tiling, since there is a 2-to-1 correspondence 
between the tile configurations (depending on which sublattice of
``small'' vertices is designated ``even'').  

The Binary HBS tiling, like the cluster  orientations, has only
a fivefold symmetry, implying  a pentagonal space group for the quasicrystal.
\LATER{TELL name of space group.}

%%%%%%%%%%%%%%%%%%%%%%%%%%%%%%%%%%%%%%%%%%
\begin{table} 
\begin{tabular}{|l|rrr|cc|c|}
\hline
\multicolumn{1}{|c|}{2.45\AA{} tile} & \multicolumn{3}{c|}  {Content} & 
\multicolumn{2}{c|} {In 10.5\AA{} tiles}& Al nbrs. \\
               & Al & Co & Ni  & ~Fat  & Skinny & (each TM) \\
\hline \hline
Decagon           &  10  &  5 & 0   &  0.6   &  0.2 &  4+6 \\
Even Star Cluster &  10  &  5 & 5   &  0.2   &  0.4 &  3+4 \\
Hexagon           &   3  &  1 & 0   &  0     &   0  &  3+6 \\
Boat              &   5  &  1 & 0   &  3     &   0  &  4+6 \\
Star              &   6  &  1 & 0   &  0     &   1  &  5+4 \\
\hline
\end{tabular}
\caption{ \footnotesize  Atom content for decoration in
Fig.~\ref{fig:RulesWideal}.  The names are for tile
objects in the 2.45\AA{}-edge DHBS tiling. 
For the counts in column 1, Al on the tile corners are 
apportioned according to the corner angle.  
The numbers of each
tile object in the 10.5\AA{}-edge Binary tiling rhombi
are also given. The last column gives the Al coordination
of the TM atom(s), $m+2n$ where $m$ Al neighbors are in the same layer
at 2.45\AA{}, and $2n$ Al neighbors are in the adjacent layers
at 2.54\AA{}.
}
\label{tab:deco-content}
\end{table}

\LATER{Nan computed for the ideal decoration of Fig.~\ref{fig:RulesWideal}, 
in the W cell, the atom content is 
Al 188, Co 60, Ni 20   (Al 71.1, Co 22.3, Ni 7.5), which is 268 atoms.
And of the 2.45A objects, he has: 4 Decagons;  8 Stars; 12 Boats; 20 Hexagons
This to compare with counts in Table~\ref{tab:deco-content}.
I would count the 20 Hexagons  as four even-Stars.}

\subsection{The 2.45\AA{} Decagon-Hexagon-Boat-Star tiling}
\label{sec:small-DHBS}

\remark{Nan's  original analysis had mis-counted the Al neighbors
in other layers -- only once,  when they should be doubled,
and said the Star was optimal.  CLH says the correction is
good since his picture is actually the configuration that 
maximizes 2.45-Boats, not Stars. Nan pointed out (1/06) the
importance of 2.45 vs 2.54.}

\LATER{Although the Boat has  one more Al-Co contact, 
the Star has five in-plane contacts, as opposed to four for the
Boat (see Table~\ref{tab:deco-content}). 
The in-plane distance 2.45\AA is more favorable than the
out-of-plane 2.54\AA, so it is unclear which 
 has the better energy when this is taken into account.}

Now we introduce yet another tiling. 
Its edges are $a_0=2.45$\AA{},
as in the initial stage single-layer rhombus tiling, but 
these tiles are 8\AA{} diameter Decagons, 
as well as Hexagons, Boats, and Stars, 
so we call this the ``DHBS'' tiling.
(See Fig.~\ref{fig:RulesWideal}).
The vertices are decorated with Al atoms, in the even (odd) layers
for even (odd) vertices.  
The 8 \AA{}~decagon (with edge 2.45\AA{}) 
is a subset of the \Decagon; its perimeter (vertex) atoms
are the ring 2 Al from the \Decagon.
Each even \Star~ is represented by five 2.45\AA{} Hexagons 
in a star arrangement; since these Hexagons are decorated differently
from the regular kind, this combined unit will be
treated as a separate kind of tiling object 
called ``Even Star cluster''.  
(An Odd \Star~ center
is just a corner where three 2.45\AA{} Boats or Stars meet.)
The 8\AA{} decagons
and Even Star clusters, which are fixed once a 10.5\AA{}
Binary-HBS tiling is specified, are shown in
white in Fig.~\ref{fig:RulesWideal}.
\remark{Every tip of an even \Star~ touches either another tip,
or an  8\AA{}decagon.}

The remainder of space  -- that is, the \Decagon~borders 
-- becomes tiled with 2.45\AA{} Hexagon/Boat/Star tiles 
(shown shaded in Fig.~\ref{fig:RulesWideal}.
The external vertices of the HBS tiles represent all Al(2) [ring 2
of the \Dec], all Al sites in the \Star, and all Al(3) [ring 3]. 
Each HBS tile interior includes one Co on its ``internal vertex'' 
(formed when the HBS tile is subdivided into rhombi), and 
also Al site(s): one per Hexagon, two
in each Boat or Star.   These last Al sites represent all Al(2.5) in
the \Dec~ and all Al on candidate-TM sites of the \Star.
Thus, the placement of HBS tiles directly determines that of
the ring 3 Al, but not of the ring 2.5 Al.
The Even Star type hexagon is a special case:
its two internal sites are Co-Ni in the decoration
of Fig.~\ref{fig:RulesWideal} but in others 
(see Subsec.~\ref{sec:alt-deco}) would be Ni-Ni.

It should be emphasied that the above description is not just 
a reformulation of the observations in Sec.~\ref{sec:fixed-site} 
but is, in fact, an additional insight into the motifs
emerging from the lattice-gas Monte Carlo on the 4.0\AA{} rhombi.
The 2.45\AA{} DHBS tiling is not just used to describe the 
nearly ground-state structures (which are the focus of this
section), but also the less optimal configurations that  were
our typical best snapshot from a Monte Carlo run (at the 4.0\AA{} stage), 
or the configurations found when density and composition are 
somewhat changed, such as in Fig.~\ref{fig:Ni-arrow}.
Despite many irregularities, almost the entire space between Decagons~
decomposes into HBS tiles. One difference from the
description given above is that, in these imperfect configurations,
the Even \Star~ grouping of 2.45\AA{} Hexagons is seen less; also,
either Hexagon filling (TM-TM or Al-TM) may occur anywhere.

It will be noticed that all our decorations of the HBS tiles
are identical to those in the ``basic Ni'' structure~\cite{alnico01}.
The important difference is that in ``basic Ni'', there were no 
8\AA{} Decagons: the HBS tiles filled space by themselves.  
This suggests that, as Ni content is increased, conceivably the ``basic Co''
structure evolves smoothly to the ``basic Ni'' structure by filling less of
space by 8\AA{} decagons, and more of it by HBS tiles.

\subsubsection{Optimization among HBS tilings}

The next question is to single out the DHBS tilings with the
lowest energies.
The particular Al configuration depicted in Fig.~\ref{fig:RulesWideal}
was obtained by adjusting Al corresponding to different 2.45\AA{}
HBS tilings to optimize the energy in this (40$\times$ 23) unit cell.
All the tilings being compared had equal numbers of Al-Co first-well bonds,
as well as TM atoms in the same positions, 
so any energy differences must be due to
the second well of $V_{\rm Al-TM}$ (which is about 1/9 as strong as the first
well, see Table~\ref{tab:potentials}).  
The total energy difference between two of these states is
estimated to be of order 10 --  50 meV.

\remark{Only the Al placement was varied by hand; however, 
Nan did run MMC tests of Al-Al swamps, TM-TM swaps, or with all swaps
allowed freely...
In the tests by hand, apparently records were not kept of the original
energies.  MM (9/05) wrote ``I have now encoded this decoration, 
so it can be applied easily to an arbitrary 10.4A binary tiling.''}

We can interpret the result in the light of
Guideline 2 from Subsec.~\ref{sec:deco-guidelines}, together with
the last column of Table~\ref{tab:deco-count}.
The largest energy term is proportional to the number of
Al-TM (especially Al-Co) bonds; with the fixed sites 
available, the bond distances are either $2.45$\AA{} (in the
same layer) or $2.54$\AA{} (interlayer); the Al-Co potential 
is stronger at the former separation, leading in principle to
smaller energy differences even with the same number of Al-Co
bonds.  Now, Co centering any HBS tile has a good Al coordination 
(9 or 10), but this is best in the 2.45\AA{} Boat cluster -- 
mainly because that has more Al atoms.  Hence, the number of
Boats should be maximized, as is the case in Fig.~\ref{fig:RulesWideal}.
(Recall that tile rearrangements allow us to trade
2 Boats $\leftrightarrow$ Hexagon + Star in an HBS tiling.)

The TM in the 2.45\AA{} Hexagon tile has a smaller number
$Z_{\rm Al}$ of Al neighbors.  Thus, if Ni concentration
is increased at the expense of Co, the Ni atoms will first
occupy these TM sites (on account of the strong Al-Co attraction).
Also, where TM-TM neighbors are forced, this tends to occur in
2.45\AA{} Hexagon tiles.  For example, the ``arrow'' motif 
of induced  by increased Ni concentrations
just consists of three successive 2.45\AA{} Hexagons on the border 
of the \Decagon, each of them having a TM-TM interior occupation
(See Fig.~\ref{fig:Ni-arrow}).

\subsubsection{Pentagonal bipyramid motif?}
\label{sec:PB}

The comparison of nearest-neighbor Al coordinations 
missed one important fact: a 2.45\AA{} Star tile is
generally part of a larger motif with pentagonal symmetry.
Empirically, it is invariably surrounded by a 
pentagon of TM atoms (at 4.46\AA{}) in the other layer
than the central TM.  This means that Star tiles are
strongly biased to be on the five \Decagon~ corners that line 
up radially with a Co(1) (of the core), and not the other
five corners.
[That Co(1) is needed to complete the outlying TM pentagon.]

In projection, the five TM atoms surrounding the Star, together with 
the five Al atoms at its outer points, form a decagon of radius 
8\AA{}. The other five Al atoms on the 
Star's border turn out to lie in ``channels'', in the
terminology of the following section (see Sec.~\ref{sec:channels}), which 
implies that in a relaxed (and more realistic) structure, 
these atoms displace out of their layer. 
The 5 Al + 5 TM atoms  forming the outlying decagon all sit
in the same layer which turns out to become a mirror 
(non puckering) layer upon relaxation.  In the end, the total motif 
is simply the ``pentagonal bipyramid'', 
a familiar motif in decagonal structures~\cite{Hen93,Wid96}.

\subsubsection{Alternate description using 4\AA{} DHBS tiles}
\label{sec:4.0-DHBS}

The decoration depicted in Fig.~\ref{fig:RulesWideal}
has 5Ni + 5 Co on the internal sites of the Even Star Cluster,
which ensures that the \Decagons~ have purely Co atoms (never Ni) 
on their outer vertices (ring 3).
The 2.45\AA{} Stars and Boats are the most favorable locations for TM (Co).
The Ni site in the even \Star~ is the least favorable of the TM
  sites in this decoration.

We pause to express the results in the language of the 4\AA{}-edge 
DHBS tiling. This tiling has been studied in less detail, for it
is less handy than the 2.45\AA{} DHBS
or the 10.5\AA{} Binary HBS tilings, for the following reasons:
(i) different 4\AA{} HBS tilings, in some circumstances, can 
correspond to the same atomic configuration;
(ii) We lose all hope of systematically describing the Al(2.5) atoms.
(iii) the Al(3) variability is now represented by arrows along tile
edges, the rules for which are unclear.
(We might impose Penrose's matching rules, on the edges in HBS tiles
-- leaving the \Dec~edge as a ``wild card'' that matches anything
-- however that probably disagrees with the energy minimization.)

\begin{figure}
\includegraphics[width=3.4in,angle=0] {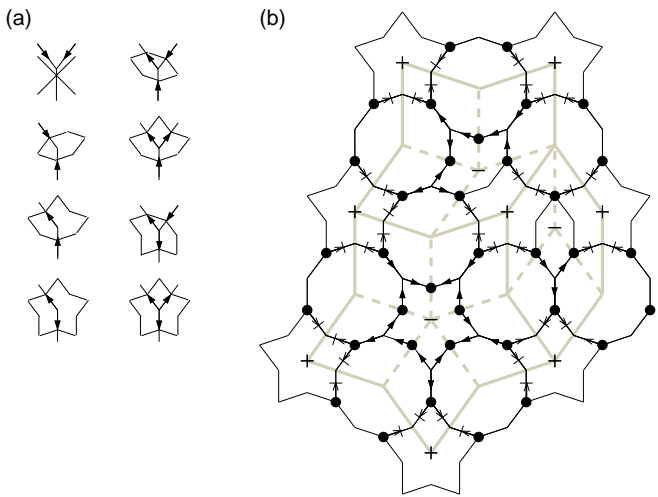}
\caption{ (a). Mapping from an arrow configuration to the 
2.45\AA{}-edge HBS tiles in the DHBS tiling.
  The $\times$ means that two incoming arrows,
  72$^\circ$ apart, are never allowed.  The arrowed edges belong to
  the 4.0\AA{}-edge tiling.  (b). The 10.5 \AA{}-edge Binary HBS
  tiling (gray edges, internal edges dashed) with \Decagons 
  placed on it. 
(Note that, to accomodate examples of all three HBS tiles, a cell
would be needed considerably larger than the W-phase cell of
Fig.~\ref{fig:RulesWideal}.)
Solid lines mark edges of the 4.0\AA{}-edge DHBS tiling.
  Even \Stars, marked with ``$+$'', get represented here by edge 
  4.0\AA{} Star tiles; an odd \Star, marked by a ``$-$'', is found 
  on the internal vertex of every 10.5\AA{} Hexagon, Boat, or Star tile, 
   and is represented by a 4.0\AA{} tile of the same shape.
  The direction of a light arrow is forced by the orientation of the
even \Star~ next to it; the bar blocking that arrow marks a boundary of 
the independent arrow subnetwork on that 10.5\AA{} Binary HBS tile.
The heavy arrows have variable direction, as described in the text, 
and determine 2.45\AA{} HBS tiles,  as shown in (a).
The black disks mark sites which are favored (by the core orientations
in adjacent \Decagons) to be the root from 
which a tree of arrows emanates, i.e. to be the center of a 
2.45\AA{} Star tile. 
\LATER{Ideally, rearrange it all to make wider, not high...}  }
\label{fig:arrows}
\end{figure}

The 4.0\AA{} HBS tiles are of course combinations of 4\AA{} rhombi. 
The \Dec~ is a tile object, while
the space between \Decs~ gets covered by 4\AA{} Stars, Hexagons, or Boats.
The tiles -- at least, with the decoration of 
Fig.~\ref{fig:RulesWideal} --  have 
Co on  {\it every} exterior vertex (in alternate layers).
Each 4\AA{} HBS tile contains, centered on its ``interior vertex'',
one \Star.  The 4.0\AA{} HBS tiling is shown in Fig.~\ref{fig:arrows}
decorating the 10.5\AA{}

It is appropriate here to review what our decoration does 
in terms of the originally identified 11-atom \Star~motif, 
which (roughly speaking) goes with the 4.0\AA{} DHBS tiling.
The decoration of Fig.~\ref{fig:RulesWideal} places Ni  on all five
of the candidate-TM sites of the  Even \Stars;
Odd \Stars~ receive two, one, or zero Co according to whether
they occur (see Fig.~\ref{fig:arrows} in a Hexagon, Boat,
or Star of the 10.5\AA{} Binary HBS tiling; this Co appears
next to each neighboring Even \Star.

\remark{This is not so good: we said a \Star~ always wants at least two TM?
  On the 32x23 tiling, where each \Star~ has just one \Star~ neighbor, we
   have one TM; which agrees with the fact it is all 10.5 A Boats in the 
   Binary HBS tiling.
Nan: ``I was never content with the 23x32 ferro idealization I found.
  The model I started off with had 5 TM in one star cluster and zero
  in the other, I just had to take on TM and move it to the other star
  cluster. My work on that stopped there.''}

\subsection{Enumeration of Al placements}
\label{sec:deco-enumerate}

The packing of space by HBS tiles, which can be done in many ways, is
a convenient way to enumerate (while automatically enforcing neighbor
constraints) all possible ways of placing Al atoms in rings 2.5 and 3.
This is seen even clearer using the abstraction in Fig.~\ref{fig:arrows}.

\subsubsection{Enumeration of 2.45\AA{} HBS tiles (and Al(3) placements)}

In this idealization (see Fig.~\ref{fig:arrows}(b)), {\it every} edge
of a \Decagon~ has one Al atom (which is also a vertex of the
2.45\AA{}~ HBS tiles) dividing it (in projection) in the ratio
$\tau^{-1}:\tau^{-2}$.  The choice on each edge is represented by an
arrow pointing towards that Al atom, and Fig.~\ref {fig:arrows}(a)
shows the translation from the arrows to the language of HBS tiles.
Every even \Star~ is represented by five 2.45\AA{} hexagons,
which in the arrow language translates to a boundary condition of a fixed
arrow direction (indicated by light-headed arrows in
Fig.~\ref{fig:arrows}(b)).  The network of arrowed edges has corners
of coordination 2 or 3, the latter being where two \Decagons{} share.
At the latter corners, it is forbidden for both arrows to point
inwards (the corresponding Al atoms would be too close).

In enumerating the possible 2.45\AA{} HBS tilings, there are several answers,
because we may place varying degrees of constraints on those tilings.
First, if we permit any mix of 2.45\AA{} H/B/S tiles, then on every Fat
10.5\AA{} rhombus in Fig.~\ref{fig:arrows}(b) we could independently 
orient the three free arrows in any of the six ways allowed by the 72$^\circ$
constraint: that would give $6$, $6^3=216$, or $6^5=7776$ choices 
on the 10.5\AA{} Hexagon, Boat, or Star, respectively.

Let us, however, maximize the number of 2.45\AA{} Boats
as justified earlier, which means there are {\it no} 2.45\AA{} hexagons
(apart from those combined into the Even \Star~ object).
Then, at every vertex in Fig.~\ref{fig:arrows}, either all arrows 
point outwards (which makes a 2.45\AA{} Star); or one arrow points inwards 
and the rest point out (a 2.45\AA{} Boat).  
Now, each 10.5\AA{} Binary HBS tile has exactly one connected 
subnetwork of arrows. 
Hence, in every subnetwork, exactly one vertex must have its arrows
all pointing outwards, and serves as the root of a tree; at the other 
vertices, the arrows point outwards from that root.  
Thus, the remaining freedom in Boat/Star placement amounts to which
vertex has the ``root'' vertex, or equivalently where the unique
2.45\AA{} Star gets put.  (On the 10.5\AA{} Hexagon, a second 
2.45\AA{} Star gets forced near the tip with a \Decagon.)
There are four choices to place the ``root'' per
10.5\AA{} Hexagon and ten choices per 10.5\AA{} Boat. But on the 
10.5\AA{} Star, there are just two choices, since there is
no ``root'' in this case -- the  only freedom is whether the arrows
run clockwise or counterclockwise in a ring around the center.

We have not yet taken into account an additional factor that
reduces the degeneracy of the HBS network: namely, the 2.45\AA{}
Star is preferentially located on the five vertices of the \Decagon
-- marked with black dots in Fig.~\ref{fig:arrows} --
that are aligned with the core Co(1) pentagon.  Counting the black dots
in each large (10.5\AA{}) HBS tile, we find three choices in the large Hexagon,
seven choices in the Boat, and (for the same reason as before)
just two choices in the Star.  
A Star and a Hexagon  together would thus have $2\cdot 3 = 6$ choices;
but the same area converted into two Boats has $7^2= 49$ choices.
Thus, if the entropy of these nearly degenerate arrangements
plays a role -- and it must at $T>0$, in a fixed-site lattice
gas simulation -- it assuredly favors the maximum possible content
of Boats in the 10.5\AA{}-edge Binary HBS tiling.

\subsubsection{Enumerating internal Al (ring 2.5) placements}

The internal Al's in the Boat have always been placed in the
(unique) symmetrical arrangement. 
In the 2.45\AA{}-edge Star, there are five possible placements
of the two internal Al; we insist on the rule that
there be one internal Al
near to each 8\AA{}~Decagon that the 2.45\AA{} Star adjoins,
since this adds one Al-Co bond. [The bond is with Co(1) from 
the \Dec's core;  such Al's were counted in the coordination 
4+6 listed for the Co(1) in Table~\ref{tab:deco-content}].
In the case of a 2.45\AA{} Star between two 8\AA{}~Decagons, 
this rule still leaves freedom among three of the five placement choices;
we think they are virtually degenerate, since they all have exactly
the same counts of nearest-neighbor distances.

The freedom in the 2.45\AA{}~-edge Hexagon is somewhat different,
being associated with the two ways of breaking it into rhombi.
Our Co placement rule would say the \Dec~ corner must be the Co site;
in the specific version of that decoration 
illustrated in Fig.~\ref{fig:RulesWideal} and tallied in
Table~\ref{tab:deco-content}
there are no Hexagons at all, so their internal decoration
is a moot question.
The Co placement rule also applies 
to the Even \Star~ type hexagon,  even though its other 
interior vertex is Ni. (In reality, the Even \Star~ hexagon
would more naturally be occupied by Ni-Ni rather than Co-Ni:
see Sec.~\ref{sec:alt-deco}.)

The resolution of the remaining near-degeneracy of the 2.45\AA{}
structures not only depends on tiny energy differences, but
quite likely the optimal placement of the ``root'' vertex
breaks the local mirror symmetry  of the Boat or Hexagon tile:
then the absolute ground state would depend on interactions
(at an even more minuscule energy scale) between the ``root'' 
vertex placements on neighboring tiles.  

It would not make sense to pursue these intricate details, for
the real behavior of the ring 2.5/3 Al atoms (which dominated
this section) is actually governed by ``puckering'' as 
explained in Sec.~\ref{sec:relax}.  Since the puckered structure
is still built out of 2.45\AA{} DHBS tiles, the general 
\MM{to CLH, I changed HBS to DHBS above}
framework remains valid, but our detailed enumeration is not, 
since a different subset of the 2.45\AA{} DHBS tilings may be 
preferred.  We have not investigated that as far as we took
the fixed-site case in this section, but we can guess that
the degeneracy resolution is at least equally intricate.

\MM{
[From MM 9/05.
We have commented in Sec.~\ref{sec:Co-Ni} how the substitution
Co$\to$Ni produces ``Ni(2.5)'' sites and makes a distorted 
Al$_6$TM$_5$ cluster inside the \Dec.]
This modification can alternatively be analyzed as the
replacement of the 2.45\AA{} decagon 
by two Hexagons and one Boat at the 2.45\AA{} scale.
%%% they have identical boundary.
``This somewhat decreases atomic density (11 atom 
\Dec~ core for 10 atoms; notably Al3Ni
is looser than Al13Co4), and somewhat increases TM content (5/11 to 5/10).''
}
\CLH{to MM: I don't see the 2H+B here.}

\subsection{Stoichiometry of the decoration}
\label{sec:deco-stoich}

It is easy to find the number of Fat and Skinny binary tiling 
rhombi at the 10.5\AA{} scale for a unit cell, and also in 
an infinite fivefold symmetric tiling (where the number ratio
of Fat to Skinny is $\tau:1$).
Then if we know what atoms are contained in each 2.45\AA{} DHBS tile
and how many of the latter are contained in each 10.5\AA{} rhombus
(both of which are given in Table~\ref{tab:deco-content}), we
can obtain the total atom contents.

To calculate the number of small (2.45\AA{}) DHBS tiles on each
large (10.5\AA{}) Binary tiling rhombus,
we decompose both of these into small (2.45\AA{}) rhombi.
The edges of the two kinds of rhombii are in the 
ratio $\tau^3:1$, so their areas are in the ratio 
$\tau^6= 8\tau+5 = 13+8\tau^{-1}$; furthermore, the area
of a Fat and Skinny rhombus on the same scale
are in the ratio $\tau:1$.
For example, each large Fat rhombus decomposes into 
13 small Fat rhombi +  8 small Skinny rhombi.
The small (8\AA{}) Decagon accounts for 5 Fat + 5 Skinny small rhombi;
the Even Star combination of five special small Hexagons accounts for 
5  Fat + 10 Skinny small rhombi.
When those contributions are subtracted, the remaining small rhombi
are assigned to small Star (5 small Fat rhombi) and small Boat 
(3 Fat + 1 Skinny small rhombi).
Remember it is possible to convert  two Boats $\to$
small Star + small Hexagon, 
which converts their atom content to Al$_{10}$Co$_2$ $\to$ Al$_9$Co$_2$;
that freedom was resolved in Table~\ref{tab:deco-content} 
by minimizing small Hexagon content (thereby maximizing Al content).
The net decoration of 10.5\AA{} tiles is then
Al$_{23}$Co$_{7}$Ni$_1$ on the Fat and 
Al$_{12}$Co$_{4}$Ni$_2$ on the Skinny.

If applied to a Binary tiling with fivefold symmetry
(that is, a quasicrystal having no perp-space strain), the overall
stoichiometry would be Al$_{0.722}$Co$_{0.225}$Ni$_{0.053}$.
That is obviously poorer in Ni than intended, even though
the same decoration gives the desired stoichiometry when 
applied to the large ($40 \times 23$) approximant in
Fig.~\ref{fig:RulesWideal}. 
The reason an unusually large
approximant is necessary, in order that both the decoration 
rule and the stoichiometry agree with that in the quasicrystal limit,
is that the Ni and Co placements
are inhomogeneous at relatively large scales.

\remark{MM 9/'05: We can estimate the upper limit on density variations, 
using 2.45 DHBS tiling. Conversion HS$\rightarrow$2B increases
density by 2 Al atoms. If {\em all} 2.45 Stars
are paired with H and converted to Boats, 
this should give the upper limit on the density.}

\subsection{Alternative  decoration rules}
\label{sec:alt-deco}

How should we fix the  unreasonable stoichiometry of the above-specified
decoration (when applied to general tilings)?
If we review the guidelines from Sec.~\ref{sec:deco-guidelines}, 
it makes sense to convert half of all the 2.45\AA{} Boats into 
Stars and Hexagons (the Al atoms are a bit overpacked when 
Boats are neighbors).
It also makes sense to convert much -- say half -- of all Co on the
Even Star Cluster into Ni 
(we know TM pairs are strongly favored to be Ni-Ni).
Now the atom content is Al$_{22.25}$Co$_{6.5}$Ni$_{1.5}$ on the Fat and 
Al$_{12}$Co$_{3}$Ni$_3$ on the Skinny, giving a more reasonable net
stoichiometry of Al$_{0.717}$Co$_{0.186}$Ni$_{0.097}$.

\LATER{URGENT. Check the above: 
I think the stoich. is for converting 80\% of Co.}

\LATER{Good to generate a structure of this version, 
and one of the other version, and compare energies?}

In Sec.~2 of Ref.~\onlinecite{Gu06-ICQ9}, we specified a distinct ideal 
decoration, similar to the variation just outlined.
Its purpose was not only to accomodate a larger Ni fraction 
among the TM atoms, but especially to decorate \Decagon~ clusters of
arbitrary orientation. 
\LATER{Shall we call it? (``Decoration II'')}
It still uses the 10.5\AA{}-edge Binary tiling with \Dec~ clusters 
placed on the ``large'' vertices, and (possibly overlapping) \Stars~ 
placed on ``small'' vertices;  but unlike the decoration
of Fig.~\ref{fig:RulesWideal}, each \Dec~ has an orientational
label which is an independent variable of the tiling.
If we limit ourselves to clusters oriented the same way,
that rule says (in this section's language)
the Even \Star~ object has Ni-Ni occupation on all
five of its Hexagons (this includes both those that 
connect to an Even \Star~ center, and other 2.45\AA{}~Hexagons 
that reach into a \Decagon, so Co(3) on some of its corners are
converted to Ni(3).  The decoration in Ref.~\onlinecite{Gu06-ICQ9} 
was incomplete, in that no attempt was made to specify the 
Al(2.5) and Al(3) positions.

\remark{When two \Stars~ overlap, their near-neighbor pair of candidate TM
sites becomes NiNi; otherwise, a candidate TM site in the \Star~ is
filled with Ni if and only if it is {\it not} lined up with the ring-1
Co in the adjacent \Decagon.  [This observed tendency seems to be
explained by the third-well Ni-Co interaction.  The distance from
\Star~TM to ring 1 Co, in the lined-up case would be 5.5\AA{}, near to
a maximum in $V_{\rm CoNi}(r)$, whereas in the not-lined-up case it is
6.4\AA{}, a potential minimum.  [See Table ~\ref{tab:potentials}.]
Finally, Co in ring 3 of the \Dec~ is converted to Ni if and only if
it is a nearest neighbor to a Ni-occupied site in the \Star.  [That
will occur if and only if that ring-3 Co site is not lined with a
ring-1 Co, and also is not on a shared edge of \Dec's.]}

\subsection{20 \AA{} decagons?}
\label{sec:ideal-Burkov}

\CLH{New subsection added (from colong-y.tex) on 1/25/06.}

Our story till now has skipped over the possibility of decagonal
clusters larger than our \Decagon. The question is pertinent,
as 20\AA{} diameter decagons have often been identified in 
electron micrographs as the basis of a cluster network.
In fact, reexamination of
Fig.~\ref{fig:RulesWideal} reveals that around every \Dec, there is 
another nearly perfect decagon larger by a factor $\tau$,
so its edges are $\tau^2 a_0$ and its vertex-to-vertex 
diameter is $2\tau^3 a_0 = 20.8$\AA{}; these 20\AA{} decagons, of
course, overlap, wherever the \Decagons~ just shared an edge.
Each vertex of the outer decagon has an Al: this is either the 
center of a \Star, or a ring 2 Al atom from an adjoining \Dec. 
Every edge of the outer decagon has two atoms in different layers, 
dividing it in the ratios $\tau^{-2}: \tau^{-3}: \tau^{-2}$;
these are usually both TM, but are Al/Co where they belong to
ring 1 of the adjoining \Dec.

\remark{MM commented (email 7/22/05)
``About 12A vs 20A cluster size. One possible reason why
12A linkages are rather rare in experimental HREM's could be 
flipped configuration of the cluster inside the (binary) 10.5A Hexagon tile.
I think we never tried to think about such configurations.''
[CLH did not understand this remark.]}

It should be noted that our fixed-site model -- described this
way, via overlapping clusters that cover all of space --
is practically identical to Burkov's model~\cite{bur91}.
This was inspired by an early structure solution~\cite{steu90}
as well as a conjectured real-space cluster~\cite{hir91}, 
 (based on electron microscopy), for
$d$(Al$_{65}$Co$_{15}$Cu$_{20}$).
Burkov's decoration is based on a 
Binary tiling of edge 10.5\AA{}, the same as ours.
This is decorated by overlapping 20\AA{} decagons, known as Burkov clusters,
which share a decagon edge when situated at the far tips of 
a Thin rhombus (here they are separated by 19.7\AA{}),
or overlap when situated across the short diagonal of a Fat rhombus.

Burkov's atom sites are nearly the same as ours, but
the chemical species are somewhat different (note he made
no attempt to distinguish among TM species.)
Most importantly, Burkov's ring 1 consists of ten TM atoms, and
furthermore the Small vertices of his Binary tiling (our \Star~ sites)
are generally decorated by Al$_5$TM$_5$, whether Even or Odd:
thus, his structure model is 10-fold symmetric where ours is
pentagonal.  (The Small vertex decoration must be modified where the 
clusters overlap, and thus ring 4 deviates a bit from 10-fold  symmetry.)
The main other difference is that Burkov has no ring 2.5 atoms, 
but has two ring 3 Al atoms on every edge of the \Dec; 
if those atoms were allowed to escape the fixed ideal sites,
as in Sec.~\ref{sec:relax}, they will probably run to exactly
the same locations (within ``channels'')
as they did from our different ideal sites.
After our studies (of density variations, and relaxations
as in Sec.~\ref{sec:relax}, it is clear that Burkov's model
is unphysically ``overpacked'' with Al atoms in the last-mentioned
places.

\CLH{to MM CLHNEW: Above note added in response to
your comments 1/25/06.}
 
\section{Relaxation and Molecular Dynamics Annealing}
\label{sec:relax}

\MM{(9/05): [CLH notes: this comment was made in the context
of the Decagon linkages from viewpoint of CoAl9 clusters, 
has been moved to here.]
"In the 8A puckered structure,
the Co$Al_9$ are the crucial stabilizing cluster (and I will add numbers
for energies supporting this claim). Looking at W-structure, you can
see Co$Al_9$ organized symmetrically around \Dec, but also around two
other Stars. Further, the puckering analysis suggests that Co pentagon
around central Co atom is integral part of the cluster. So at least
these additional 5 Co, if not another 5 Al forming usually 5-gon 36deg
rotated with respect to Co 5-gon, should be considered being a part
of the cluster."  [Where to insert this observation?]}

Up to this point, we have reported analyses of the simulations using
rigid site positions. This section addresses more realistic
configurations of atoms found when the final results are put through
relaxation and molecular dynamics (MD).  Our approach is similar to
relaxations on the ``basic Ni'' phase~\cite{alnico02}.  However, the
present case differs in that the fixed-site stage, did not resolve
certain alternative configurations that are nearly indistiguishable in
energy, thus we have not yet settled on a set of fixed-decoration
large tiles.  In devising a realistic idealized structure for the
``basic Co'' case, study of the relaxed structures and energies is
inescapable.

In this section, we briefly review the results of relaxations on a
bilayer structure, and then consider the effect of relaxations when
the simulation cell is doubled to $\approx 8$\AA{}.  A subset of atoms
undergo significant displacements out of the planes (``puckering'');
the structure (at least, many Al sites) undergoes a symmetry breaking
to the 8\AA{} period.  The remaining subsections are devoted to
characterizing this ``puckering'', and explaining its origin
theoretically.  The puckering will be the key ingredient of the
explanation for the ordering of cluster orientations
(Sec.~\ref{sec:decLRO} Studies of longer-range correlations of the
puckering will be left to Sec.~\ref{sec:puckerLRO}.

Our standard cycle for these off-ideal-site simulations begins with a
relaxation to $T = 0$, in twelve stages of $\Delta T=$50K each.
We then perform MD with temperature around $T = 600$K; this is rather
low, as our purpose is not to heat the system so much that the gross
structure can change, but only to anneal a subsystem of relatively
loose atoms.
just fine tuning the details.  After MD, we once again relax the
structure to $T = 0 $. This cycle as a whole is called
relaxation-MD-relaxation (RMR).

\subsection{Results of relaxations}

Upon relaxation to $T = 0$ [for both 4\AA{} and 8\AA{} period], we
find that the TM lattice is quite rigid and displaces only slightly
from the ideal positions.  The Al atoms, however, are subject to
displacements as large as $\sim$1.5\AA{}. After RMR, a few of the Al
atoms diffuse a comparatively large distance of $\sim 1$\AA{}
from their original sites, but their new
environments are similar to the original (relaxed) ones.
%%%%%%
\remark{Nan: The atoms diffuse about 1.5\AA{} at most, which is about half an
  interparticle spacing.}

\subsubsection{4.08\AA{} periodicity}
\label{sec:relax-bilayer}

As a preliminary, we relax the same bilayers (cell thickness
$c=4.08$\AA{} in the $z$ direction) as were used in the fixed-site
simulations of Sec.~\ref{sec:fixed-site}.  This excludes {\it most}
possibilities of puckering, and prevent the associated energy
reduction.  Thus, no reliable conclusions can be based on energy
differences that appear in this stage.  
\OMITb{energy differences at
  this stage that are comparable with the energy gains due to RMR.}

\OMITb{ The Al-TM nearest neighbor interaction radius wants ring 1 Al
  to be right over the line joining two TM atoms of a \fiveand
  cluster.  But then, the Al-Al nearest neighbor distance would be too
  short.  Thus, the \fiveand shape after relaxation and MD.  is a
  compromise between a large pentagon and a decagon in projection.}

The \Decagon~ evolves as follows under relaxations: (i) In ring 1, the
Al atoms move inwards towards (but not all the way to) the lines
joining the projections of the Co atoms. Thus, in projection, ring 1
-- initially a decagon (with fixed sites) -- becomes more pentagonal.
(ii) The Al atoms forming ring 2, unlike most other Al sites under
relaxations, retain their positions quite rigidly.  (iii) In rings 2.5
and 3, some Al atoms completely change position. These moves usually
occur so as to increase the number of nearest neighbor
($\sim$2.5\AA{}) Al-Co bonds.

\OMITb {The maximum number of Al atoms every Co atom can be nearest
  neighbors with is 7.  This formation leaves 5 Al atoms in the same
  layer as the 3rd ring Co atom with two in the alternate layer.}

Under relaxation, the \Stars~ are subject to numerous adjustments
which adapt to defects in the \Decagons~ or to deviations in the
stoichiometry from ideal.  The ideal occupation in the \Dec~ involves
about a total of twenty Al atoms in rings 2.5 and 3.  If any \Decagon~
is lacking these Al atoms on an edge adjacent to a \Star, Al atoms
from the \Stars~ tend (under relaxation) to move towards the vacanccy
in the \Dec's 2.5th/3rd ring. Presumably this is favored because it
forms the maximum possible number of Al-Co nearest neighbor bonds to
take advantage of the strong attractive potential.

\subsubsection{8.16\AA{}  Periodic Structures}
\label{sec:doubled}

Relaxing an $8.16$\AA{} periodic structure will cause the same general
relaxations as described in the $4.08$\AA{} periodic simulations.  In
addition, Al atoms in ring 2.5/ring 3 tended to run to new locations,
in which they are displaced in the $z$ direction out of the layers.
This puckering develops as a spontaneous symmetry breaking,  
local or (usually) global, wherein all displacements occur
in two of the atom layers (identical, except that all the $z$ displacements
are reversed) while the other two layers are mirror symmetry planes. 
The atomic arrangement in either mirror layer looks virtually
identical to a 4.08\AA{} structure after RMR, but some Al sites
differ between the two mirror layers.
On the other hand, the TM atoms stay very close to ideal sites, 
they pucker very little even in layers where symmetry permits it,
and their positions remain practically identical in the two mirror 
layers (i.e. the TM lattice preserves the 2-layer periodicity)
Usually, the layer in which a nearby \Dec~ has its 
central Al atom becomes a puckering layer,  
whereas the layer in which the ring 1 Co atoms sit 
becomes a puckering layer, as will be
justified in Subsec.~\ref{sec:channels}
and Appendix~\ref{app:channel-anal}.

%%% REMOVED FIGURE (Fig 7 in 9/9/05 draft)
% \includegraphics[width=3.1in,angle=0]{Wpuck.eps}
% \caption{
% (a).  
% Puckering pattern in simulated W(AlCoNi) cell.
% [This is the same configuration
% compared to the real W(AlCoNi) phase in
% Fig.~2 of Ref.~\onlinecite{Gu-letter}]
%%%  
% ~\ref{fig:WA-WB}.  The view is along the {\it WHICH} axis.  Light gray
% circles are Al atoms, dark circles are TM.  (b). {\bf Nan: to be added
%   sometime}.  Similar plot for the experimental W(AlCoNi) cell of
% Ref.~\protect\onlinecite{Su02}.  \CLH{Maybe some remarks are needed --
%   in a footnote? -- about Sugiyama's convention for setting the
%   crystal structure of his cell.  Maybe remove this.}  }
% \label{fig:Wpuck}
% \end{figure}

\subsection{Aluminum Potential Map}
\label{sec:Alpot}

\begin{figure*}
\includegraphics[width=5.6in,angle=0]{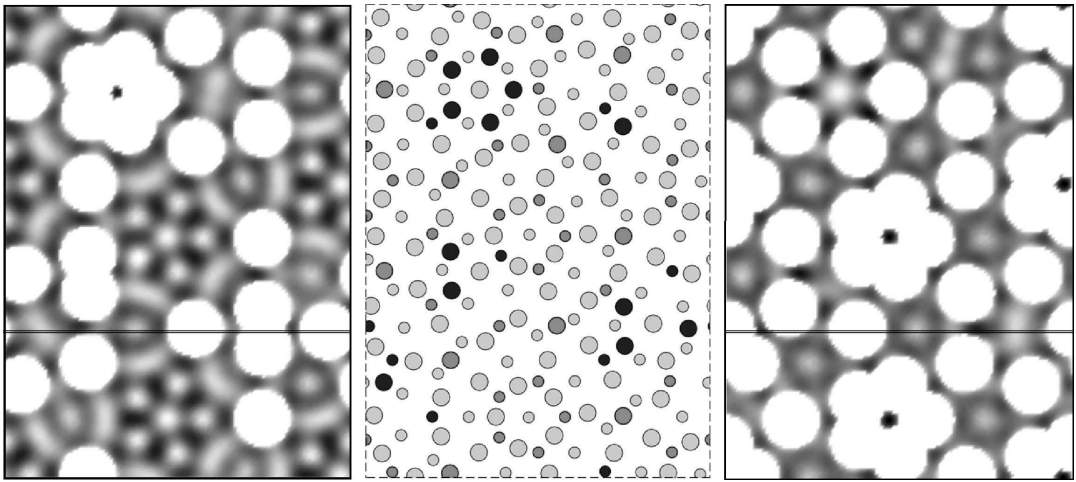}
\caption{(a) Aluminum potential map $U_{\rm Al}(\rr)$
  in the TM poor layer. The double line shows the intersection with
  the $z$-slice in Fig.~\ref{fig:ycut53}.  (b)The central image shows
  the actual atomic coordinates under RMR.  Small circles represent
  atoms in the TM rich layer, which will become the mirror later under
  period doubling. Larger circles are in the TM poor layer, which will
  pucker under period doubling.  (c) Al potential map in the TM-rich
  layer.  \LATER{add: ``Edges of 4A tiling are shown.''}  }
  \remark{Nan Gu email 1/28/06:  The entire potential
  map exercise was executed in Angstroms, but Rydbergs. 
  Fig. 11 has this problem as well (but CLH relabeled axes).}
\label{fig:b4a}
\end{figure*}

%\begin{figure}
%%% \includegraphics[width=3.1in,angle=0]{zcut15.eps}   %% Original size
%\includegraphics[width=2.4in,angle=0]{zcut15.eps}
%\caption{}
%\label{fig:zcut15}
%\end{figure}

\begin{figure}
%%% \includegraphics[width=3.1in,angle=0] 
%% {ycut53.eps}  Original size
\includegraphics[width=2.0in,angle=0]{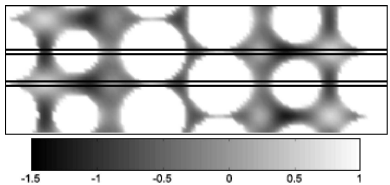}  
\caption{
A slice of the Al potential map along the $c$ direction of  the 
same configuration as Fig.~\ref{fig:b4a}.
The middle portion is aligned on an  edge shared by two \Decagons. 
The top double line shows  intersection with the cut
in the right image of Fig.~\ref{fig:b4a} while the bottom
double line shows the intersection with the left image.
At bottom, the color/shade  scale (in eV) is given for
all our Al potential  maps.}
\label{fig:ycut53}
\end{figure}

Here we introduce a general framework to predict or rationalize
the optimum positions of Al atoms, independent of the fixed-site
list.  It relies on the assertion made in Sec.~\ref
{sec:pairpot}: one first places the transition metals (with their
long-range interactions) into a sort of rigid quasilattice, 
and then optimizes the arrangement of Al (with their weak mutual interactions)
around the TMs.  To make this scenario quantitative we introduce 
the {\it Al potential function} $U_{\rm Al}(\rr)$:
   \begin{equation}
           U_{\rm Al}(\rr) \equiv  
           \sum _{\rr'} V_{\rm AlCo}(\rr-\rr')
            + \sum _{\rr''} V_{\rm AlNi}(\rr-\rr''), 
   \label{eq:Alpot}
   \end{equation}
where $\{ \rr' \}$ and $\{ \rr'' \}$ are Co and Ni sites.  This is
directly analogous to the potential (for a test charge) in
electrostatics, with the replacement electron $\to$ Al atom, and
Coulomb potential $\to$ pair potentials.  It is convenient to study
the potential based on ideal positions for all transition metals
while omitting any Al.~\cite{FN-AlpotwithAl}

Figs.  \ref{fig:b4a} and ~\ref{fig:ycut53} show 
two dimensional slices 
(in planes of {\it local} mirror symmetry) with the energies 
$U_{\rm Al} (\rr)$ depicted grayscale.  
The $U_{\rm Al} (\rr)$ functions plotted in this paper
were produced for $\rr$ on a discrete grid of points covering 
the unit cell, taking the TM positions in a low-energy 
configuration from the 4.0\AA{}-edge
(bilayer) fixed-site Monte Carlo simulation (Sec.~\ref{sec:fixed-site}).
This had first been put through RMR: the only effect 
on the Al potential map is to make it slightly more realistic, 
on account of the small displacements the TM atoms undergo 
in response to the ``typical'' Al distribution.
The gray scale representing energy was cut off at a maximum around 
$+1$ Ryd to hide the large (and irrelevant) variations
of $U_{\rm Al}(\rr)$ inside the hard core of each TM.

How is the potential function interpreted to yield a
set of Al sites?  We start by noticing $U_{\rm Al}(\rr)$ has a set
of rather sharp and deep local minima; each is where several
spheres coincide, representing minima of Al-TM potentials around
different TM atoms.  Each sharp well, starting with the deepest,
should get filled with one Al atom. (When minima are separated by 
less than $\sim 4$\AA{}, we must take into account the significant
Al-Al interaction, but this is not a serious worry for
this stage, since these deep minima are well isolated.)
The Al sites filled in this fashion include the central atom and 
ring 1 of the \Decagon~ cluster (at center of Fig.~\ref{fig:b4a}(c) and
(a), respectively), or the analogous atoms in the \Star~ (upper
left in Fig.~\ref{fig:b4a}(a) and (c), respectively), 
as well as five of the ring 2 Al atoms 
(Fig.~\ref{fig:b4a}(c) center).~\cite{FN-anisowell}

The sites associated with single, deep wells were easily
discovered without the help of the Al potential map: they
are the unproblematic Al atoms in the fixed-site ideal 
structure (e.g., Al in ring 1 and 2 of the \Dec~) that
were obvious even in our first stage simulations.
The potential map offers the following 
advantages over simulation: 
(i) it helps explain the structure from microscopics;
(ii) it shows the energy barriers for an Al atom to pass between
different local minima, which illuminates how Al atoms
diffuse between sites during MD and relaxation;
(iii) it can locate potential
minima that lie outside of the atomic layers;
(iv) it reveals potential wells which are moderately deep, 
but extended rather than sharp, which require more sophisticated
treatment (Sec.~\ref{sec:channels}).

The Al potential map has a complementary
relationship to another diagnostic, the ``site energies'' described
in Sec.~\ref{sec:siteE}, below.  The former identifies good sites that
are currently {\it not} occupied; the latter identifies unfavorable
sites that currently {\it are} occupied.  Together, they may be
used to guide modifications by hand of idealized structures, so as
to improve the energies.

\subsection{Channels and puckering}
\label{sec:channels}

The isolated deep wells of $U_{\rm Al}(\rr) $ do not accomodate all
the Al atoms.  Indeed one-dimensional ``channels'' are evident, along
which the Al potential is low and comparatively flat.  Channels appear
between two columns of TM (especially Co) sitting in alternate layers,
as shown in Fig.~\ref{fig:channel-Al} (a); the TM are the
white disks in the middle of Fig.~\ref{fig:ycut53}.  These TM columns
typically lie (in projection) on adjacent vertices of the 4\AA{}-edge
tiling.  Looking at Fig.~\ref{fig:ycut53}, 
a vertical slice through the periodic layers,
we see how the Al has a potential trough
which appears in the center as a vertical chain of dark triangles, 
pointing in alternating directions.  
The track of the channel bottom roughly 
consists of line segments forming a ``zigzag'' pattern,
so as to connect the ideal Al sites that fall between
the Co chains in each layer. 
[Fig.~\ref{fig:channel-Al} (a) shows how, wherever the 
channel crosses an atom layer, it passes through an ideal Al
site that is nearly at the minimum of three Al-Co potentials.]
%%%%%%%%%%%%%%%
\remark{Each segment is nearly the bisector of
the Co-Co bond, and where the channel crosses that bond it is only
2.24\AA{} from each Co, so it is not surprising this is a local
maximum along the length of the channel; the summed potentials from
the two more distant Co atoms also have a maximum at this point.}
%%%%%%%%%%%%%%%
One expects Al atoms would be comparatively free to slide
along such a channel.
Our plots of $U_{\rm Al}$ are complementary to those of the
time-averaged Al density in a molecular dynamics simulation in
Ref.~\onlinecite{alnico02}, from which ``channels'' were originally
inferred to occur (in the ``basic Ni'' structure.)  

\LATER{YES. Refer to diffusion as in Gaehler papers.}

\subsection{Origin of puckering in channels}

\begin{figure}
\includegraphics[width=0.6in,angle=0]{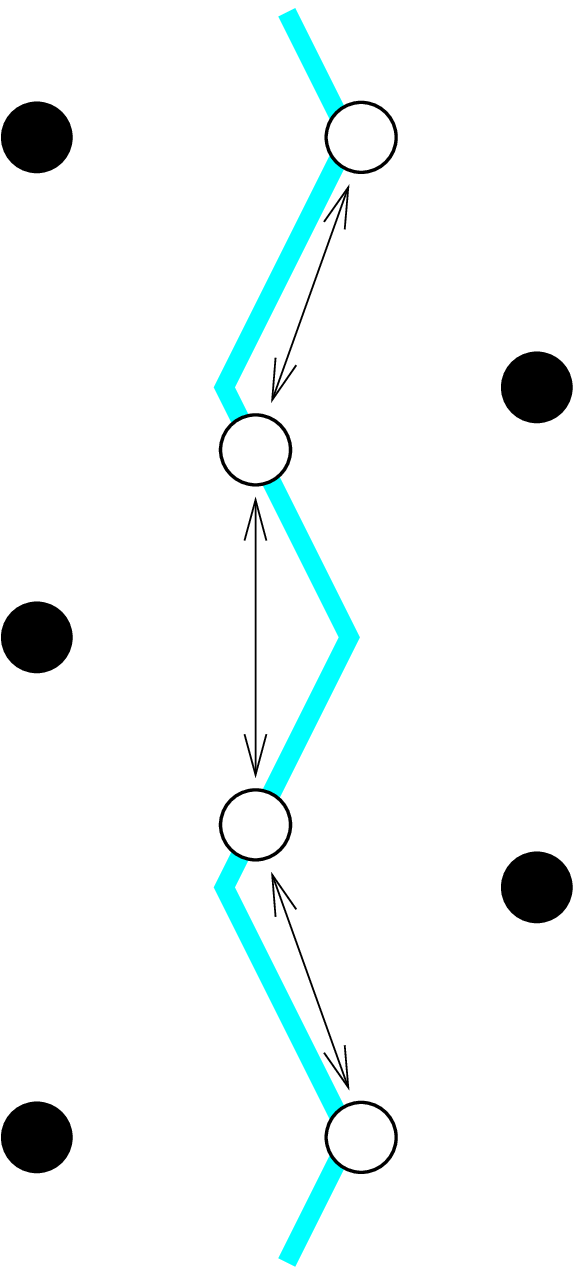}
\includegraphics 
%%% [width=2.2in,angle=0]{channel6363.eps}  
[width=2.2in,angle=0]{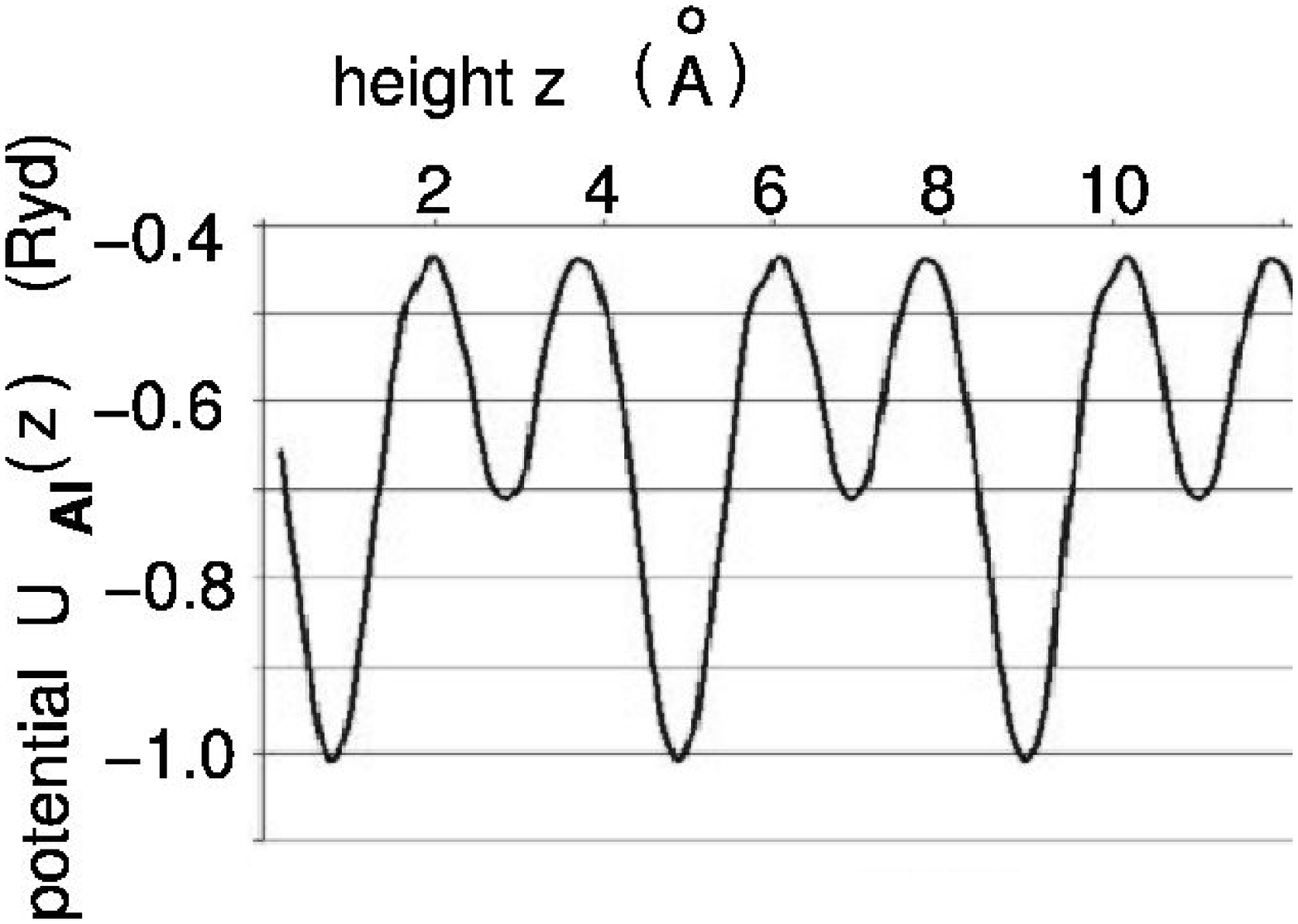}  
\caption{(a). Schematic of a channel  
(shaded zigzag line) between two
columns of Co atoms.  The Al in each channel are
close enough to feel a strong mutual repulsion (arrows).
\remark{Ideally, to place this  
sketch alongside the graph  
of the z-cut potential map?}
(b). The Al potential energy (in Rydbergs)] along
the bottom of an actual puckering channel.
\LATER{I fixed figure to read in Ryd, but need to
make a new figure that is not bitmapped.
Also, combine in xfig to add (a) and (b).}
}
\label{fig:channel-Al}
\end{figure}

In such a ``channel'', the TM interactions do not suffice
to fix Al sites.  We must take Al-Al
interactions into account in order to predict the Al occupation.
We start with the Al potential function 
$U_{\rm Al}(\rr)$ defined in Eq.~(\ref{eq:Alpot}).
Let us approximate a channel with a one-dimensional vertical track
parametrized by $z$.  As evident in Fig.~\ref {fig:channel-Al}(b), the
Al potential variation along the trough is well modeled by
\begin{equation}
    U(z) = U_0 - U_{c/2} \cos (4 \pi z/c)
     + \sigma U_{c} \cos  (2 \pi z/c)
  \label{eq:Uz}
  \end{equation}
Here $\sigma \equiv +1$ (resp. $-1$) in (\ref{eq:Uz}), 
if the distant Co atoms are in even (resp. odd) layers.  
From plots like Fig.~\ref{fig:channel-Al}(b) one can read off
  $U_{c/2} \approx$ 3 eV and $U_{c}\approx$ 2.5 eV~\cite{FN-U0value}.

\LATER{CLH must explain $\sigma$ better here.
Yes, improve definition $\sigma$!}

Let us explain the coefficient $U_{c/2}$ in (\ref{eq:Uz});
for simplicity, we neglect $U_c$ until Subsec.~\ref{app:channel-Uc}.
If the {\it adjacent} TM
columns were all Co, and we included only interactions with them,
their twofold screw symmetry would guarantee $U(z)$ has period
$c/2$, modeled by the first term of (\ref{eq:Uz}).  (Nearby Al in
non-channel sites have the same symmetry.)  Note that along the
track, the minima of $U_{\rm Al}(z)$ lie right at the level of each
atom layer.  
(Those locations are equidistant from three Co atoms at $R\approx 2.5$\AA{}, 
the very strong minimum of $V_{\rm Al-Co}(r)$.)
This explains the period and sign of the first non-constant
term.

\LATER{subsubsectiondivision here?}

Conceivably, in some materials puckering could arise 
because the single-Al potential would have 
minima out of the atom layers; but that is not the case in
Al-Co-Ni, so puckering must indeed be
a consequence of the short range Al-Al repulsions ({\it combined}
with $U_{\rm Al}(\rr)$).

If both local minima were occupied in each bilayer , the Al-Al spacing
would be not much more than $c/2 \approx 2.0$\AA{}, which is far too
close; on the other hand, if only the best minimum in a each bilayer
was occupied, and no other Al sat close to a channel, the total Al
content would be too small.  The solution is that there is room for
{\it three} Al to fit in every {\it two} bilayers, as shown in
Fig.~\ref{fig:channel-Al}(a).  Since this makes the period to be $2c$,
it is a symmetry breaking in each channel.  The mean vertical spacing
$2c/3 \approx 2.72$\AA{} is a bit closer than the Al-Al hardcore
radius (see Table~\ref{tab:potentials}]) so the Al-Al
forces are probably dominant.

{\sl Mathematical details are worked through in 
Appendix~\ref{app:channel-anal}.}
%%%%%%%%

Mathematical details -- how the collective energy of 
three atoms in a channel depend on their collective  position
-- are worked through in Appendix~\ref{app:channel-anal}.
The $U_{c/2}$ term favors atoms to sit in layers. The $U_c$ term
(as shown in the appendix) favors an individual atom to
avoid the layer which is (locally) TM-rich, but when there
are three atoms it favors one of them to sit in that layer, 
which is a point of local $z$ mirror symmetry in that channel.
The result is that if a layer is globally TM-rich, it becomes
a global mirror  symmetry plane. 

We have been deriving the configuration of a channel
assuming it has a fixed number of Al atoms.
Actually, of course, this number is variable. The optimum  occupancy of
each channel must be a function of the Al chemical potential;
equivalently (in our simulations with fixed Al content) it
is the result of competition with competing kinds of Al site
(as measured by the site-energies). It will certainly change
as a function of TM composition (changing the number of
Co columns) and the total Al density (since a home must
be found for every Al atom).

\remark{We could compare site-energy of Al in an isolated
  well of $U_{\rm Al}(\rr)$ with the value of (\ref {eq:Etot-zbar}) to
  decide whether to move Al's from the former type of site to the
  latter.}

\subsubsection{Comparison  to fixed-site results}

It is profitable to revisit the ideal-site models (Sec.~\ref
{sec:ideal-deco}) with the ``channel'' picture of the Al placement.
All of those variable Al's, e.g. those constituting ring ``2.5'' in
the \Dec, were in fact channel Al.  However, in the ideal-site models
they were accomodated with a periodicity $c$ everywhere: how could
that work, seeing that some channels would have to fit in four Al
atoms? The answer is that the mirror-layer Al's are {\it all} in
positions offset from  channels, 
like the merged-Al site to be discussed in Sec~\ref{sec:pucker-units}.  
They are never in
line with the Al in puckering layers (which don't pucker
in a fixed-site approximation), and the extra $xy$ displacement allows the
mirror-layer Al to be accomodated without puckering.  Such unpuckered
configurations of the puckering units are observed to compete with the
puckered configurations in actual simulations using RMR (see
Sec.~\ref{sec:puckerLRO}).

\LATER{YES. Move this to the proper place, and make it into a footnote.
(Actually we don't say at all enough about what happens if the
orientations turn out to be different.)
If the clusters have ``antiferromagnetic'' orientations, then the {\it
local} patterns of puckering are similar to the ``ferromagnetic''
case, but there are regions (around clusters of the respective
orientations) in which there is a different choice of which layer
becomes a mirror layer.}

\section{Long Range Order of  \Dec~ orientations}
\label{sec:decLRO}

In referring to the ``orientation'' of the \Decagon,
we have always meant that of its Al$_6$Co$_5$ core, since
the rest of the cluster, as laid out in Sec.~\ref{sec:clusters}, 
is ten-fold screw symmetric: only the occupation of ring 1, 
and level of the Al atom at the center, break the symmetry.
The orientation relationship of neighboring \Decs~ is essential
because this was a prerequisite for extending our simulation results 
to a full-fledged decoration model (Sec.~\ref{sec:ideal-deco}),
and because it is tied to the differentiation of layers into
``mirror'' and ``puckered'' layers, once relaxation is allowed
in a structure with periodicity four (or more)layers
(Sec.~\ref{sec:relax}).

The ring 1 atoms from adjacent clusters are basically too distant
to have a significant direct interaction:
the shortest interatomic distance between the respective 
first rings is $2\cos 18^\circ \tau a_0 \approx 7.7$\AA{}, 
whereas our potentials were cut off around 7\AA{}.
Hence we must look for more subtle, indirect mechanisms to favor a 
relative orientation.  Indeed, we have already encountered 
various ways the positions of the Al atoms in rings 2.5 and 3, 
or the substitution of Co(3) by Ni(3), is modulated by the core and
thus reduces the symmetry of the outer portion of the \Decagon
(see Secs.~\ref{sec:MC-4A}, \ref{sec:PB}, and
\ref{app:channel-Uc} in particular).

The interaction must be mediated by other atoms in one of two possible ways.  
Firstly, neighboring TM atoms in the \Stars~ interact with each
other and also respond to the first-ring orientations of adjoining
\Decagons: we suggest this is the origin of the ``antiferromagnetic''
term (Subsec.~\ref{sec:decLRO-AF}) Secondly, the variable Al atoms in
rings 2.5 and 3 are within range of ring 1 of both clusters. This
contribution appears to favor ``ferromagnetic'' order
(Subsec.~\ref{sec:decLRO-F}) and is probably the more important one,
both for the fixed site-list and for the relaxed, puckered structures
with $c'\approx 8.32$\AA{}.

\subsection{Effective Ising Hamiltonian for orientations}
\label{sec:decLRO-Ising}

Let us formulate the problem as an Ising model.
We take as given a fixed network of $N_D$ \Decagons~ placed on 
the ``large'' vertices of some configuration of binary-tiling rhombi.
Each \Dec~ may have either orientation, which is labeled by an 
Ising spin $\sigma_i=\pm 1$ associated with that \Dec.  
This does not unspecify all the atoms: there are options
in the ring 2.5/ring 3 Al atoms, explained at length
in Secs.~\ref{sec:small-DHBS} and \ref{sec:deco-enumerate}, as
well as TM atoms (especially in the \Stars).
For each of the $2^{N_D}$ possible combinations of $\{ \sigma _i \}$, 
we define the orientation effective Hamiltonian,
$\HHor(\{\sigma_i \})$, to be the minimum energy
after all those other degrees of freedom are optimized.~\cite{FN-rotateDec}
We presume the orientation effective Hamiltonian 
is well approximated by an Ising model,
  \begin{equation}
       \HHor = - \sum _{\langle i j \rangle} \Jor \sigma_i \sigma_j , 
  \end{equation}
where $\langle ij \rangle$ means each nearest-neighbor pair
is included once.

If $\Jor >0$, the ground state obviously has $\{ \sigma_i \}$ 
all the same (clusters oriented identically), which we call ``ferromgnetic'' 
(FM) in the Ising model language. If $\Jor < 0$, it favors an 
``antiferromagnetic'' (AF, also called ``alternating'') arrangement 
in which neighboring clusters always have opposite orientations;
that is possible, however, only if the \Dec~ cluster network is
bipartite.  That is not always the case on the Binary tiling -- 
e.g. groups of five \Decs~ can form pentagons -- but in both our
simulation cells, the network happens to be bipartite.

In the rest of this section, we first report numerical studies 
of the energy differences between different arrangements, and 
then give physical explanations in terms of the pair potentials 
and of structure motifs (identified in previous sections).
There is one story for the fixed-site simulations, and a
different one for the (physically pertinent) relaxed and 
MD-annealed simulations.  Another complication is that
the answers depend on the overall density.
Finally, the results show a strong dependence on the particular 
simulation cell being used. [We used both the standard 
32$\times$23 cell and also the ``W-phase'' (40$\times$23) cell.]

\subsection{Orientation dependent energies (numerical)}

For the numerical calculation, our procedure was to perform a 
series of Monte Carlo runs given  FM orientations and 
a similar series under the same conditions for AF orientations,
recording the lowest energy from each run (which is our
empirical approximation of $\HHor$, as just defined).
We average over tens of runs, since the run-to-run fluctuation
usually exceeds the AF/FM energy difference.
Simulations were done for fixed-site (4\AA{} tiling)
Monte Carlo on the 32$\times$23 tiling as well as the W-phase
tiling, and also with the ``RMR'' procedure (relaxation after
MD annealing). 

It was simple to constrain the orientations of each cluster.
Recalled that in the rhombus decoration for our 4\AA{} 
MC simulations (Sec.~\ref{sec:MC-4A}), every \Dec~cluster is forced 
in a particular orientation by the five Fat 4\AA{} rhombi that 
(with five Thin rhombi) make up the decagon of 4\AA{} tiles. 
In particular, only one of the two layers is even available as a 
candidate site for the central Al atom.

\LATER{URGENT! rewrite the text so as to tell correctly 
what config we tried.  Need to redo arithmetic of 
extracting bond energy from the table.}

\remark{
Nan's names for config's on the 23x32 tiling 
(from an old email, and confirmed 1/28/06).
`SameO', all 4 decagons same orientation. 
`DiffO' where two decagons have the same orientation if they 
   have the same $y$ coordinate; in otherwords, they are
stacked in layers which alternate.
   `DiffO2' alternates every other decagon orientation.
With a nearest-neighbor interaction, `DiffO2' would be
the case favored if it is negative, but Nan's viewpoint
was that `DiffO' was more reasonable, hence it appears
only `DiffO' was tried in the simulations.}

The results are given in Table~\ref{tab:Eor}.
Let $\Eav\equiv (\EAF+\EFM)/2 \Nat$,  where 
$\EAF$ and $\EFM$ are the total energies
for AF and FM orientations, respectively.
Also, $\Ediff \equiv (\EAF-\EFM)/N_D$, 
where $N_D$ is the number of \Decagons per cell 
For these particular cells, in which neighbors are always 
opposite in the AF case, we can immediately extract 
$\Jor = 2 \Ediff/ \Zbar$, where $\Zbar$ is the average
coordination number of the cluster network.
(Note $\Zbar=3$ in the 32$\times$23 tiling and $\Zbar=2$ in
the ``W-phase'' tiling, so $\Jor$ is equal to the numbers
in columns 5 or 6, or the number in column 3 divided by 1.5).

Consider first the fixed-site simulations. In the 32$\times$23 cell
we had $\EFM < \EAF$ for $0.068$AA{}$^{-3}< \rho < 0.074$AA{}$^{-3}$ 
That is the whole range of physically reasonable
densities; at higher or lower densities, $\EAF < \EFM$ apparently.
(Of our cells, the 32$\times$23 is the closest approximant 
to fivefold symmetry, i.e. zero perp-space strain.)
On the other hand, in the W-phase cell, we see $\EAF$ was
always lowest -- though for $\rho \approx 0.070$AA{}$^{-3}$, $\EFM$
was nearly as low. In other words, the concentration dependence is 
similar in both cases, except 
   \begin{equation}
     \Jor(\text{W cell}) \approx\Jor(32\times 23 \text{cell}) - 0.2{\rm eV} .
   \end{equation}
On the other hand, when the relaxed energies are compared, 
we found $\EFM < \EAF$ in all cells and at all realistic densities.
In all cases, the interaction $\Jor$ is of order $0.1$ eV.

%%%%%%%%%%%%%%%%%%
\begin{table}
\begin{tabular}{|lrrrrrr|}
\hline
    &
32$\times$23 &  &    W-cell & & W-cell & (rel.) \\
%%%%%%%%%%%%%%%%%%%%%%%%%%%%%%%%%%%%%%%%%%%%%%%%
$\rho$ & $\Eav$ &  $\Ediff$  &  $\Eav$ & $\Ediff$ & $\Eav$ & $\Ediff$ \\
(\AA{}$^{-3}$)  & (eV/at) & (eV) & (eV/at) & (eV) & (eV/at) & (eV) \\
%%%%%%%%%%%%%%%%%%%%%%%%%%%%%%%%%%%%%%%%%%%%%%%%
0.66 & $-0.471$ &  $-0.101$ & -- &  -- & -- & -- \\
%%%%%%%%%%%%%%%%%%%%%%
0.68 & $-0.453$ & $-0.259$ & $-0.447$ & $-0.007$ & -- &  -- \\
%%%%%%%%%%%%%%%%%%%%%%
0.70 & $-0.430$ &  0.181 & $-0.431$ &  $-0.123$ & $-0.561$ &   0.1 \\
%%%%%%%%%%%%%%%%%%%%%%
0.71  & $-0.414$ &   0.197 & $-0.410$ & $-0.083$  & $-0.5445$ &   0.2  \\
%%%%%%%%%%%%%%%%%%%%%%
0.72  & $-0.395$ &   0.142 & $-0.394$ &  $-0.089$ & $-0.5365$ &   0.2  \\
%%%%%%%%%%%%%%%%%%%%%%
0.74  & $-0.348$ &   0.012  & $-0.335$ & $-0.143$ & -- &  -- \\
%%%%%%%%%%%%%%%%%%%%%%
\hline \hline
\end{tabular}
\caption{ \footnotesize
Energies depending on \Decagon~ orientations,
as a function of number density $\rho$.
Here $\Eav$ is the mean energy/atom
[averaged over the cases of alternating (AFM)
 and identical (FM) orientations].
Also, $\Ediff$ is the energy cost (per \Decagon)
of opposite orientations.
The number of atoms in the simulation cell
is $\Nat=207$ for the 32$\times$23 simulation cell
and $\Nat=268$ for the W-phase (40$\times$23) cell,
while the number of clusters $N_D=4$ for both cells. 
The first four columns are fixed-site simulations,
the last two columns were relaxed after MD.}
\label{tab:Eor}
\end{table}

\LATER{YES, CHECK. 
Plots were runs Nan handed CLH in 4/21/05, 5/3/05, and 5/10/05;
The ``relaxed'' columns are eyeballed from this graphs 
-- my reading uncertain by +-0.0003; but I noted 
for the relaxed case Nan told the means too -- look
that up to replace this.}

\remark{
Nan always simulated W-cell with 268 atoms or 32x23 with 207 atoms.  
MM points out that $(\tau/2) 268 =\sim 216$, which shows the
difference between the (higher) W-phase density and
0.068at/$\AA^3$ density. [The best VASP-refined model of W has 267 atoms 
though, not 268]}

\subsection{Explanations of orientation interactions}

Now let us try to explain the above results.
In the fixed-site case, the data indicate the sign of $\Ediff$
-- i.e., the effective interaction --  varies with composition.
It suggests $\Jor$ (for the fixed-site case) is a sum of
competing terms of opposite sign, and we indeed identified 
both an FM and an AF contribution (below).
The relaxed case is simpler, since the result is more straightforwardly FM.

The enormous difference, in the fixed-site simulations, between 
the 32$\times$23 cell and the W-phase cell, is ascribed to the
quite different relative ratio of \Decs~ to \Stars~ in the
respective cells.  That means that, if the stoichiometry is
constrained to be the same, the actual TM content of the \Stars~
is quite different, which presumably affects the interaction
term described next.

\subsubsection{``Antiferromagnetic'' cluster interaction via \Stars}
\label{sec:decLRO-AF}

We use Decoration II (Sec.~\ref{sec:alt-deco}).  Let us assume
each nearest-neighbor pair of Ni atoms in a \Star~ has a repulsive
energy $V_{\rm Ni}$.  (This distance is around 2.9\AA{}, which is not
very good with the Ni-Ni potential: see Table \ref{tab:potentials}.)  
How does this energy depends implicitly on the orientations 
of nearby \Decagons?

First, where a NiNi pair is present on an overlapping of two \Stars, 
it always sits at the center of a Thin rhombus of the 10.5\AA{}-edge 
Binary tiling. In this environment it can be shown that we get exactly 
$2 V_{\rm Ni}$ from Ni pairs in the respective rings, independent of 
the orientations of the \Dec's centered of the far tips of that 
Thin rhombus, so this contribution is an uninteresting constant.  

Otherwise, it can be shown
that every pair of adjoining \Decagons, with the same (FM) orientations,
creates one Ni-Ni nearest neighbor costing an additional
$V_{\rm Ni}$ (not present in case of AF orientations).  That gives
$\Jor= - V_{\rm Ni}$, favoring AF arrangement.

Notice that if the TM content were to be changed, there would be
additional opportunities for optimizing the Ni arrangement in the
\Star. Thus the effective interaction of \Decagon~ orientations may
involve the $\sigma_i$ for {\it all} \Decagons~surrounding the \Star.  
In that case, it is not clear if the effective interaction
remains pairwise, nor whether it remains AF in sign.

\subsubsection{``Ferromagnetic'' cluster interactions via Al channels}
\label{sec:decLRO-F}

Clusters want to have the same orientation for about the same reason
that two steel balls, rolling on a mattress, want to be at the same
place.  (Here TM's in the clusters distort Al atoms in channels
in the same way the steel balls distort a mattress.)
We can understand it mathematically in terms of Eq.~(ref{eq:Uz}).
When there are {\it two} distant TM columns near a channel in layers
labeled by $\sigma_i$, $\sigma_j$, then the second coefficient in
(\ref{eq:Uz}) becomes $(\sigma_i+\sigma_j)U_c$.  In the fixed-site
case, when each channel has only one Al atom per bilayer strictly
speaking, its position will be determined by minimizing (\ref{eq:Uz}).
The lowest energy is a term nearly independent of $\sigma_i$ plus
$-U_c |\sigma_i+\sigma_j|$, which is the same as $-U_c(1+
\sigma_i\sigma_j)$ when $\sigma_i=\pm 1$, so we read off 
$\Jor=-|U_c|$ favoring the ``FM'' relation.

In the puckering case, a generalization of the last term of
(\ref{eq:Etot-zbar}) is proportional to $-(\sigma_i+\sigma_j)^2$, so
we obtain a cross-term proportional to $-\sigma_i\sigma_j$ again
favoring ``ferromagnetism.''

\LATER{By expanding the cross-term in we could obtain the actual Ising
  coupling $\Jor$.}
Now, ${V_1}''$ depends very sensitively on how close $2c/3$ 
is to the Al-Al hardcore radius, and consequently so does
$\Jor$.  A corollary is that small changes in the layer spacing 
can have large effects on the orientation order.
%%%%%%%%%%%%%%%%%
\remark{This might be observed in a comparison of different
decagonal phases, all built from similar motifs at the atomic level.}

\section{Systematics of the puckering  pattern}
\label{sec:puckerLRO}

\LATER{As Nan reminds in email 1/28, our conclusions are based on 
multiple relaxations, but only two or three initial fixed-site simulations.
Really, need to do this better!}

We now return to the thread of Sec.~\ref{sec:relax}:
there we understood puckering within an isolated ``channel''
between two columns of TM (usually Co) atoms in alternating layers 
A filled channel contains three Al atoms per four atomic layers.
one in a mirror plane and two atoms 
assigned to the ``puckered'' layers above and below it. 
This picture does not specify {\it which} of the two
mirror layers gets occupied (which determines the out-of-layer
displacements of the other two Al):
this is a {\it local} twofold symmetry breaking. 
In this section, we address the puckering 
correlations,  in particular the relation of the local
pattern to the local geometry of tile packing/cluster network, 
and whether long range order of the symmetry breaking can be propagated.
An effective Ising model helps define the question, but
is inadequate to answer it.  Instead, we focus on ``puckering units''  
defined as the (up to five) channels surrounding a Co column, 
and their Al atoms, which are subject to strong steric constraints.

% This figure was Fig 12 in draft of June 2005, but cut
%   in favor of new way to show puckering.
%------------------------------------------
%  \begin{figure}
% \includegraphics[width=1.79in,angle=90]{nn8A2rmr245.eps}
% \hskip  0.05 truein
%%% \vskip 0.15 truein
% \includegraphics[width=2.00 in,angle=90] {nn8A2puck245.eps}
% \caption{Global puckering pattern in a case that 
% \Decs~ have opposite orientations. The local patterns are similar
% to those shown before, but there is a domain wall.
% (a). One bilayer unit cell (of the 20$\times$23 cell)
% that contains two decagons with opposite orientations.
% The configuration has been put through relax-MD-relax.
% (b). The same unit cell configured to show puckering 
% ONLY in the puckering layer.  
% [The other puckering layer is identical, but 
% the signs are all reversed.]
% \CLH{The code (inset) should
% be explained.  Also, I think the squares in the inset NEED
% to be made bigger.  And if  ALL squares are made even 
% bigger than that, the whole figure can be shrunk 30 percent.}
% \label{fig:af-pucker}
% \end{figure}

To explore the puckering patterns, we performed RMR
on two bilayers, starting from a (4\AA{}~ periodic) fixed-site
configuration (of Sec.~\ref{sec:ideal-deco}), using the
decoration of Fig.~\ref{fig:RulesWideal}
%%%%%%%
\remark{In this decoration, every even \Star~
has five Ni, every odd \Star~ has two Co, and every TM on the
\Decagon~ borders is a Co.}
%%%%%%%%
Several independent relaxations 
(with MD annealing from $T=$700 K to $T=0$ in stages of 50K)
in the ``W-phase'' unit cell.  
gave  substantially similar arrangements (Fig.~\ref{fig:puck-Wsim}).
%%%%%%%%%%%%%%%%%%%%%%%%
\remark{Nan Gu: c 8/06: 50K per step, so 14 steps. 
The default settings were used for what was done at each step.}  
%%%%%%%%%%%%%%%%%%%%%%%%
(Note many channels had $n_\alpha=4$ Al, suggesting either the Al 
content was too large, or the MD time was insufficient to allow 
Al atoms to diffuse between channels. )
We will analyze (in Subsec.~\ref{sec:pucker-stats})
the typical patterns in the puckering units, and
discover the key role of \Stars~ in organizing longer-ranged
correlations.

\subsection{Ising-spin variables and channel occupancies}
\label{sec:pucker-ising}

\begin{figure}
\begin{minipage} {8.66cm}
\includegraphics[width=3.1in,angle=0]{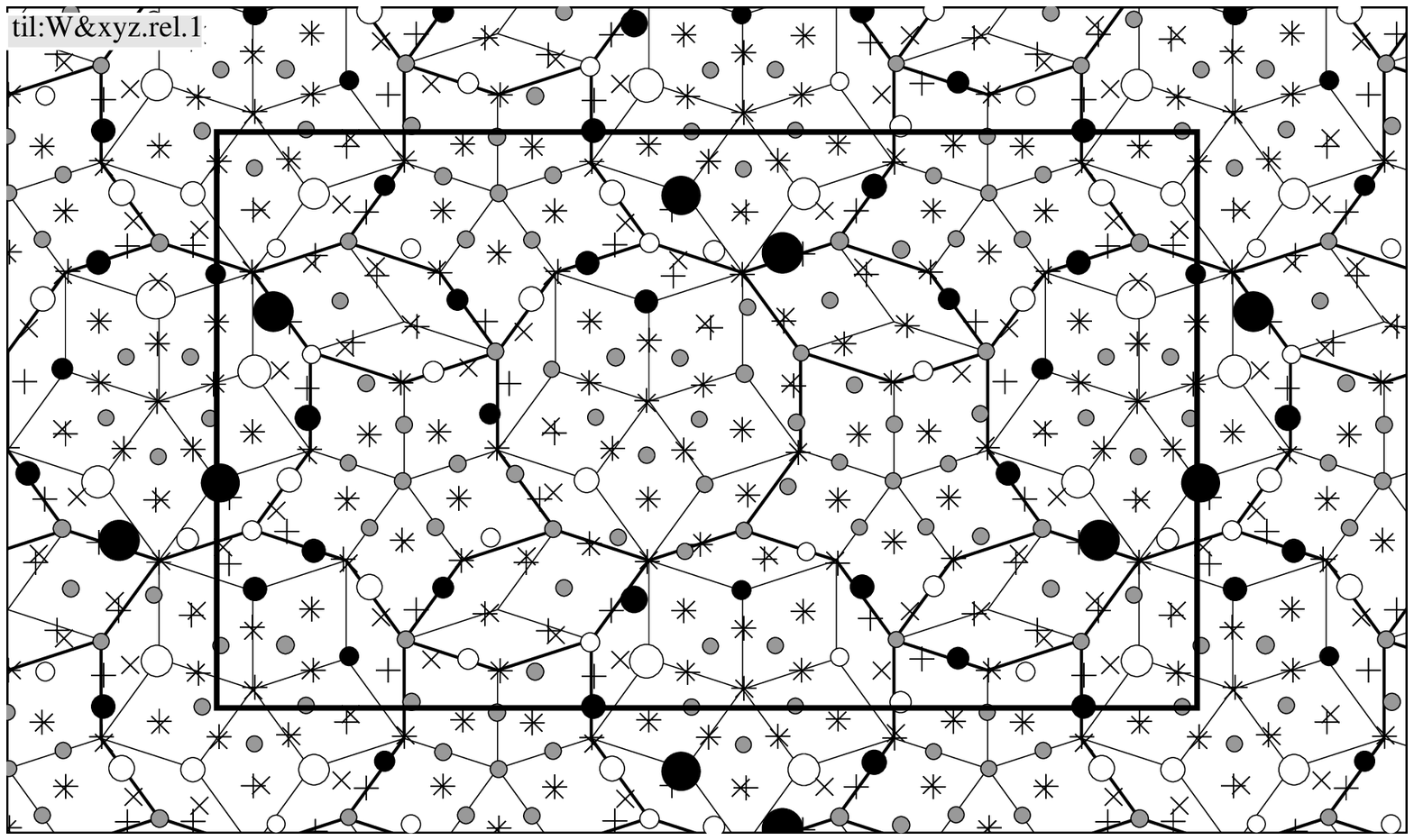}
\vskip 0.15 truein
\includegraphics[width=3.1in,angle=0]{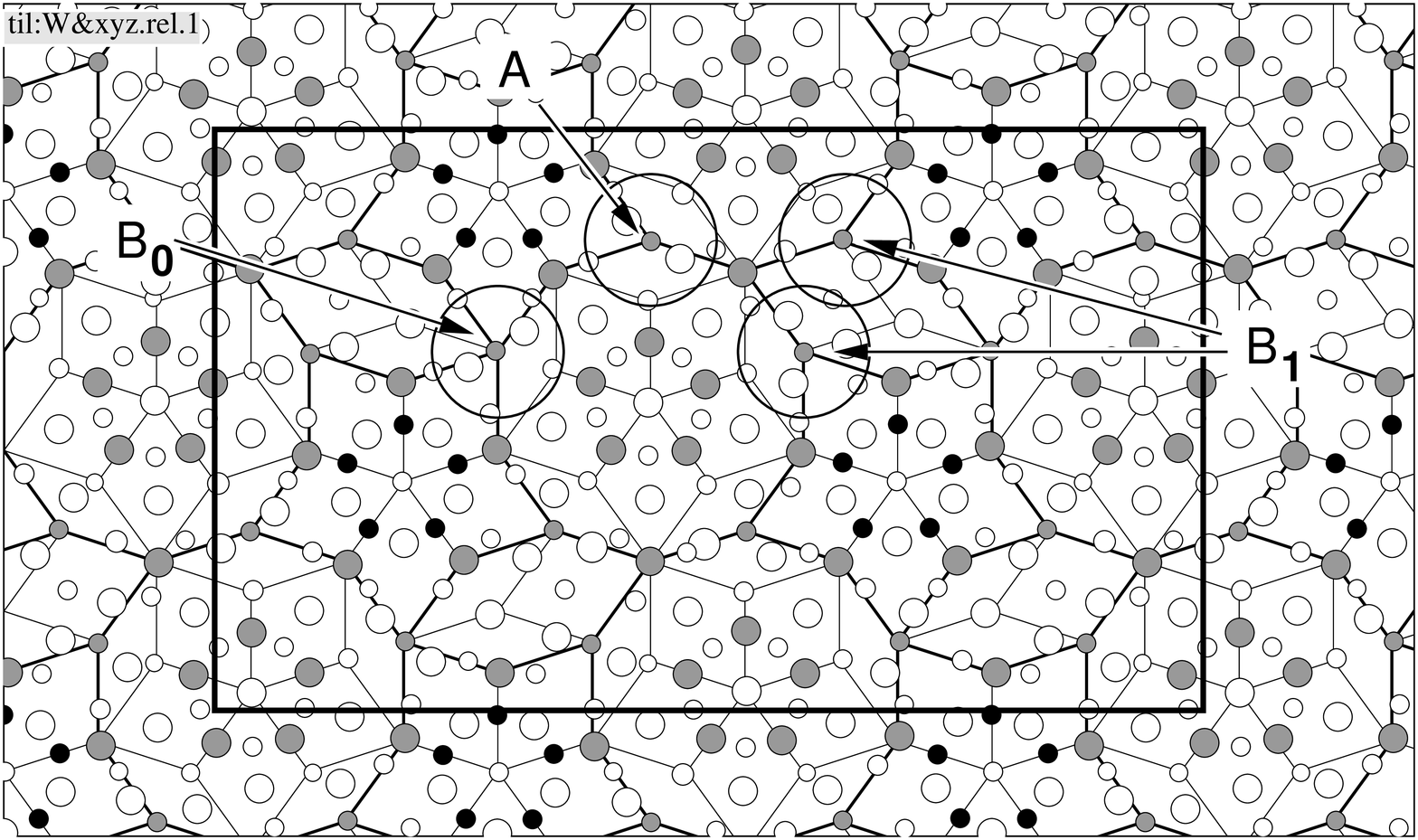}
\caption {Puckering of 8\AA{} structure in a simulation 
in the W-phase cell (after RMR relaxation 
from a 4\AA{} periodic structure, a realization of
the same decoration as in Fig.~\ref{fig:RulesWideal}.
\remark{MM confirms, the original model is not quite the same
configuration as Fig.~\ref{fig:RulesWideal}.}
(a). Puckering pattern, showing three of the four layers.
The ``$\times$'' and ``$+$'' symbols represent, respectively, atoms 
in mirror layers 0 and 2.  White (resp. black) filled circles 
are atoms in puckered layer 1, deviating in
  the plus (resp. minus) sense, where the circle radius is
  proportional to the displacement;  gray circles have small puckering
  displacements.  Puckered layer 3 has atoms in the same places 
(apart from a handful of defects), each deviating in the opposite 
sense from layer 1.  
(b). One bilayer, showing relaxed positions and atom chemistry with 
our usual conventions.  The other bilayer is similar, except at atoms where the
  mirror layers differ; those are always Al and can be identified from
  places in (a) where ``$\times$'' and ``$+$'' do not overlay.
Locations of puckering units are shown by circles,
labeled $A$, $B_0$, and $B_1$ according to their environment
in the tiling, as described in Sec.~\ref{sec:pucker-labels}.
}
\label{fig:puck-Wsim}
\end{minipage}
\end{figure}

One way to formulate the puckering problem is to represent 
the symmetry breaking in each Al channel by an 
Ising spin-like variable~\cite{FN-pucksigma} $\puck_\alpha =\pm 1$.
(Here the index $\alpha$ runs over all $\Nch$ distinct channels.)
Arbitrarily designate one of the mirror layers as layer 0, 
consistently throughout
the system.  Where layer 0 is occupied and layer 1 puckers upward,
we define $\puck_\alpha\equiv +1$; where layer 2 is occupied and layer 
1 puckers downwards, $\puck_\alpha \equiv -1$.  
(Layer 3 always puckers in the direction opposite to layer 1.)
An Ising value $\puck_\alpha=+1$ corresponds on Fig.~\ref{fig:puck-Wsim} 
to a $+$ symbol and a blackened circle, usually on a 4\AA{} tile edge  
and always between a pair of Co (identified in the lower panel)
in different layers; 
similarly $\puck_\alpha =-1$ appears as an $\times$ symbol and a white circle.

\remark{A precedent for the Ising effective Hamiltonian comes
from the approximant Al$_{13}$Co$_{5}$. In that case, 
the geometry is a Hexagon-Boat-Star tiling with a 
lattice constant $\tau^2 a_0 \approx$ 6.4\AA{}, and
stacking period.
Over each vertex sits a column of `pentagonal bipyramid' (PB) clusters
(The pentagon orientation
depends on whether it is an even or an odd HBS vertex: 
the pentagon's edges are normal to the HBS edges.)
Each such column is assigned an Ising spin variable $\puck=\pm 1$
according to which mirror layer the PB is centered in vertically
(layer 0 or layer 2).
The tile Hamiltonian for the interaction of
these columns has the form of an Ising spin model. 
The resulting pattern -- e.g. the orthorhombic vs. the monoclinic 
versions of Al$_ {13}$Co$_5$ -- is determined by the ratio of the 
bond strengths ${\mathcal J} _1$ and ${\mathcal J} _2$ for the
two nearest kinds of neighbors.}

Now imagine computing the total energy for every one of the
$2^{\Nch}$ channel configurations. The result, one hopes, is
well approximated using pair interactions between nearby channels,
giving an effective ``Ising model'' Hamiltonian 
   \begin{equation}
         \HHpuck = 
         \sum _{\alpha \beta} \Jpuck{\alpha\beta} \puck_\alpha \puck_\beta .
   \label{eq:Jpucker}
   \end{equation}
The effective interactions $\{ \Jpuck{\alpha\beta} \}$ would depend on the 
locations of channels $\alpha$ and $\beta$ relative to \Decagon~clusters, and
also on the choice of TM occupancy of nearby sites (Ni versus Co in 
many places, but also Ni versus Al in \Stars.)
The final ground state would be determined by minimizing (\ref{eq:Jpucker}).

The real story is more complex. 
We have presupposed a fixed set of channels, each containing $n_\alpha=3$ Al
and thus having an Ising ``spin'' degree of freedom.  
But if an Al atom is moved in or out of a channel (so $n_\alpha$ = 2 
or 4), the atoms are locked in unpuckered layers.
In the case $n_\alpha=2$, both Al go into a puckering layer
since we found (see Appendix\ref{app:channel-anal}) the Al 
potential is lower there, but they need to pucker only 
negligibly (in response to distant Al).
In the case $n_\alpha=4$, 
some of the Al must deviate sideways and the atom sites are 
essentially an arrangement (using ``ring 2.5'') of the fixed-site 
structure of Sec.~\ref{sec:ideal-deco}, so again there is no
local symmetry-breaking by puckering; in either case,  there
is no longer a spin $\puck_\alpha$  at that place. 
(Of course, a new $\puck_\beta$
will have appeared somewhere else, if $n_\beta =3$ now
as a consequence of the move.) 

\remark{The situation here is amusingly reminiscent of real
spin models.  E.g., in a Hubbard model, if you have zero or
two electrons on a site, you have no net spin, just like
the channel with two or four Al.  It is like filling  atomic
orbitals in a mixed-valent compound, and asking where the
ions are that have a moment.}

Thus, the channel occupation numbers $n_\alpha$ are a separate 
degree of freedom. 
We presume that, in most channels, the optimum 
is $n_\alpha=3$ and the energy cost of $n_\alpha$ deviating is 
much larger than the $\puck_\alpha$-$\puck_\beta$ interaction.
But when the total Al available to channels is (say) less
than $3\Nch$, this forces a ``doping'' by 
$n_\alpha=2$ channels, and there are many nearly degenerate
ways to place them.
Since the puckering effective Hamiltonian depends on 
the configuration $\{ n_\alpha \}$, we may very easily find
that two separated $\puck_\alpha$ variables are favored to
be the same or opposite, depending on the occupancy of 
some intervening channel.
(The location of channels with puckering also depends on
the presence of Ni neighbors to the Co atoms in the central column;
that is also highly sensitive to composition and density, 
see Sec.~\ref{sec:Co-Ni})

Furthermore, under our protocol -- MD simulations at moderate temperatures,
starting from an arrangement on ideal sites -- the occupancies
$\{ n_\alpha \}$ are mainly quenched, after the ring 2.5/ring 3 Al 
atoms have found their way into nearby channels;  diffusion of Al
from one puckering unit to the next seems to be suppressed. 
%%%%%
\remark{Thus these atoms subsequently do not efficiently explore 
alternate arrangements of the puckering.}
Consequently we cannot trust MD simulation to discover the
optimum arrangement; since the $\{ \Jpuck{\alpha\beta} \}$ are
not only  random but frustrated, the puckering effective Hamiltonian 
in fact describes a spin glass. 

\LATER{Alternate:
Of course, it is possible too that a puckering unit gets stuck 
metastably with a particular set of Al even though that is not the 
optimal number in
view of the local geometry and the values of the chemical potentials.
In such a case, there would be quenched random effective interactions
of {\it random sign} between adjacent $\puck$ variables -- the recipe
for a ``spin glass'' Hamiltonian.}

\remark{The above worries can be expressed using the language of 
puckering units.  Imagine a certain
puckering unit has four mirror-Al and that it propagates a negative
sign along the decagon perimeter.  [That is, $\puck_\alpha$ has opposite signs
on the two decagon edges involved.]  Next, change the puckering unit
to hold five mirror-Al atoms: perhaps it rearranges itself so as to
propagate the sign positively along the perimeter.  Finally, change it
to hold six Al atoms in mirror layers: now there is no puckering and
the sign is not propagated at all via this puckering unit.}

\begin{figure}
\includegraphics[width=1.0in,angle=0]{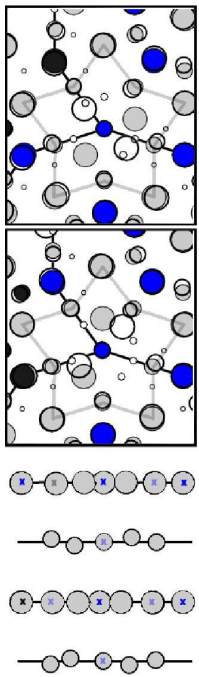}
\caption{Puckering in a typical 2.45\AA{} star.  Colored circles
  (using the same conventions \CLH{as where ?}  for species and layer)
  are the optimum configuration assuming a bilayer ($c \approx
  4$\AA{}) period, given ideal quasilattice sites as used in
  lattice-gas simulation.  Note that channel interpretation is valid
  here as well.  If we draw lines from the center of the 2.45\AA{}
  star to the nearest TM atoms (3 are already drawn as \Dec~edges,
  four channel formations can be seen.  (a,b) Each image is two layers
  in a structure with 4-layer ($c\approx 8$\AA{}) stacking period.
  The large empty circles represent locations achieved after
  relaxation. This particular 2.45 \AA{} star is on a vertex shared by
  two 4.0\AA{} decagons (solid lines).  (c). Side view; the
  $z$-direction scale is increased by a factor of $\sim$1.44.  Six TM
  atoms are depicted from each bilayer as X's. Faded indicates that
  the TM atom lies behind the Al atom in this projection.
%%%%
  \LATER{Notes on Fig.  \ref{fig:Puck}: 1) Q-- Need to make Co not in
    color!  Also add (a),(b), (c) to graphics for ease of reference
    elsewhere.  2) it was the right size to put 3 squares side-by-side
    across a page, but we never found any good way to do that.  3)
    Plan to replace this by a similar figure compatible with Marek's
    graphical conventions for puckering?  Perhaps there's a way to
    place two such stacks of figures side-by-side... to illustrate
    different cases etc.  } }
\label{fig:Puck}
\end{figure}

\subsection{Puckering units}
\label{sec:pucker-units}

\remark{
Here is a recipe to define a puckering unit, in the absence of
\Decagons.  Take a channel: of its Al atoms, 
it is the mirror-layer Al that comes closest to the axis of a Co column. 
We assign the channel to that column, and group all the channels assigned 
to the same column as a ``puckering unit.''}

\remark{ The five TM atoms forming the large pentagon
in Fig.~\ref{fig:Puck} consist of one ring 1 and two ring 3 Co from
the \Decagon, and up to two TM (often Ni) from a neighboring \Star.
But if the puckering unit is centered on a corner shared by two
\Decagons, all five surrounding TM atoms are Co from the \Dec.}

A description in terms of independent channels is problematic
not only because of their variable occupation, but also 
(as we shall see) not every set of $\{ \mu_\alpha \}$ values
is meaningful owing to steric constraints.

Instead, our approach to make sense of the puckering pattern
is to define the ``puckering unit'' (Fig.~\ref {fig:Puck}),
which consists of all channels (usually five) centered on the same 
Co column, and their Al contents.  
In our models, each puckering unit is centered on a \Decagon~
vertex and occupies one of the 2.45\AA{} HBS tiles
introduced in Sec.~\ref{sec:small-DHBS}, which encircle the 
decagon center in the DHBS picture.
In fact, the puckering units are always on
\Dec~ corners aligned (in projection) with the
five core Co atoms, and the 2.45\AA{} tile is usually a Star.
(This is a consequence of how the latter Co atoms
determine mirror layers, as expained in Sec.~\ref{sec:channels}.
%%%%%%%%%%%%%%%%%%
\remark{Being a 2.45 \AA{} star means all five
channels are present ; a Boat or a Hexagon would have four or three.
Fig.~\ref{fig:Puck} shows a Star.}
%%%%%%%%%%%%%%%%%%
The strong interactions between 
adjacent channels a small menu of configurations 
for each puckering unit, from which one
can build the larger-scale patterns of $\{ \puck_\alpha \}$ in the structure.

Now, the channels around one Co column come very close in
the ideal mirror-layer Al sites are $2\tau^{-1} \sin(2\pi/10) a_0 =$ 1.79\AA{} 
from each other and cannot simultaneously be occupied.
-- adjacent channels must have opposite signs of $\puck_\alpha$.
%%%%%%%%%%
\remark{This follows since the ideal site for the mirror-layer Al atom 
in a channel is $\tau^{-1} a_0 \approx 1.52$\AA{} from the Co column. }
%%%%%%%%%%
But if (as usual) there are five channels, 
this alternation is frustrated.
The resolution is that two adjacent $\puck$ 
values are the same, but the two mirror-layer
Al sites get merged into one Al at the midpoint.  
There is practically zero cost in the Al potential 
for such a deviation: Fig.~\ref{fig:b4a}(c) shows the 
channels are actually connected by ring-shaped troughs in
the mirror layers, which include the fused Al sites.
As for the puckering-layer Al atoms, since they sit farther 
from the central axis, there is no steric rule against
adjacent ones having the same puckering sign $\puck_\alpha$.

Thus, in a fully puckered configuration, a puckering unit
has room for only two Al atoms in either mirror layer, a total 
of four. 
These atoms generally arrange themselves into a motif we
call a ``crooked cross'' (see Fig.~\ref{fig:Puck}).
In projection, one arm of the cross (consisting of two
channel Al sites) is bent to an angle $2(2\pi/5)$;   
it is bisected by a straight arm,  consisting of a channel
site on one side and a merged site opposite to it.
[If there were just four channels, both arms of the cross are
liable to be bent at angle $2(2\pi/5).]$

\remark{ When the mirror Al atom sits in a ``merged''
site between from two channels, 
its transverse displacement (from the channel line) 
is greater than otherwise, hence the puckering
displacement in those channels is much smaller.}

\LATER{We might represent this state in the Ising language 
by $\puck_\alpha=\puck_\beta$ on the two channels, 
with the proviso that in translating from spins to atom positions some atoms
are merged.  But these merged atoms, being at a different distance
to neighboring channels than unmerged ones would be, must have changed
interactions with those channels. That could only
be represented if three-body interactions were included in $\HHpuck$.}

\subsection{Configurations in puckering units}
\label{sec:pucker-stats}

In this subsection, we first classify puckering units 
according to their environment with respect to the \Decagon-\Star~ 
geometry (or equivalently the 10.4\AA{}-edge Binary tiling);
we also classify the resulting patterns of Al occupancy
and puckering in each puckering unit. Then, studying
plots like Fig.~\ref{fig:puck-Wsim} from separate RMR relaxations,
we count the frequency of puckering patterns in each kind
of location;  indeed, the patterns are nearly determined by 
the environments (especially if the placement of TM atoms in nearby 
\Stars~ is taken into account).

Besides the W-cell, we also performed RMR relaxations
(starting from a different decoration) in the 32$\times$ 23 cell (not shown);
in this case, the initial fixed-site structure was the result of
a lattice-gas MC simulation (using 4\AA{} tile decoration), rather than
an ideal decoration rule.
%%%%%%%%%%%%%%%%%%%%%%%%%%%%
\LATER {Being a raw result, it can have irregular Co/Ni substititutions.
MM reran from the scratch, since that was summer 2005 before Nan restored
the old 2004 samples.  It would be good to try this with a more systematic
structure as starting point.}
%%%%%%%%%%%%%%%%%%%%%%%%%%%%
The behavior of puckering centers was different in the two
approximants; one reason is that our 32$\times$23 
structure (with a total point density 0.068\AA$^{_3}$)
is packed with a much lower density of Al atoms than the 
W-cell (at density 0.070\AA$^{_3}$).

\subsubsection{Nomenclature for puckering units}
\label{sec:pucker-labels}

We categorize the puckering-unit
centers on the \Decagon~ edge as type $A$, $B_0$, $B_1$, or $B_2$,
according to where they fall 
on edges of the 10\AA{}-edge Binary Tiling.
Type ``$A$'' sits on the interior of a Fat rhombus of the Binary tiling;
more important, it is a vertex shared between two \Decagons.
A type $B$ site is along a ray connecting the \Dec~ center to that of 
a \Star, which is always an odd \Star~ in the assumed scheme of orienting the
cluster centers. 
That ray is an edge of the 10.4\AA{} Binary tiling;
the cases that it goes between Thin/Thin, Fat/Thin, or Fat/Fat rhombi 
define environments $B_0$, $B_1$, and $B_2$, respectively.
%%%%%%%%%%%%%%%%%%%%%%%%%%
\remark{In the Binary HBS tiling, the odd \Star~ is the internal site,
so that ray becomes an ``internal edge''.}
%%%%%%%%%%%%%%%%
The environment $B_k$ has at least $5-k$ columns of TM neighbors 
(in the other layer) from the \Dec~ itself, each of which 
creates one channel in between.
It will have more channels (up to five),  whenever TM occupying the
right candidate-TM site(s) of the neighboring \Star~ supplies the 
necessary second TM column.
%%%%%%%%%%%%%%%%%%%%%%%%%%%%%%%%%%%%%%%
\remark{Every environment, except  $B_0$, has a vertical local mirror plane:
in the $A$ case it bisects the \Decs along their shared edge,  while 
in the $B_0$ and $B_2$ cases it runs radially from the \Dec's center.}
%%%%%%%%%%%%%%%%%%%%%%%%%%%%%%%%%%%%%%%

In projection, the
positions of the $m$ mirror-layer atoms next to the central Co 
(appearing as $+$ or $\times$ symbols in Fig.~\ref{fig:puck-Wsim})
are the best way to visualize the configuration adoped by a puckering unit;
So, we label the possible Al configurations in a puckering unit 
by a letter ``$p$'' or ``$u$'' for ``puckered''
or ``unpuckered'', followed by the number $m$.
``Unpuckered'' means all the $+$ and $\times$ symbols are superposed
in pairs (sometimes the pairs are not quite lined up);
``puckered'' means the Al in one mirror layer is missing, 
in at least one place.  
Farther out from the puckering unit's center, there are always two 
puckered-layer Al sites in every channel, each of which follows
the closest mirror-layer Al site(s):
displaced in a determined sense 
(large black or white circles in the figure) 
if the latter is puckered, undisplaced if it is not.

Finally, we sometimes add a $+$ or $-$ superscript
to the label, to record the parity of the puckering sense under
the (vertical) mirror plane of the \Decagon that passes through 
the \Dec~ center and the Co puckering-unit center.
(The $-$ parity appears more frequently.)
Thus, a typical shorthand symbol is ``p$4+$''.
Still, some of our labels refer to more than one configuration.
A unique way to name any puckering-unit configuration
is given in Appendix~\ref{app:pucker-labels-details}.
%%%%%%%%%%%%%%
\remark{As explained there, we can't assign a parity to a type
``$A$'' on account of the frustration of the puckering sense 
along the \Decs' shared edge.}

\subsubsection{Results: statistics of puckering units}
\label{sec:pucker-results}

\remark{Notes from email mvic-pucker-anal.out0721;
it is further analyzed in CLH's file mvic-pucker-anal-summ 8/2006.}

Our observations need to be prefaced by a caution.
The idealized decoration, when applied to different
approximants, will lead -- in view of the locally inhomogeneous
densities of species and of the binding energy in our
model -- to differing densities and compositions 
for the approximants, and may destabilize some
finite approximants with a decoration that would be stable 
in the thermodynamic limit.  As a corollary, if the
same composition and density is forced on the 
different approximants, it may occur e.g. that one
of them is overpacked with Al atoms and the other
one is underpacked.  

%%%%%%%%%%%%%%%%%%
\begin{table}
\begin{tabular}{|cccrrrr|}
\hline
Location & Cell & Number   &  \multicolumn{4}{c|} {Frequency} \\
type &   & in cell &  ~Ni  & \quad $p4$  &  $\quad p5$  &\quad  $u6$  \\
\hline
$A$   & W-cell        &  4 & 0 &   0  &  0.6  &  0.4 \\
$A$   & 32$\times$23  &  6 & 0 &  0.6 &  0.4  &  0   \\
$B_0$ & W-cell        &  4 & 0 &  0.9 &  0.1  &  0   \\
$B_1$ & W-cell        &  8 & 0 &  0.1 &  0.5  &  0.4 \\
$B_1$ & 32$\times$23  &  4 & 0.25 & 0.5 &  0.2  &  0.3 \\
$B_2$ & 32$\times$23  &  4 & 0.75 & 0  &  0.2  &  0.8 \\
\hline
\end{tabular}
\caption{ \footnotesize
Frequency of local puckering configurations ($p4$, $p5$, and $u6$) 
in puckering units, classified according to location type
in the large tiling; these add to 1.
Column ``Ni'' gives fraction where the central Co has 
a nearest-neighbor Ni atom.
%% Trials were done in both unit cells listed.
The total number of distinct puckering units was 16 in the 
W-cell and 14 in the 32$\times$23 cell. 
Each frequency is based
on 20-50 examples of the puckering unit, in different places
within the cell and/or from different runs.  
\LATER{CLH to MM, to Nan: This table is based on lots of counting
I did looking at pucking config's generated by Marek in
6/05-7/05.  Both MM and CLH thought that the way channels got resolved 
depended strongly on random details (e.g. initial velocity realization 
in MD), that's why a statistical description seemed best.
But in 1/06 CLH realized that, if we control
for the Ni neighbors, we could have a mostly deterministic
answer for each environment.  
(MM's comment: it seems MD was at sufficiently high temperatures that
the initial condition didn't really matter?)
We definitely need to follow up the physical question.
The writing question is whether this table is necessary, is it
the best way to present this information.}
}
\label{tab:pucker-patterns}
\end{table}

Table~\ref{tab:pucker-patterns} summarizes the statistics 
we found; they should be taken only as rough numbers, especially 
as runs taken under different conditions were combined.
Each column lumps together several distinct patterns,
distinguishable by the long names from
Appendix~\ref{app:pucker-labels-details}
(if not by parity).

In the ``$A$'' environment, half the units were $p5$, 
and the rest were $p4$ or $u6$, depending on Al density.
Both $B_0$ and $B_1$ environments show a ``crooked cross'' pattern, 
in two variants oriented differently with respect to the \Dec: 
$B_0$ has $p4^-$ while $B_1$ has $p4^+$ 
Actually, in the $B_1$ case, Al in the neighboring candidate-TM 
site in an Odd \Star~ (which counts as a merged mirror-layer 
channel site)  strongly  tends to be unpuckered: thus the crooked-cross
gets modified to $p5^+$.
Finally, the ``$B_2$'' environment is typically an unpuckered $u6$. 

However, the overwhelming factor affecting puckering is whether 
the central Co has a Ni neighbor in the candidate-TM site of an 
adjoining Odd \Star, which is a merged-type site
if there are channels on that side.  In any case, the Ni
always occupies both mirror-layer sites, so that tends to 
favor unpuckered channels all around this puckering unit.  
A ``$B_2$'' environment is typically unpuckered mainly because 
it typically has a Ni neighbor (at least in the 32$\times$23 cell).
Note also that if the neighboring candidate-TM site of the
odd \Star~ is not TM, then one or both of the candidate-TM sites 
one step away probably is TM, which increases the number of 
channels in this puckering unit and (probably) makes it likelier
to adopt a puckered configuration.

In Table~\ref{tab:pucker-patterns}, both $A$ and $B_1$ environments 
are packed with more Al in the case of the W-phase cell, reflecting
its higher overall packing.
Despite this, the mean occupancy $m$ is practically the same
(5.0 in the W-phase cell, 4.9 in the 32$\times$23 cell).
The reason is that the W-phase cell contains another environment
$B_0$, which usually has $m=4$, while the 32$\times$23 contains
$B_2$, which usually has $m=6$.
If the overall density of Al (and hence its effective chemical potential)
were set the same, we iamgine each environment type would 
show similar behavior in both cells.
 
\LATER{It would be nice to report about site energies, 
classified in terms of the environments B$_1$ etc.}

\subsection{Puckering around \Stars}
\label{sec:star-pucker}

In any Even \Star, there tend to be TM atoms on all ten 
vertices of its 4\AA{}-edge Star tile (of the DHBS tiling): most 
of them are \Decagon~ vertices, while the others are candidate-TM
sites where two \Stars~ touch; as noted in Sec.~\ref{sec:fixed-site},
the latter often have a TM-TM pair.
If furthermore the latter sites are Co, and some candidate-TM sites
of the \Star's interior are {\it also} occupied by TM, then 
the Star vertices not on a \Dec~ become puckering centers too, and
all ten exterior edges have channels.  
This happens in the special decoration of Fig.~\ref{fig:RulesWideal}.

\remark{An old observation of Nan Gu's 
seems to say the Star puckering pattern also can develop even
when the \Star~ has five NiNi pairs. (The Ni in the \Dec, recall,
is not the puckering-unit center.)
Such \Stars~  are uncommon in raw configurations  emerging from MC.}

On the two edges meeting at a $2\pi/5$ corner, 
the puckering sense should be opposite due to
the steric constraint.
At an indented [angle $3(2\pi/5)$] corner, the sense is also opposite,
i.e. the parity (with respect to the adjacent decagon) is $-$,
consistent with the usual tendency (noted above).
Thus, the puckering sense alternates as one passes all the way
around the Even star's exterior edges, producing a
striking pattern in images of the puckering.  
(Such patterns are even more prominent in the real $W$(AlCoNi) structure:
see Sec.~\ref{sec:W}.)

Furthermore, when there is a  chain of \Stars~ (as in the W-cell), the
interaction between successive Even \Stars~ is such that
their patterns have the same puckering sense.   This 
accounts for most of channels around every \Decagon
(all those in puckering units of types $B_0$ or $B_1$).  
It leaves unspecified, however, what happens
along the edge shared by two \Decagons.  
The non-channel atoms nearby are perfectly symmetric 
under the (vertical) local mirror plane that includes the shared edge.
And, following the star rule just described, the puckering sense 
will be {\it opposite} on the adjacent unshared edges
of the respective \Decagons.  
Thus the puckering sense on the shared edge is necessarily
given by a local symmetry breaking, and cannot propagate
the pattern from the \Star~ chain on one side  to the
\Star~chain on the other side.

\remark{
The interior ring 2 Al atoms pucker in response to
the pattern described, and in principle the ring 2 Al's
could interact (via the ring 1 Al's) to propagate a sense
across the \Dec.  However, for the pattern of \Decs~ in the
W-cell, even this interaction is frustrated.}

With the alternative decoration of Subsec.~\ref{sec:alt-deco},
the 4\AA{} edge star would not have Co on every vertex 
(nor would the \Dec, for that matter), and the puckering
patterns just mentioned would, one expects, be disrupted.
On the other hand, in a model built from disjoint 20\AA{}
decagons (see Appendix~\ref{app:20A}), the \Star~ chains
are more extensive and might propagate a puckering sense
globally.  Conceivably, the puckering interactions might
be strong enough to tip the balance between different
placements of TM atoms (e.g. alternative decoration)
or between different basic structures (e.g. the 20\AA{}
decagon structures).  The approach we followed in the
present work could not answer such questions, since the
positions of all TM atoms (and some Al) are permanently
determined at the fixed-site stage of modeling.

\CLH{to MM 1/29/06:  In this paragraph, I already made the point 
that puckering could decide the kind of cluster nerwork, 
before your email message last week urging the same point.}

%% \begin{figure}
%% \includegraphics [width=3.1in,angle=0]{123puck.eps}
%% \caption{Relative puckering strength on the  23$\times$32$\times$8
%% tiling. Note the super-patterns in the form of rings of  puckered
%% atoms.}
%% \label{fig:123puck}
%% \end{figure}

\section{Simulation of  Experimental Approximant W(AlCoNi)}
\label{sec:W}

\NAN{(Old remark)
I think since we mentioned this topic in the letter, it should be
  mentioned only slightly or not at all here.}

In this section, we compare our prediction with the 
approximant structure $W$(AlCoNi), currently the only refined 
Al-Co-Ni structure on the Co-rich side.  
The solution of of atomic positions was done by
Sugiyama {\it et al}~\cite{Su02} using direct methods
(the SIR97 package).

\subsection{Attempted prediction of W(AlCoNi) by simulation}

For our simulation, we used the same 4\AA{} rhombus tiling 
which optimizes the decagon density, 
as explained in Subsec.~\ref{sec:MC-4A-tileH}.
As inputs, we took the experimental lattice parameters
$23.25$\AA{}$ \times 39.5606$\AA{}$ \times 8.16$\AA{} and the
experimental reported point density and composition.
(This differs from the standard composition and point density 
from Sec.~\ref{sec:setup-contents} that we have used
up till now in this paper.)
By comparison, the decoration of Fig.~\ref{fig:RulesWideal}
has atom content Al$_{188}$Co$_{60}$Ni$_{20}$, 
which is too rich in Co compared to
real $W$(AlCoNi), while its density 
of $n=0.071$ \AA{}$^{-3}$ is 
slightly denser than real $W$(AlCoNi)
(See table~\ref{tab:deco-count}).

The result of our discrete-site simulation --
which our Fig.~\ref{fig:RulesWideal} was devised to idealize --
looks quite similar (in $c$-axis projection) 
to the experimentally determined $W$(AlCoNi) cell.~\cite{Su02}
However, a significant number of Al atoms present in the diffraction refinement
could not be found in our simulation result.
Also, the TM arrangements in our \Stars~ do not agree with those
in the W-phase.  

\CLH{to Nan: Can I say the decoration of Fig.~\ref{fig:RulesWideal}
was actually intended to reproduce the experimental composition of
W(AlCoNi), and not of the basic-Co phase?  I want to clarify the
relation to that, since it has been so fundamental in the
discussions throughout this paper.}

We next apply the ``relaxation-molecular dynamics-relaxation'' (RMR)
protocol defined in Sec.~\ref{sec:relax}; in the molecular dynamics 
portion, the temperature was initially $T$=600K 
and was then cooled in gradual stages to $T=$50K.  
The RMR structure shows the usual puckering (Sec.~\ref{sec:relax})
similar qualitatively to the prominent puckering of the 
actual approximant. 
\CLH{Nan wrote ``Our simulation matches the structure refinement to the 
extent that the correct layers pucker symmetrically??''
to MM, to Nan: by ``symmetrically'', did you mean just that one 
puckering layer puckers up wherever another puckering layer puckers
  down?  If that is the best sense it can be said we matched it, then
  I'm disappointed.}

\LATER{CLH notes: The W(AlCoNi) portion is sort of missing 
a punchline.}

\LATER{
MM reminds the following energy data is available [SOMEWHERE];
need to add a paragraph.
MM had re-run our procedure (4A site list) 
using precisely W-phase composition, and compared relaxed
energies with those of actual W-phase refined structure. 
The ``actual''
structure was found to be more favourable [not sure, I recall some 15
meV/atom??] - again confirmation, that the accuracy of our pair
potential is surprisingly good.}

\MM{[CLH did not understand these, see email 2/3/06].
The energy appears very high - but the
real question is - what is the competing motif? For example, by how
much is HS pair better than 2B (2.45A tiles) after puckering? (I dont
know at the moment...)  Second: such a high energy could imply that
the atomic density is constrained to be less than what allows optimal
ordering inside the channels.  (I mean - at too high density the
channels get stuck).  Our idealized decoration seems to suffer this
pitfall (it doesnt maximize number of 2.45A Stars, as is the case in
actual W-phase.)  
}

\begin{figure}
\begin{minipage}{8.66cm}
\includegraphics[width=3.1in,angle=0] 
{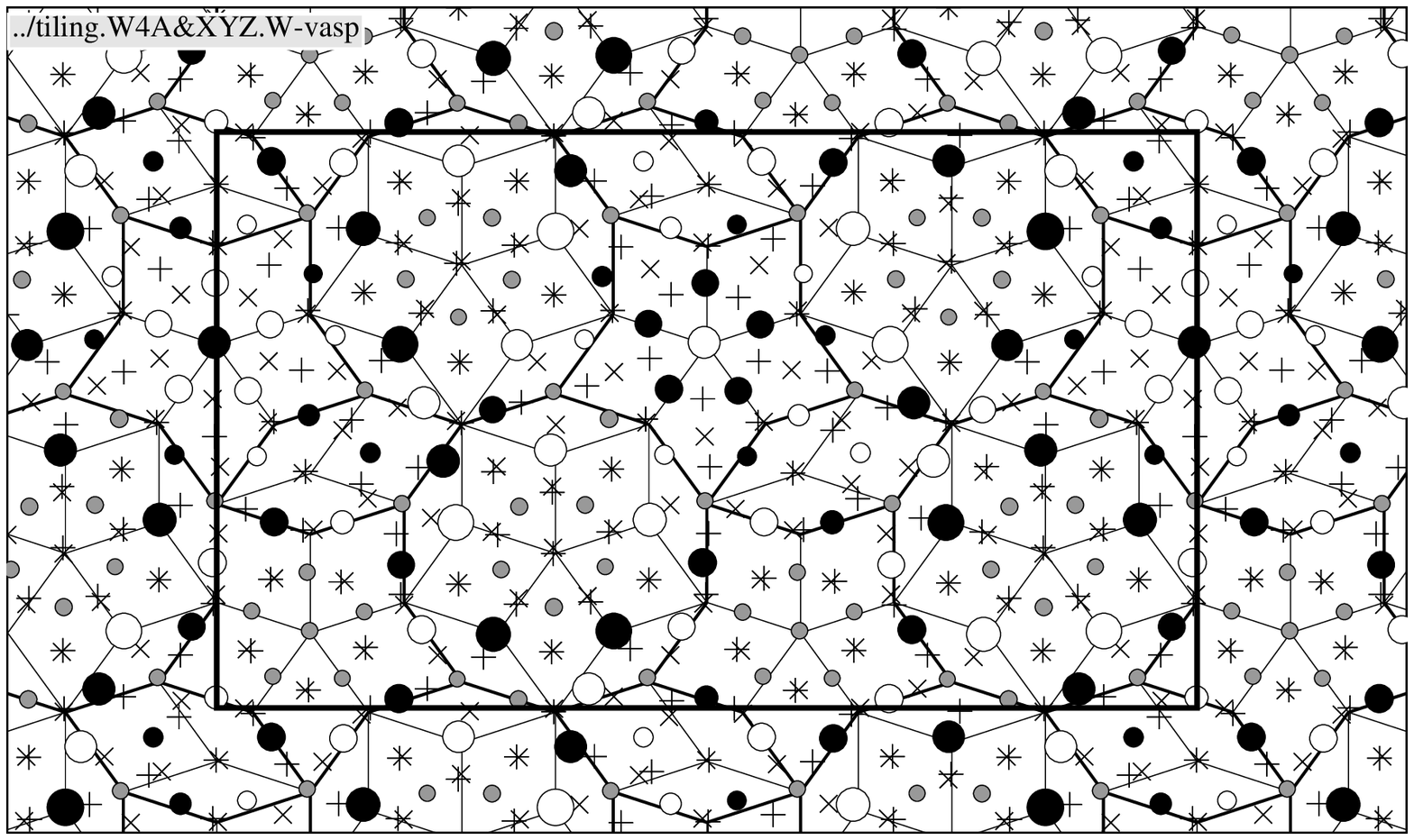}
\vskip 0.15 truein
\includegraphics[width=3.1in,angle=0] 
{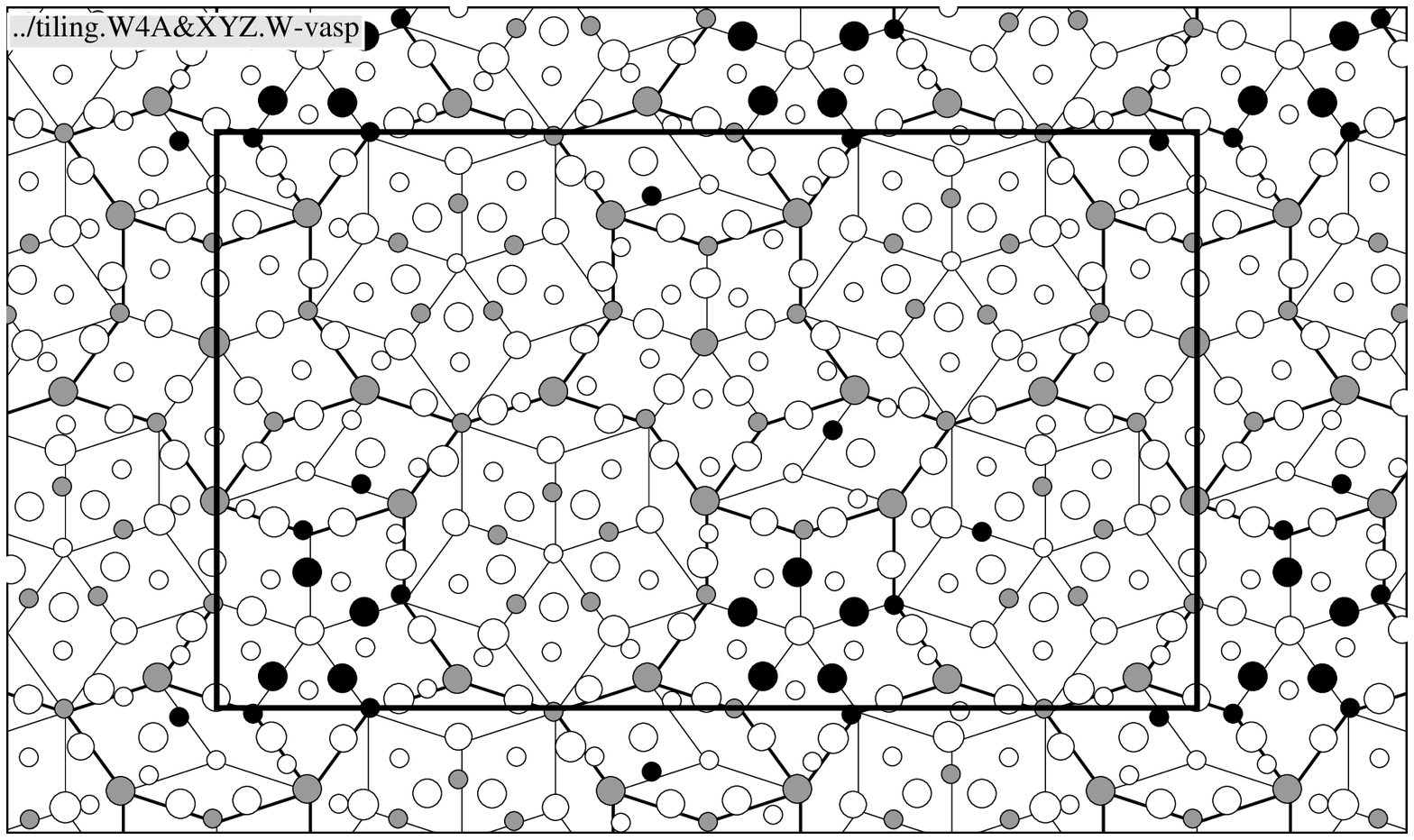}
\caption {
Puckering in the W(AlCoNi) structure, using same  
conventions as 
Fig.~\ref{fig:puck-Wsim}
}
\label{fig:puck-Wreal}
\end{minipage}
\end{figure}

%% \begin{figure}
%% \begin{minipage}{8.66cm}
%% \includegraphics[width=3.1in,angle=0]{WA.eps}
%% \vskip 0.15 truein
%% \includegraphics[width=3.1in,angle=0]{WC.eps}
%% \caption {
%% One of our simulations of the W (AlCoNi) cell, compared
%% with the experimental structure; (a),(b) are respectively
%% the mirror layer and the puckered layer. The empty circles are 
%% experimentally determined atom sites, the filled circles are those
%% extracted from our simulations. Dark gray filled circles are TM, 
%% light gray filled circles are Al. Empty circles with 'teeth' are
%% TM, smooth empty circles are Al. The large empty circles are 
%% experimentally determined atomic sites with mixed Al-TM occupancies. 
%% }
%% \label{fig:WA-WC}
%% \end{minipage}
%% \end{figure}

\subsection{Pentagonal bipyramid cluster}
\label{sec:W-PB}

\remark{from MM email exchange 7/26/05}
%%%%%%%%
The actual W phase differs from our decoration 
(such as Fig.~\ref{fig:RulesWideal}) essentially by the following
modification: half of the Even \Stars~ are replaced by a new
cluster which is just the ``pentagonal bipyramid''
(PB) cluster~\cite{Hen93,Wid96}
identified long ago in Al$_{13}$Co$_5$ (and other
decagonal approximants).  
Whereas the original (Even) \Star~ 
is Ni-rich and 4\AA{} periodic in the stacking direction,
the \PB~ is Al-rich and strongly puckered (so it
is 8\AA{} periodic). 
%%%%%%%%%%%%%%%%%
\remark{MM speculated (7/26/05) that in a decagonal model built
using the enlarged \PB~ of Ref.~\onlinecite{del06},
there would be a significantly higher Al content. Furthermore,
since our fake version of W-phase has no \PB's, 
maybe it should be richer in Ni than the real W(AlCoNi) phase.}

\LATER{Say this better about the PB, from the viewpoint of
channels and puckering centers. Didn't CLH write it up
somewhere, circa 1/25/06? Say what makes room for
the five mirror Al to be in same layer.}
%%%%%%%%%%%%%%%%%%%%%%%%%%%%%%%%%%%
One novel feature is that 
the apex TM atom (on the PB axis) puckers noticeably, 
which has not been true for any TM atoms up till now
(even the ones in puckering layers, which are allowed
by symmetery to pucker).
In fact, the arrangement around the PB center is quite
similar to that around a 2.45\AA{} Star; indeed, the
central TM column is a new kind of puckering center 
that is not on any \Decagon.  The mirror-layer channel 
Al adopt a new puckering pattern, in which {\it all}
five go to the same layer and sit in merged sites.

The model of Ref.~\onlinecite{del06} 
(elaborated here in Appendix~\ref{sec:deloudi-W5D})
adopted the PB as its fundamental cluster because, 
in the real $W$ phase, fivefold symmetry extends 
quite far from its center.
We must disagree with their assignment of atoms of the innermost
PB ring as mixed Al/TM: these are right in a channel, and should
in fact be Al.  Conceivably in a real structure, there is 
some disorder in the placement of the two ways of decorating
Even \Stars.  Since the TM ring of the ``standard'' Even \Star
(of Fig.~\ref{fig:RulesWideal}) occurs in the same place as the
Al ring of the PB, that would give an averaged structure as if
there were Al/TM substitutions.

\subsection{Puckering propagated by \Stars~ and \PBs}
\label{sec:W-PB-pucker}

Fig.~\ref{fig:puck-Wreal} shows that the actual $W$-AlCoNi structure
has a more pronounced and better propagated puckering pattern, 
as compared to our model structures such as Fig.~\ref{fig:puck-Wsim}.
The key to this is the PB which is, in a sense, one big 
puckering unit.  The three Al pentagons in the middle all
belong to channels along the interior 4.0\AA{} edges in the
five-rhombus star containing a PB (the Al pentagon in a
mirror plane consists of all merged sites).  Although the
edges of the 4.0\AA{}-edge star have alternating puckering
senses, just like the ordinary Even stars did, these are
not the key atoms for propagating the sense,  The key atoms
are those mirror-layer Al visible (in the top panel of 
Fig.~\ref{fig:puck-Wreal}) just inside the tips of that
five-rhombus star, which lie in the same layer  as the
mirror-layer Al pentagon in the middle.  

These tip Al atoms correlate the \PB~ puckering with the adjoining
\Star~ cluster, which puckers in the same pattern discussed
in Sec.~\ref{sec:star-pucker}, and thus propagates a well-defined
puckering pattern along each chain of \Stars~ in the underlying
tiling.  Inside the \Decagons, unlike Fig.~\ref{fig:puck-Wsim}, 
the ring 2 Al atoms (not in a mirror layer) pucker strongly.
Each (along with ring-2.5 Al in the same channel) adopts a 
sense opposite to the nearest puckering atoms from the \Stars
or PB's: the ring-2 Al facing \Stars~ alternate while those
facing \PBs~ all have the same sense, producing the complicated
pattern of white and black circles inside the \Decagons in]
the top panel of Fig.~\ref{fig:puck-Wreal}.  
However, just as in the \PB-less  case of Sec.~\ref{sec:star-pucker}, 
it is hard to see how the pattern actually propagates from
one chain of \PBs and \Stars, through the \Decs, to the next
chain over.

\remark{
In the \Dec~ packing of the W-phase cell, the hypothetical
interaction mediated by the ring-2 puckering is no longer 
frustrated: if you divide the five ring-2 pucker layer sites
of each decagon into two and three associated with the
Star chains on the respective sides, then there are more
places where circles of the same coloring are on vertices
across the short diagonal of a 4\AA{} rhombus.  Yet I
don't trust that this propagates anything, because these
Al atoms are too far to have much of a direct interaction,
and the ring-1 Al atoms (which they could affect) are
hardly puckering at all.  (Meanwhile, it still
seems impossible to propagate puckering past a \Dec~
edge shared by two \Decs.)}

\LATER{Important: As pointed out in MM 2/2/06 colong edit, and implicitly
many times before, the Co2Al9/puckering-center/2.45A-Star motif is really
a larger motif, and more closely related to PB, than is recognized
in the current edit of the text.  We have commented this in fragmentary
ways:  e.g. when first introduced in Sec III or IV, it is commented
how there is a further pentagon of TM; also it is commented how the
PB is a new kind of puckering center.  But we didn't specifically 
note that the 2.45A-Star commonly has an outer ring of 5 Al along
with the 5 TM (and in the same layer), which is exactly like the 
outer ring of PB.  Of course, this is messier since those overlap.
(CLH adds, you can see an imperfect 13\AA{} diameter ring even.
this description is ``complementary'' to the 13\AA{}, much as
Bergmans are to Mackay icosahedra.  A model which is a perfect
packing of one kind, is often a defective packing of the other,
and vice versa.).}

\MM{9/05 
[in comments on the relation of Ni density and composition variations
to puckering units.]
``I have studied carefully the way W-phase manages to have
five pentagonal puckering units 
all being the same nice ``$p4^+$'' type centers.
(``crooked crosses'' lined up with two Co atoms, going along the ray from
the \Dec~ center.)
In particular: the fat (Binary tiling) 10\AA{} rhombus has one 
$A$ type center, plus two other $B$ type centers
located on the edges on the other side of the rhombus.
%%%%%%%%%%%%%%%%%%%%%%%%%%%%%%%%%%%
MM analyzes this by extending [as CLH elaborates in preceding note]
the 2.45A Star cluster by TM pentagon
in the 2nd ring around Star center, 
in other words looking at it as on a kind of PB, 
with the outer decagonal ring [of Al5TM5 in mirror layers].
Now, at a place like this, one outer-TM-pentagon tip falls on the mirror
channel site of the nearby PB (associated with the other Decagon 10.4A
apart)  -- an overlap that [at first glance] seems be an
unavoidable conflict. 
But can see in our Fig.~\ref{fig:puck-Wreal}
how this is resolved [by Ni placement]: roughly speaking,
these sites which are simultaneously outer-ring of the PB's and should
be TM, and also should be mirror channel sites (Al),
become 50:50 Al/Ni.}

\CLH{CLHNEW to MM:  You also wrote here, 
``This also agrees with Nan's proposal that Ni at 
some point occupies flat-layer channel sites, at increased Ni content.''
I wanted to track down what this refers to.
In Sec.~\ref{sec:pucker-results}, Ni in the
Star cluster TM sites is always occupying mirror-layer channel-type sites;
we don't need to treat it as a defect.  Are you relating this to the
story of the ``arrows'' of Ni (Fig.~\ref{fig:Ni-arrow}? 
In that figure, it seems to me the added Ni(2.5) is puckering layer;
only the Odd-Star Ni just outside the 13A Decagon is mirror layer.
}

%%%%%%%%%%%%%%%%%%%%%%%%%%%%%%%%%%%%%%%%%%%%%%%%%%%%%%%%%%%%%%
\section{Discussion}
\label{sec:discussion}

%% Summary of the paper
\remark{Part of this is from MM email 1/19/06} 

In conclusion, we have carried out the most extensive
prediction of a quasicrystal structure that makes minimal
assumptions and combines lattice-gas Monte Carlo with
relaxation and molecular dynamics.  
The overall story of this project is that our approach, on the one
hand is fruitful at producing atomic structures 
with very good energies and very good local
order (i.e. consistent with structural experiments).
On the other hand, its application is an art rather than an algorithm, 
and there is no guarantee of discovering the absolute best solutions. 
%%%%%%%%%%%%%%%%%%%%%%%
\remark{MM's email 1/19:
To be sufficiently general, we are pushing the method to the limits 
(working far from equilibrium), so that one can expect difficulties.}
%%%%%%%%%%%%%%%%%%%%%%%
Difficulties are to be expected especially when we 
must discern between variants that have similar energies, yet
cannot easily transform to each other: brute-force Monte Carlo
is not sufficient to overcome this energy barrier.

The ``basic Co'' structure
turned out to involve substantially more complications than the 
``basic Ni'' case studied previously.
One measure of the complexity is that, 
at different places, it was convenient 
to introduce tilings on  five different length scales
(related by powers of $\tau$). Table~\ref{tab:tiling-index}
gives sort of index to the sections where they were defined or used.

There are three features of ``basic Co'' that made it more complex 
than ``basic Ni'', even at the stage of modeling
limited to discrete ideal sites (sections \ref{sec:intro} -
\ref{sec:ideal-deco}):
(i) a larger cluster unit (\Decagon), introduced in Sec.~\ref{sec:clusters};
(ii) a set of sites (the ring 2.5 and 3 Al) which break the symmetry 
of the basic cluster; to first order, these give rise to a high
degeneracy, which is broken in fairly subtle ways 
(Sec.~\ref{sec:ideal-deco}).
(iii) the orientational ordering pattern of the clusters, which
strongly affects the decoration even though it involves 
relatively small energy differences, that depend
on composition and density in complex ways (Sec.~\ref{sec:decLRO}).

%%%%%%%%%%%%%%%%%%%%%%%%%%%%%%%%%%%%%%%%%%%

%%%%%%%%%%%%%%%%%%
\begin{table} 
\begin{tabular}{|lll|}
\hline
Tile edge & Tiling    &   Sections  \\
\hline
2.45\AA{}  & rhombus   & 
   \ref{sec:methods} (Fig.~\ref{fig:tileflips}), 
   \ref{sec:MC-a0} (Fig.~\ref{fig:smalltiles}),  \\
       %%%%%%%%%%%%%%%%%%%%%%%%%%%%%%%%%%%%%%%%%%%
           & \quad (two layers)         & 
  %%%  IIA (Fig. 1), III A (Fig. 2), III B (Fig. 3a) \\
   \quad \ref{sec:clusters} (Fig.~\ref{fig:dds}(a) only)  \\
%%%%%%%%%%%%%%%%%%%%%%%%%%%%%%%%%%%%%%%%%%%%%%%%%
2.45\AA{}  & HBS (basic Ni)      &    \ref{sec:bNi-compare} \\
%%%%%%%%%%%%%%%%%%%%%%%%%%%%%%%%%%%%%%%%%%%%%%%%%
2.45\AA{}  & DHBS              &
   %%% III D, IV B 1 (Fig. 7)
   \ref{sec:struct-pot},
   \ref{sec:small-DHBS} (Fig.~\ref{fig:RulesWideal}), \\
         %%%%%%%%%%%%%%%%%%%%%%%%%%%%%%%%%%%%%%%%%%%
   %%% IV C (Fig. 8) \\
     &     &
   \quad \ref{sec:deco-enumerate} (Fig.~\ref{fig:arrows})  \\
%%%%%%%%%%%%%%%%%%%%%%%%%%%%%%%%%%%%%%%%%%%%%%%%%%%%
4.0\AA{}   & rhombus               &
   %%% III B (Fig. 3(a), only), III E, 
   \ref{sec:clusters} (Fig.~\ref{fig:dds}(b) only),
   \ref{sec:MC-4A},                \\
    %%%%%%%%%%%%%%%%%%%%%%%%%%%%%%
           &                       &
   \quad \ref{sec:struct-pot}(Fig.~\ref{fig:siteE-ideal} only), \\
   %%% III D (fig. 5 only)
    %%%%%%%%%%%%%%%%%%%%%%%%%%%%%%
           &                       &
   %%% VII A  (Fig. 12), VII D \\
  \quad \ref{sec:puckerLRO} (Fig.~\ref{fig:puck-Wsim}), \ref{sec:star-pucker}\\
%%%%%%%%%%%%%%%%%%%%%%%%%%%%%%%%%%
4.0\AA{}   & DHBS                 & \ref{sec:4.0-DHBS} \\
%%%%%%%%%%%%%%%%%%%%%%%%%%%%%%%%%%
6.5\AA{}   & DHBS                 & App.~\ref{app:20A-bNiapp} \\
%%%%%%%%%%%%%%%%%%%%%%%%%%%%%%%%%%
10.4\AA{}  & Binary      & 
%%% II C (Fig. 4(c)), IV A 2, IV C (Fig. 8) , IV F 
%%% VI A, VII C 1, App.B (Fig 16)
    \ref{sec:Dec-relation} (Fig.\ref{fig:d-adjoin}(c)), 
    \ref{sec:binary-HBS},  \\
    %%%%%%%%%%%%%%%%%%%%%%%%%%%%%%
         &  ~(rhombus  &
    \quad \ref{sec:deco-enumerate} (Fig.~\ref{fig:arrows}),  
    \ref{sec:ideal-Burkov},
    \quad \ref{sec:decLRO-Ising},   \\
    %%%%%%%%%%%%%%%%%%%%%%%%%%%%%%
         &   \quad or HBS) & 
    \quad \ref{sec:pucker-labels}, 
    App.~\ref{app:rhombus-linkage} (Fig.~\ref{fig:clink}) \\
%%%%%%%%%%%%%%%%%%%%%%%%%%%%%%%%%%
17\AA{}   &  any                       &  
%%% IV F,  App. E (Fig 17)
    \ref{sec:ideal-Burkov}, 
    App.~\ref{app:20A} (Fig.~\ref{fig:20A})  \\
%%%%%%%%%%%%%%%%%%%%%%%%%%%%%%%%%%%%
\hline
\end{tabular}
\caption{ \footnotesize 
Different tilings used in this paper to describe
$d$(AlNiCo), with the sections (or figures) where
they are referenced.}
\label{tab:tiling-index}
\end{table}

\LATER{If we split up the paper, cut the Discussion HERE.
Also, Subsec.~\ref{sec:disc-alconi-phases} and \ref{sec:disc-discrete}
will go to paper I, 
Subsec.~\ref{sec:disc-puckersites} and \ref{sec:disc-puckerLRO}
go to paper II.}

Sections \ref{sec:relax}-\ref{sec:W} developed a second layer
of simulation, the use of relaxation and molecular dynamics to obtain
more realistic configurations.  This turns out to make a fundamental
difference in the ``basic Co'' case, because many of the Al atoms
deviate from their fixed positions to break the 2-layer ($\sim 4$\AA)
stacking periodicity down to 4-layer periodicity.  This happens
(Sec.~\ref{sec:relax}) in ``channels'' due to columns of Co atoms that run 
perpendicular to the layers and are filled with three Al atoms each
that define mirror planes and puckered layers.  We have explained this 
behavior in terms of the potentials (Appendix \ref{app:channel-anal})
-- this work~\cite{Gu06-ICQ9} appears to be the first time any 
explanation has been given for such period-doubling, a very common 
phenomenon of period-doubling in decagonal quasicrystals.
To understand the {\it correlations} of the puckering deviations
(which create the structure in well-known layers of diffuse scattering
seen in decagonals), yet another framework was needed of the
``puckering center'', (up to) five mutually constrained channels
around a single column of Co atoms (Sec.~\ref{sec:puckerLRO}).

In this latter half of the paper, two sections are included 
that are not specially focused on relaxation and puckering, 
but which could not be formulated in terms of just the
fixed sites. First, in Sec.~\ref{sec:decLRO} we found that 
puckering drives the clusters' orientational order (which 
breaks the symmetry down to pentagonal). Second, in Sec.~\ref{sec:W}
we show that our approach goes a long way towards successfully 
predicting the structure of the phase $W$(AlCoNi), and in turn
$W$(AlCoNi) offers additional clues for future modeling of
Al-Co-Ni decagonals.

Massive as it is, this study is still far from a definitive
answer about the $d$(AlCoNi) structure.  Although the atomic
structure we presented is unquestionably a good one, 
we suspect there exist competing structures (built from similar
local structure) that are just as good.  In part, our
failure to study these is an intrinsic weakness of the initial 
approach via discrete site lists, when we know puckering is a key
feature of the structure.  But to a greater extent, it stems 
from small misapplications of the technique.
Although we used much larger cells for the discrete simulation 
than in previous work~\cite{alnico01}, we should have used 
even larger cells;
furthermore, the degrees of freedom were too quickly reduced 
when we passed to a description based on 4\AA{}, or
really 10.4\AA{},  tiles (Sec.~\ref{sec:MC-4A}).
These adjustments would
have revealed the alternative framework based on
20\AA{} decagon clusters (Appendix~\ref{app:20A}).
Based on this experience, we anticipate that future applications 
of the method will evade these pitfalls.

In the rest of this section, we examine some of
the implications or future possibilities in more detail.

%%%%%%%%%%%%%%%%%%%%%%%%%%%%%%%%%%%%%
\subsection{Pitfalls of discrete site list}
\label{sec:disc-discrete}

From some viewpoints, one may be surprised that
constraining sites to atiling works at all, 
or suspicious whether it is transferable to
pther quasicrystals. Perhaps it is that, in order 
to form a high-quality quasicrystal, the atomic configurations already have to
  be tiling-like. [The identical local pattern has to be compatible
  with different environment patterns.]
%%%%%%%%%%%
\LATER{Nan, MM: Is the above OK?}

We still believe we it was effective to initially simulate 
using a 4.08\AA{} period, in order to discover the main features, 
and to refine this later on.  In part, this was justified 
by some simulations using 8.16\AA{} periodicity, in which we 
saw that 4.08\AA{} periodicity persists for a large subset 
of the atoms.

However, the dangerous step is eliminating sites: without care,
an unjustified assumption can get built into later stages.
In particular, there are subtle issues in connection with the density
The candidate site list for MC lattice-gas
simulations on the 4\AA{} edge rhombi was constructed to eliminate
sites that were observed to be unused in the previous stage of
simulation using 2.45\AA{} rhombi.  This is valid, so long as we
retain the original composition and density in the final model.
Usually, however, as we grow to understand the structure better,
different compositions and densities recommend themselves for the
idealized model, because the modified atom decoration (i) is simple to
prescribe, or (ii) is favorable energetically.  In the present work,
the initial explorations were conducted at a density 
$0.068$\AA{}$^ {-3}$, 
which is a bit loosely packed, whereas the idealized model of 
Fig.~\ref{fig:RulesWideal} at $\sim 0.070$\AA$^{-3}$ 
may be somewhat overdense.  
Thus, when MC annealing of Fig.~\ref{fig:RulesWideal} fails to find
any better configuration, it might be an artifact of the poverty of
the candidate site list {\it for this higher density.}  
The moral is
that the initial exploratory runs ought to be done with (at least) two
densities; to ensure a conservative choice of site list in later
stages, one of the densities might be higher than the expected real
one (though not too high, as that would slow down the lattice-gas
annealing).

\LATER{Meld this into Discussion.
The remedy for this should be discussed in the
Discussion -- CLH would say we must eliminate degrees of
freedom gingerly,  trying some intermediate level
of description, such as the 2.45A DHBS tiling.}

\CLH{To remarks. (adapted from Nan's report 2004).
One must always test the sites removed, e.g. in going
from the 2.45 \AA{} to the 4.0 \AA{} tiling.
Sites we had too hastily left behind did come back to haunt 
us at the next level:  large-scale patterns can indeed change due to 
a small perturbation in the site list (see Appendix \ref{app:20A}).}

\LATER{MM 9/05: I wanted to calculate lower/upper DENSITY limit, assuming
that we convert all 2.45A Boats to HS pairs, or vice versa.
(In the formulation of Sec.~\ref{sec:deco-stoich}.)}

\FUTURE{MM c 9/05. There is a striking
$\tau$-inflation self-similarity, which also tells us the way to set up
two kinds of 6.4A simulations (on a Binary tiling).
One has 2.45\AA{} Co-rich-decagons on L sites,
the other has them on (every other) S site. 
This thought was implemented
in MM's REMARKS file for 10\AA{} binary tiling, including figures.}

\LATER{TO MELD! CLH improved this comment and moved it
from IIIA; it duplicates the above.}
%%%%%%%%%%%%%%%%%%%%%%%%%%%
Although natural, doing (almost all) our fixed-site simulations
with a ``standard''  density and composition
(see Sec.~\ref{sec:methods-input} and \ref{sec:MC-a0})
equal to those of ``basic Co'' $d$-AlCoNi was an
unfortunate choice.  For one thing, the nominal composition
and density are unlikely to match exactly those in 
our ultimate idealized decoration.) More importantly, 
even if they do match, one cannot rely on such simulations
to infer the appropriate site-list for later stages, since
the later stages will explore variations in density and
composition.  There is a chance the preferred sites for the variation 
already got eliminated at the earlier stage, since they 
were not being occupied in 
the initial small-tiling simulations.
Instead, the initial exploratory simulation should
run at a density chosen {\it higher} than the expected
value, indeed higher than
the largest density variation to be tried in subsequent runs.
(Or, even better, at densities and compositions bracketing 
the expected ones.)
%%%%%%%%%%%%%%%%%%%%%%%%%%%%%%%%%%%%%%%%
\remark{Sites would then be retained on the next stage sitelist 
(specifically, our 4\AA{} rhombi), if they
seemed necessary at {\it either} density -- or composition.}
%%%%%%%%%%%%%%%%%%%%%%%%%%%%%%%%%%%%%%%%

\subsection{Adapting the method to puckering?}
\label{sec:disc-puckersites}

Yet another reason that the ``basic Co'' story is more complex is 
that relaxation and the formation of ``channels'' that violate the layering
(Sec.~\ref {sec:relax}) have more dramatic effects than they
did in the ``basic Ni'' case~\cite{alnico02}.  Perhaps 
the reason is simply that ``puckering centers'' form around 
columns of Co atoms; they are present in both phases, but 
since ``basic Ni'' has half the density of Co atoms, its puckering
centers are sufficiently separated that their interactions are 
unimportant.
%%%%%%%%%%%%%%%%%
\LATER{to Nan, MM:  (CLHNEW) does this speculation really make sense?
I know MM agreed with the sentiment here, but I'm realizing that 
(i) in basic-Co, half the Co is locked up in Decagon core.
So maybe the density of puckering-centers is about the same?
(ii) we certainly should visit basic Ni and look at the
correlations of puckering centers there!}
%%%%%%%%%%%%%%%%%%
\LATER{Moved from Sec. V.  MELD with above. Note that
``channels'' seem to occur with a higher density in ``basic Co'' than
in ``basic Ni''; presumably that is because channels are associated
with Co atoms and there is more Co in the former phase.  It follows
that channel-channel interactions should be more important in the
``basic Co'' case.}
%%%%%%%%%%%%%%%%%%%%%%%%%%%%%%%%%%
The most serious issue here is that relaxation might
reverse the sign of a small energy difference between 
competing variants of the detailed atomic structure --
we encountered such a sign reversal when comparing 
different cluster orientation orders (see Table~\ref{tab:Eor}).
Thus, one must worry whether our recipe may converge to
a non-optimal answer, having discarded the correct one
in the early fixed-site stages.  
Are there any technical ways to incorporate puckering, 
while still using discrete Monte Carlo simulation?
One may distinguish three points in our story at which one 
could ask for such a remedy.
\LATER{Merge this:
``Is it possible to prepare a better ideal site list to predict relaxed
and post-MD energies -- in particular, to allow for relaxation and
puckering to some degree?''}

The first point is in the initial small-tile stages of MC, where
we would worry that we might miss a nice form of local order,
due to the unphysical fixed-site and layering constraints.
Obviously, this should be performed using a four-layer unit cell,
but that is insufficient by itself: if atoms cannot reduce their
energy by deviating off layers, the 4\AA{} symmetry remains unbroken
(as we verified by some trials).
The key to improvements must be the understanding that
puckering is built from an alphabet of in discrete entities 
-- channels (Sec.~\ref{sec:relax}) or puckering units 
(Sec.~\ref{sec:pucker-units}) -- which are put together,
somewhat as tiles are put together in a tiling.  

One approach, at the raw 2.45\AA{} rhombus level, is to add a
correction to the Hamiltonian which models the energy reduction
that would occur under relaxation.  This would have important 
negative contributions only in cases where atoms in adjacent layers
are stacked nearly on top of each other (e.g., one of the 
``short bonds'' recounted in Appendix~\ref{app:short-bonds}),
but are {\it not} thus constrained by other atoms on the opposite
side from the close neighbor.  This would have exactly the
form of a three-atom interaction.  The coefficients in this
effective Hamiltonian  could be fitted to the relaxed energies
for a database of random 2.45\AA{} configurations.

Alternatively, we could approach the problem at the level of the 
2.45\AA{} DHBS tiling (see Sec.~\ref{sec:small-DHBS};
this tiling has not yet used for MC for the present model system.)
We found that the 2.45\AA{} HBS tiles, each centered by a 
Co column, correspond closely to the puckering units.
Thus, we might incorporate e.g. 2.45\AA{} Star tiles 
of several different flavors, corresponding to the
common puckering patterns (e.g. Table~\ref{tab:pucker-unit-long}).
Within each tile, the puckering-layer Al would be displaced,
but other Al would lie strictly in layers.
This would undoubtedly be a crude way to represent the continuum
of possible Al positions, but the existing method is much
cruder (in forcing them to lie in the layers).

\remark{CLH notes: the above scheme is much messier than
described here! In particular, not every 2.45A-DHBS tile
is a puckering unit; mainly the even ones.  Will the
decoration presuppose particular layers are mirror layers, 
Or is there some way of extending it to e.g. the case where
the Decagon centers are oriented oppositely and there
is no global mirror layer?}

\remark{1/'06, MM suggested 9/05 to introduce spin-like
states for groups of atoms with optimal puckering, and then have an
update move to shift by 4A the group of atoms, and/or do a rotation.
(Or, presumably, to change flavor to another discrete flavor.)
This seems to be the same idea.  CLH now thinks the number of
flavors is not quite so bad as it seemed.  MM also wrote (9/05)
``I now have an idea how to include puckered sites
even for 2.45A tiles. Of course, the particular implementation is 
a technicality, but it is clear that this kind of simulation should 
be done in the future for BOTH Co-rich and Ni-rich phases.''}

A second point where we need a technical adaptation 
was the stage where we conducted molecular dynamics 
and relaxations,  to obtain configurations such as 
Fig.~\ref{fig:puck-Wsim}, or relaxed energies such as 
those in the right columns of Table~\ref{tab:Eor}.
We were hampered by using starting configurations that 
always have the wrong number of atoms in every channel: there 
ought to be three, but two copies of a bilayer necessarily 
have an even number (two or four).  We worry that the
channels may get stuck with random, non-optimal patterns 
of occupancy (see Sec.~\ref{sec:pucker-ising}) and this
may obscure any pattern that would emerge.

At this stage, it doesn't matter greatly how well 
the model positions approximate the real ones, since we are
not comparing energies of the {\it unrelaxed} configurations.
Instead, we just need more of an ideal decoration model 
similar to Sec.~\ref{sec:ideal-deco}, 
but having four layers, such that the two mirror layers 
differ in places.  The model should admit variants, so
that we could discover which rule allows for the best
relaxed results.

A final stage where puckering should be
represented has been reached in the ``basic Ni''
case~\cite{alnico01}, but not yet for ``basic Co'':
a deterministic decoration for quite large tiles
(4\AA{} or probably larger), allowing discrete Monte Carlo
simulations in which only these tiles were reshuffled.
One approach that was used in Ref.~\onlinecite{Mi96b} to
devise such a decoration is ``constrained relaxation'', 
whereby all atoms in the same ``orbit'' (quasi-equivalent 
sites generated by the same decoration rule) are
forced to move in the same exact manner relative to the tiles they
decorate, defining a sort of consensus relaxation.

\LATER{Shorter alternate. 
A possible tool to find candidates for new (non-ideal) sites is 
constrained relaxations: to require all atoms on the same type
of decoration site (``orbit'') to move in the same manner relative
to their tiles.}

\subsection{Long-range order of the puckering pattern?}
\label{sec:disc-puckerLRO}

\remark{One source: Steurer visit (email 7/18/05 to MM);
restated in email 7/26/05 to MM.}

Our structure model develops very robust puckerings
(Sec.~\ref{sec:relax}) with a 8\AA{} period in the $c$ direction.
Assuming the motifs that we discovered and built our
description on, the puckering interactions are frustrated
and sensitive to the Al density and to the Ni placements in \Stars,
which together (see Subsec.~\ref{sec:pucker-results})
determine which channels pucker.
The disorder inherent to any real quasicrystal might 
introduce sufficient randomness that the Ising effective Hamiltonian 
of Sec.~\ref{sec:pucker-ising}
would be a {\it spin glass model} having many almost degenerate minima.
If so, our attempts to discover the true ground state are rather academic.
as the real material would probably get trapped in (somewhat higher) 
metastable states.  

Within each ``channel'', the correlations 
should extend far in the $c$ direction; 
yet our tentative conclusion is that
the puckering order propagates poorly within the $xy$ plane.
The consequence of this would be diffuse scattering
concentrated into ``pancakes'' in thin layers 
close to $q_z= \pi/c$ and (stronger) $3\pi/c$, midway
between the Bragg layers, but rather broad in the $xy$
direction in reciprocal space.

However, the observed diffuse scattering
associated with the 8\AA{} periodicity tends to show 
longer in-plane correlations~\cite{frey00,kat04}.
In fact, many $d$-AlNiCo modifications
propagate true long-range order of the puckering, as shown 
in diffraction patterns have sharp Bragg spots in the intermediate layers
(that appear between the main layers associated with $c$ periodicity).
So, when the real material does have long-range puckering correlations, 
one may wonder if it includes some motif beyond our model. 

In fact, the $W$(AlCoNi) phase does propagate a well-ordered puckering,
and manages this by replacing half of the even \Stars with another 
motif, the pentagonal bipyramid (see Sec.\ref{sec:W-PB}).
So, a plausible conjecture is that the \PB~ is the
missing motif which is responsible for extended puckering correlations
(in the more Co-rich modifications of Al-Co-Ni).

\subsection{Relation to decagonal Al-Co-Ni at other compositions}
\label{sec:disc-alconi-phases}

\LATER{YES. Cite the (several) sections of the paper which relate to bNi.}

\CLH{This next mostly about the Co15-Ni15 composition,
Was originally in MM's email ``colong-mm-phases'' (12/2005) which 
has additional refs on this composition; the rest of the
phase diagram discussion is in Intro.}
%%%%%%%%%%%%%%%%%%%%%%%%%%%%%%%%%%%%%%%%%%

A variety of small-grained, apparently metastable, crystalline
approximant phases are found alongside quasicrystals at compositions
near $d$(Al$_{70}$Co$_{15}$Ni$_ {15})$; it was suggested that the
presence of quenched-in vacancies might tilt the balance to stabilize
one of the approximants against the quasicrystal~\cite{grushko-approxts}.  
The difficulty of determining stability suggests
that these related phases are very close in free energy.
Since the decagonal domain of 
the Al-Ni-Co phase diagram is bracketed by 
phases we studied (``basic Ni'' in Ref.~\onlinecite{alnico01})
and ``basic Co'' in  this work), can we say more about 
those intermediate phases?

One clear conclusion~\cite{alnico01} is
that special compositions are stabilized, in large part,
because each species is filling a particular type of site.  Thus a
small composition difference (density or stoichiometry) can cause
certain orbits (classes of quasi-equivalent sites) to become
occupied or to change species.  At a higher level, the interactions
of these atoms will then change the tile Hamiltonian of the tiles
they sit on; and that can make a big difference in how these tiles
freeze into supertiles at even larger length (and smaller energy) scales, 
hence the variety of modifications.

We can speculate how changes in composition might change the whole
geometry of our structure, e.g. between the ``basic Ni'',
\Decagon, or 20\AA{} decagon-based structures.
The heart of our understanding of the
physical relationship between the atomic interactions and large-scale
geometry is the 2.45\AA{} DHBS tiling of Sec.~\ref{sec:ideal-deco}.
There is no reasonable way to {\it increase} the frequency of \Decs. 
(Recall the arrangements in Fig.~\ref{fig:d-adjoin}(a,b) 
violated strong interactions, namely the TM-TM second well.)  
But perhaps replacing Co with Ni in the composition would induce
replacing 8\AA{} decagons by $2.45$\AA {} HBS tiles.
Indeed, if we eliminate the \Decs~altogether, this {\it is}
essentially the structure of Ni-rich $d$-AlNiCo~\cite{alnico01}.  
So might intermediate compositions like $d$-Al$_{70}$Ni$_{15}$Co$_{15}$
be described by a smooth gradation in which the frequency of \Decs~
diminishes?

We can speculate on how the inexactness of our pair-potential description 
will distort the computed phase diagrams.  The fact that supertiles
form means that what decides the large-scale geometry is the
effective tile-tile interactions (``tile Hamiltonian'').  
In this picture, over a range of compositions the same supertiles
are valid, but the species filling certain sites on them changes
with composition and consequently so do the effective interactions
in the tile Hamiltonian.  At this level of description, errors
in the potential themselves would shift the graph of the interactions
as a function of composition, but probably not change its gross
shape.  The corollary is that the phase diagram of our toy system
might well have the same phases, in the same topology, as the
true one, but with the phase boundaries shifted.
\LATER{Alternative text: ``Yet it is a reasonable hope
that the trend -- how the tile Hamiltonian {\it changes} with
composition -- is correct.  If so, then the theoretical phase diagram
will look like the experimental one, but with shifts of composition
that we are not equipped to predict a priori.''}

A particular application is to the issue of the 20\AA{}
decagon (App.~\ref{app:20A}).
The models based on 13\AA{} and 20\AA{} decagons are very similar 
in structure, and (not surprisingly) very close in energy.
Our viewpoint is that {\it both} are physically relevant.
Slight modifications of the potentials, or of the 
assumed composition and density,  might well tip the
balance between these two models (or other related ones, 
in particular those incoporating the PB)
(Sec.~\ref{sec:W}).

\FUTURE{CLH suggests a comparison of the \Decagon~
  and the \Star~ clusters, by concentric rings. How different are they
  really (very!)?  Could there be a structure model in which the
  distinction got erased -- what would it be like?}

\begin{acknowledgments}
  This work is supported by DOE grant DE-FG02-89ER45405; computer
  facilities were provided by the Cornell Center for Materials
  Research under NSF grant DMR-0079992.  MM acknowledges support from 
  Slovak Grant Agency (grants 2-5096/25 and 51052702).
  We thank M. Widom for discussions, and K. Sugiyama for sharing the 
  coordinates from Ref.~\cite{Su02}.
%%%%%%%%%%%%
  \remark{The z components can't be seen at all from the published images, 
   can they? So this was essential.}
%%%%%%%%%%%%
\end{acknowledgments}

\appendix
\section {Code and decorations}
\label{app:code}

In this appendix, we described some technical aspects of the
code.~\cite{FN-decadeco}

We begin with a {\it tiling} file that defines a (2-D) unit cell in
terms of Penrose rhombi. ' We then assign a scale to the unit cell,
identifying the physical length of rhombus edges as well as fixing a
certain periodic lattic constant between adjecent Penrose tiling
layers. The full 3-D unit cell is then specified by providing the
number of layers we want before periodic boundary conditions.

The {\it decoration} file specifies {\it objects} by looking for
specified patterns (called ``objects'') in the tiling geometry,
specifically groupings of rhombus edges with specified orientations
relative to each other.  An ``object'' can be as simple as two edges
at $144^{\circ}$ from each other, or as complicated as full decagon
traced out by all interior and exterior edges. Atomic sites can then
be placed on each object, and all objects of a given class will get
equivalent sites.

\OMITb{ To clarify, an object is a pattern template, the outline of
  the object must be present on the Penrose tiling as the boundaries
  of the rhombi.  Objects need not be closed i.e. two adjoining sides
  of a PR can be defined as an object. A Penrose rhombus can be an
  object, as may be a cluster of these tiles -- e.g. the fat or thin
  hexagon made from three rhombi -- but not all objects are built from
  tiles.}

The decoration also allows energies to be assigned to objects. For
example, when a tile flip occurs, the energies of the atomic
interactions and the energy of any geometry created are taken into
account. We use this option only for tiling purposes (there are no
atoms on our tiles when we use object energies) as we search for
tiling which satisfy the large scale ($a_0 {\tau}^3$) binary tiling.
After we find such a configuration of Penrose rhombi, we then use a separate
decoration to create a site list.

The decoration file also lists any symmetries of the tiles and objects defined
therein. If an object has a reflection symmetry and the symmetry is defined in
the decoration then adding a site on one side of the object will add a mirror
site to the other side.  
\NAN{to MM: I only have minimal understanding of how symmetries 
are implemented in the decoration. Dr. Mihalkovic, please expand on 
this if it seems necessary.}

It is also possible to assign a label to each vertex of a tiling
through the decoration file. The label assigned is the same as the
discrete component of the `perpendicular space'of a decagonal tiling.
We shall called these labels {\it levels}.  \NAN{to MM: I have no idea...so
  Prof. Henley or Dr. Mihalkovic, please elaborate.}

The random Penrose tiling allows, in principle, an unlimited number of
levels; our other tilings typically have vertices on two to five
different levels,which are treated as different flavors of vertices.
For example, the HBS tiling has two levels~\cite{Hen91ART} the binary
tiling has three levels;and the original (quasiperiodic) Penrose
tiling has four levels, as does the HBS tiling when the interior
vertex of each tile is filled in~\cite{tang-jaric}. 
It makes sense to assign different
decorations to vertices depending on their level in the tiling, or to
tiles depending on the levels of their vertices. When these levels are
taken into account, not all rhombi of the same shape on our tilings
are equivalent.

These levels allow more specific control over the location of atom
sites.  In the tilings with a bounded set of levels, there is a
(statistical) symmetry operation which combines a $180^\circ$ point
rotation of the tiling with a reflection in level space.  In our
decorations of decagonals, this symmetry may be combined with a
vertical shift of $c/2$ to form a kind of screw symmetry. (It is
statistical in the sense that a random tiling ensemble is invariant
under it although paricular tiling configurations are not; also, it is
local in the sense that clusters can be found within which the screw
operation is an exact symmetry.

The simulation uses Metropolis Monte Carlo~\cite{FN-seed} to perform
atom swaps between nearby atoms or hops to nearby vacant sites
(generally with a smaller number of long-distance swaps/hops
included). A novel aspect of the procedure of Ref.~\onlinecite
{alnico01} is that concurrently, tile reshuffling is also performed:
this means a hexagonal configuration of three rhombi (two thin and one
fat or vice versa) is rearranged. The tile reshuffling has the effect
of a number of atom swaps and moving atomic sites around. Because a
tile flip generally causes a large change, its acceptance rate is
relatively low and virtually nil at lower temperatures. The low
acceptance rate causes the tiling to freeze at low temperatures.

A variety of other methods were used to alter and test certain aspects
of the simulations in a controlled manner. This included, but is not
exclusive to a) a series of manual swaps, b) manual tile flips, c)
analysis of atomic pair distances and pair potentials along with site
energies to determine frustrated sites, and d) direct modification of
data files to obtain custom configurations.

\section{Tests of cluster-cluster geometry}
\label{app:dec-rel-test}

\remark{Nan says this sort of test came out of 2.45A simulations.
But later tests devised  MM made use of the 4 A tiling too.}

This appendix reports tests performed to eliminate various
possibilities in Sec.~\ref{sec:fixed-site}, namely
short bonds and alternate cluster linkages  (Sec.~\ref{sec:Dec-relation}).
We treat these as technicalities since they do not enter
our final model.

\subsection  {Short bonds}
\label{app:short-bonds}

A noteworthy issue in our simulation was the ``short bond'', an Al-Co
in adjacent layers, with an $xy$ displacement of $\tau^{-2}a_0\approx$
0.935\AA{}, hence a total separation of 2.245\AA{}.
This is so short as to be up against the hardcore of the pair
potential $V_{\rm AlCo}(r)$, hence questionable.  Indeed, in Al-TM
quasicrystal-related alloys, some exceptionally short Al-TM bonds have
been noticed for a half-century~\cite{Pearson}:
in particular, Al-Co pairs exist in
Al$_5 $Co$_2$ at 2.34\AA{} and in Al$_{13}$Co$_4$ at 2.25\AA{}.
%%%%
\remark{(MM, 9/05). These distances were also confirmed by
ab--initio calculations (http://alloy.phys.cmu.edu).}
%%%%
But our pair potentials are not very trustworthy when the
closeness of the cores enhances covalent effects. 
We still suspect that our short bonds 
are artifactual as for as our simulation potentials are concerned;
the short bonds appearing in nature have a somewhat different
explanation.

\remark{Indeed, embedded-atom potentials,  which 
incorporate non-pair contributions,
were used~\cite{redfield90} to justify the
Mackay Icosahedron cluster (with its short Al-Mn bond) 
in $i$-AlMnSi.)}

Almost every configuration from the initial simulation stage
using 2.45\AA{} tiles contained some Al-Co short bonds; 
they could appear in any ring of the \Dec, but are particularly
problematic in ring 2.5/ring 3 where Al positions deviate
from symmetry in any case.
Our initial guess was that the short bonds were artifactual, being our
lattice gas's attempt to approximate a minimum-energy position that
actually fell between two discrete candidate sites.
Therefore, our canonical 4.0\AA{}-tile decoration 
omitted candidate sites that allow short bonds.  

However, as a variation we did augment the 
4.0\AA{}-tile site list so as to allow
short bonds with ring 2.5/ring 3 Al.
When we tested this in the 23 $\times$32$\times$4 unit cell, about five short
bonds appeared in each simulation run, furthermore the
total energy was {\it lower} than in the 4.0\AA{}-tile simulations
without short bond sites.  That would suggest the short-bond Al
positions are approximating the true relaxed positions {\it better}
than the non-short-bond sites did.

When relaxation was performed (Sec.~\ref{sec:relax-bilayer}), 
short-bond Al-Co distances increased, while 
non-short-bond Al-Co distances decreased.
The relaxed configurations were still distinguished, in that (with a small
sample of three runs), the energy were {\it higher} (worse)
when relaxed from the short-bond configuration;  in other words,
under pure relaxation without MD annealing,
the Al atoms apparently get stuck in shallow local minima.
%% ``RMR'' procedure 
On the other hand, after ``RMR''  simulation (in a unit cell of periodicity
$2c$, as in Sec.~\ref{sec:doubled})
seems to reach the deeper minima:
initial short-bond or non-short-bond configurations gave
results indistinguishable in energy and configuration 
(modulo some effectively  random choices of which direction
to pucker).
\remark{``Indistinguishable configuraation must
be understood statistically: since each ``puckering unit'' tends to
find its configuration independently in the RMR process, there is a
large near-degeneracy.}

The short bonds are most clearly understood using the framework 
introduced in Sec.~\ref{sec:channels} of ``channels'' -- approximately
vertical troughs in the potential function for an Al atom.
If an Al-Al hardcore distance of $\sim 2.8$\AA{} is enforced, there is
room for just two Al per four layers on the fixed sites, or three Al
per four layers once puckering is allowed.  
The ``short-bond'' observations in bilayer fixed-site simulations
suggest it may after all
be tolerable to place {\it four} Al per four layers
(in preference to the sites where the Al would otherwise be forced to go, 
at the Al density being assumed).

We conclude that allowing short bonds in simulations does not help us
to capture the true order any better.

\subsection{Overlapping cluster-cluster linkage?}
\label{app:overlap-linkage}

\begin{figure}
\includegraphics[width=2.9in,angle=0]{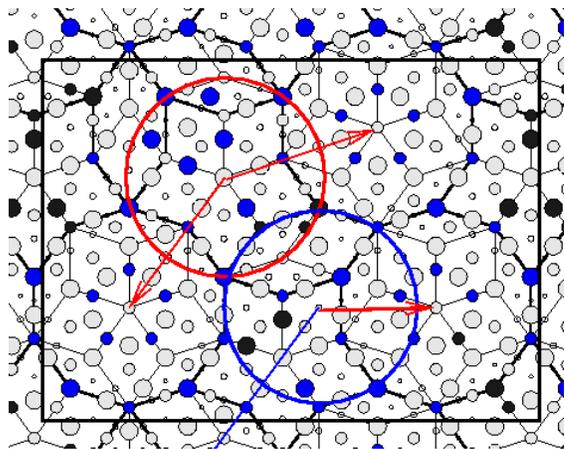}
\caption{Test of the overlapping linkage in 
Fig.~\ref{fig:d-adjoin}(a).  Tile edges marked
are 4\AA{} long; heavy edges mark \Decagons
(or 4\AA{}-edge Star and Boat tiles that fill the
space between \Decs).
Atom colors indicate species as in Fig.~\ref{fig:smalltiles}
Blue and red circles indicate 
Thin red arrows, fat red arrow, and blue arrow 
indicate hypothetical linkages of length 
$\tau^3 a_R$, $1.176\tau^2 a_RR$, and ${\sqrt 5} \tau^2 a_R$, 
respectively.
\LATER{Improve caption; trim the figure
(CLH Note 1/1/06: ps file trimmed by xv caused LaTeX error, 
even though length is not too long.)
Convert the figure to black-and-white,
and distinguish linkages w/o color. Also, mark another circle 
at the place where the decagon actually materializes? }
}
\label{fig:clink}
\end{figure}

As a test of the overlapping linkage in Fig.~\ref{fig:d-adjoin}(a),
we made use of the 4\AA{} decoration of Sec.~\ref{sec:MC-4A}, 
using the tiling shown in Fig.~\ref{fig:clink} which violates
the binary-tiling rules for placing the clusters.  Only
three \Decagons (indicated by circles) are defined by this tiling 
(four normally fit into this cell); 
the extra space has been filled by 4\AA{}-edge Stars, Boats and Hexagons. 
\remark{Site list is our "fixed 4A Decagon-packing".}

The figure shows a typical configuration that formed at low temperature. 
A fourth \Decagon~has spontaneously materialized on a grouping
of a 4\AA{}-edge Star tile and two  Hexagons
in the upper left corner.  The outer border of this tile cluster
forms a decagon, 
but its interior (and the associated site list) lacks decagonal symmetry, 
forcing ring 1 to form with a small mistake.
\remark{This demonstrates that the Decagon is not too fragile:
it does not depend on having the sitelist designed for the Decagon.}

%%%%%%%%%%
%%%%%%%%
Now, the red circle in Fig.~\ref{fig:clink} shows an alternative
place where the site list would have allowed a \Decagon to appear
instead,  overlapping as in Fig.~\ref{fig:d-adjoin} (a)
with two other \Decagons; indeed, with the tiling shown,
the site list in fact favors this alternative location.
However, in a few tries of this sort, that cluster
with the overlapping linkage never formed.

\LATER{CLH would remove the blue  circle}

The blue circle in Fig.~\ref{fig:clink} shows 
another hypothetical cluster location forming a different
linkage as indicated by  the thick red arrow.
It corresponds to edge sharing by the 2.45\AA{}-edge decagons;
in this case, the \Dec~ overlap forms a thin 4\AA{} Hexagon.
\CLH{MM wrote ``this would conflict with the ring 1 Co pentagon. 
I think the point is to explain why the 2.45 A decagons
can't share edges, in terms of some bad bond length.}

\LATER{The site list actually favors the red circle location.}

\CLH{to Nan (and to MM): The following had been clipped by Nan; I've
  restored it (rewritten) -- should it stay?}

Another test involved the simulations on the 
the 20$\times$38$\times$4 tiling mentioned at the end
of Subsec.~\ref{sec:Dec-relation}.
After short-time anneals on the small 2.45\AA{} tiling,
the configurations contained many Al$_6$TM$_5$ 
(ring 1)  motifs, but the \Decagons were imperfect and 
often interpenetrating, in contrast to the good ordering 
observed after similar annealing on the 32$\times$23 tiling.  
However, such configurations were $\sim 2$eV ($\sim 0.02$eV/atom) 
higher in energy than the $32\times 23$ tiling best energies.  
Furthermore, after 5 cpu hours of high-temperature 
($T \sim 0.5$eV) annealing, 
\CLH{to Nan: Am I correct to interpret high-T annealing
as meaning beta is about 2?}
these samples evolved to a proper configuration of \Decagons,
which was in fact lower in energy than on the $32\times 23$ cells
(with the same volume and atom content).

\subsection{Rhombus cluster-cluster linkage?}
\label{app:rhombus-linkage}

\begin{figure}
\includegraphics[width=3.1in,angle=0]{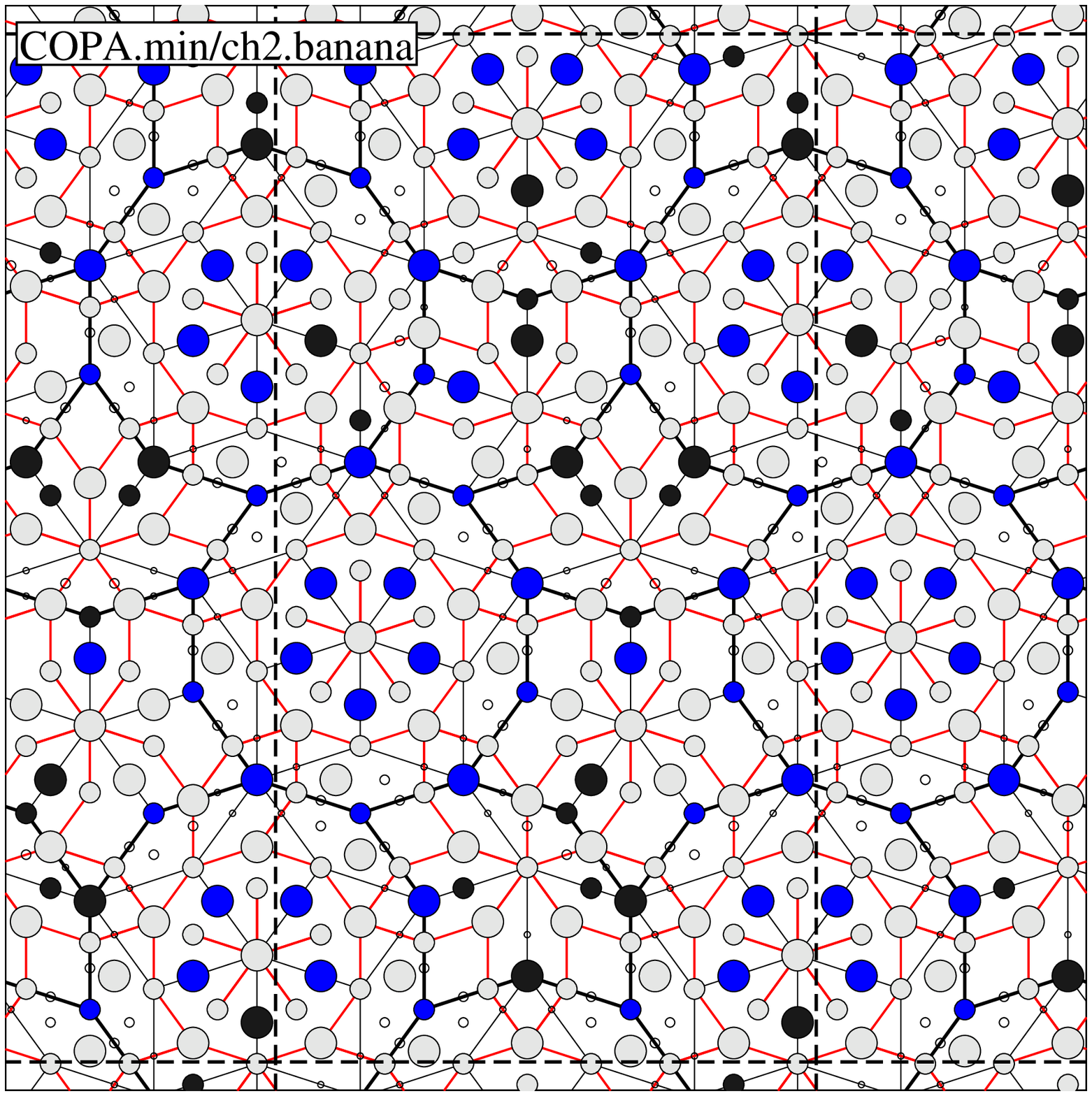}
\caption{Test of the ``rhombus diagonal'' linkage in 
Fig.~\ref{fig:d-adjoin}(b). 
The rhombus center is just below the cell's upper-right 
corner (the fourth decagon, seen at lower-right corner;
its copy by periodic boundary conditions is outside the
upper frame.)
\LATER{Improve caption; trim the figure.
The rhombus is at the corner of cell and is practically impossible
for the reader to re-assemble the 4 pieces from periodic
boundary conditions.  Need to trim it a little narrower to fit
in one PRB column (problem: CLH did this by xv, it made huge
bitmapped ps file).  Would be good to add rhombus shape on top 
to emphasize it. Also, need to convert to black-and-white.}
}
\label{fig:linkage-banana}
\end{figure}

Next we test the linkage shown in Fig.~\ref{fig:d-adjoin}(b)
%%%
In the 32$\times$23 cell,
with content Al$_{145}$Co$_{41}$Ni$_{21}$ (hence density
0.0682 \AA{}$^{-1}$), a  followup fixed-site simulation
was done,
using the standard site list which avoids short Al-Co bonds.
Three different  4\AA{}-edge tilings were
tried, each with four \Decs specified  per cell.
One of these tilings (Fig.~\ref{fig:linkage-banana} has 
four clusters at the vertices of a Fat rhombus.
%% ~marek/banana.ps
Its energy is roughly 2 meV/atom higher than the others.
(We averaged this difference over the two cases where the 
\Dec~center orientations  are all the same and where they are
alternating).
Since there is one 1.176$b''$ linkage in that cell, this
amounts to a substantial cost of roughly 0.4eV for each such linkage.

%%%%%%%%%%
%%%%%%%%

\section{Optimum configuration in channel: analytic calculation}
%%% This  was moved to appendix
\label{app:channel-anal}

This Appendix augments the mathematical details of
the story in Sec.~\ref{sec:channels}, which are all
consequences of the form of 
Al potential function $U(z)$ in a channel 
as written in Eq.~(\ref{eq:Uz}).
We assume the channel has three atoms, and derive the consequences.

\subsection{Three-Al collective coordinate and local mirror layer}

The main freedom of the three Al atoms in a channel is the collective 
$z$ coordinate $\zbar$.  Let the respective Al positions as
    \begin{equation}
z_m = (2c/3)m + \zbar + u_m
    \label{eq:zm}
    \end{equation}
with $u_{m+3}\equiv u_m$, and constraining $\sum_{m=-1}^{+1} u_m=0$. 
[The mean displacement is  
accounted in the collective
coordinate $\zbar$.]
The total energy (per Al) is~\cite 
{FN-omit2ndAl}

    \begin{equation}
        E \equiv \frac{1}{3}\sum _{m=-1}^{+1}  
[U(z_m) + V(z_{m+1}-z_m).
    \label{eq:Etot}
    \end{equation}
%%%%%%%%%%%%%%%%%%% 
%%%%%%%%%%%%%%%%
We want to find the best energy, given that $\zbar$ is fixed at a
certain value.
Taylor expanding (\ref{eq:Etot} to second order in $\{ u_m \}$  
yields
    \begin{eqnarray}
       E = E_0 + \frac{1}{3} \sum _{m=-1}^{+1}
[ U'(\zbar  
+ \frac{2c}{3}m) u_{m} + \\
 U''(\zbar + \frac{2c}{3}m) u_ {m}^2 + 
V''(\frac{2c}{3}) (u_{m+1}-u_m)^2 \nonumber
    \label{eq:Etot-quad}
    \end{eqnarray}
where $E_0  
\equiv U_0 + V(2c/3)$.  
If we omit the $U_{c}$ term in (\ref{eq:Uz}),
the minimum  of the quadratic form 
(\ref{eq:Etot-quad}) is~\cite{FN-solve-Echannel}
   \begin{equation}
       E(\zbar) = E_0 -\frac {1}{6}  \frac{B_U^2}{K_V^2-K_U^2} 
\left[ K_V + K_U \cos \frac{12\pi}{c} \zbar \right], 
   \label{eq:Etot-zbar}
   \end{equation}
where $K_V \equiv 3 V''(2c/3) \; $ $K_U \equiv (4\pi/c)^2 U_{c/2}$,
and $B_U \equiv \sqrt {3/2} (4\pi/c) U_{c/2}$.  The effective
potential $E(\zbar)$ in general has a period $c/3$, one-third of the
unit cell periodicity, since sliding the Al chain by $\pm 2c/3$ would
move each Al atom to the old position of its neighbor, and
additionally the potential is invariant under a shift $\pm c$.  
[The period in Eq.~(\ref{eq:Etot-zbar}) is $c/6$, not $c/3$, since we assumed as
a simplification that $U(z)$ has period $c/2$, not $c$.]
%%% 
The minimum configuration can be written $\zbar=0$ or $c/2$, with
$u_{-1}=-u_{1}$ and $u_0=0$, so the channel has a local 
{\it mirror layer} at $z=\zbar$.  

Although the local minima of (\ref{eq:Uz}) as a function of the
single-atom $z$ are quite strong, the effective potential $E (\zbar)$
is much flatter as a function of the collective coordinate, 
since the three atoms are constrained to sample different,
counterbalanced parts 
of the potential.  One can quantify ``much flatter'', using
the assertion above that Al-Al interactions are stronger 
(within a channel) than the Al potential due to Al-TM interactions.
i.e. that $U''(2c/3)/ V''(2c/3) \sim K_U/ K_V \ll 1$.
This implies via
(\ref{eq:Etot-zbar}) that the ratio of the energy variations of
the collective to those of the one-body potential is
$\Delta E(\zbar)/ \Delta U(z) \sim (K_U/K_V)^2$.  Thus, it is conceivable
that the Al atoms in each vertical channel have, at moderate
temperatures, considerable freedom to fluctuate (collectively) in the
$z$ direction.

\subsection{Further symmetry reduction by TM-rich layer}
\label{app:channel-Uc}

We now address the role of the $U_c$ term in eq.~(\ref{eq:Uz}).
It often happens that (in the language of the 4.0\AA{} tiling)
there are two tile edges forming a 72$^\circ$ angle, with a Co
column over each of the three vertices, and channels on both the
mid-edges. (For example, this can  involve ring 1 and
ring 3 Co atoms in the \Dec). From the viewpoint of one channel,
one Co column is distant and breaks the $c/2$ periodicity: this is 
the  justification of the last term in (\ref{eq:Uz}).  
To explain  the {\it sign} of that term, note that
if an Al in the channel is in the same layer as a distant Co,
the separation is $R=3.80$\AA{},
%%%%%%%%%%
\remark{Also it is 5.57\AA{} away from Co in the same distant column,
  is two layers above or below the Al atom [$\delta z =\pm c$],
which is another maximum in $V_{\rm Al-Co}$.}
%%%%%%%%%%
  close to a local {\it maximum} of the Al-Co potential,  
  whereas if Al is offset by one layer then $R=4.47$\AA{} to the
distant Co atoms, close to a local {\it minimum}
(see Table~\ref{tab:potentials}).

\LATER{Need to insert some of the mathematical
derivation of this here?}

Now consider the implication for the collective coordinate, 
when we include the $U_c$ term of (\ref{eq:Uz}). 
That will generate an additional contribution to $E(\zbar)$ 
of form $-\sigma {\rm const} \cos \frac{6\pi}{c}\zbar $.
As it turns out, the sign of this term favors the mirror atom 
to sit in the {\it same} layer as the distant column of TM atoms 
(this is reversed from the layer preferred by the single-atom potential).  
Simulations show it is favorable for the atom occupancies 
to arrange themselves such that one layer of the two layers 
is richer in TM, which causes the puckering to develop such that 
this layer is a mirror layer {\it globally}.
That is directly associated with the long-range order
of decagon orientations (Sec.~\ref{sec:decLRO})
and of puckering (\ref{sec:puckerLRO}.)

\subsection{Transverse displacements in channels}

In the above account, the displacements of Al atoms in channels
to balance the  ``external'' Al potential with Al-Al repulsions
were represented as purely in the $z$ direction, only to allow
a transparent analytic description.
A more exact analysis would need to consider the transverse variation
of the potential trough, for the Al-Al repulsion obviously should be
based on the total Al-Al distance, and not just its $z$ component.
In fact, the transverse undulation of the channel (bottom) line as a 
function of $z$ [see Fig.~\ref{fig:channel-Al}(a)], as well as the
$xy$ deviations of the atoms from the channel line, 
may well make a contribution to the total $E(\zbar)$
comparable to the dependence in (\ref{eq:Etot-zbar}).
For example, to make all three Al-Al separations be equal at
$R=2.87$\AA{}, the puckering displacement must be $(c-R)/2 =
0.645$\AA{}, and the $xy$ difference between the puckered-layer site
and the mirror-layer site should be $\sqrt{R^2 -(c-R/2)^2} =0.90$\AA.
For comparison, the channel's extremes are practically on
ideal sites, separated  in the $xy$ direction by 
$\tau^{-2} a_0 \approx 0.94$\AA{}. The actual $xy$ displacement
would be only 2/3 that much if Al are assumed to stay on
the bottom line, where the latter is approximated
a ``sawtooth'' pattern of straight segments and the Al 
$z$ components are approximated as equally spaced.  
\remark{We need to exaggerate the horizontal
displacements about 15\% beyond what would be expected from the
channel bottom, in order to make the Al atoms fit.}

%%%%%%%%%%%%%%%%%%%%%%%%%%%%%%%%%%%%%%%%%%%%%%%%%%%%%%%%%%%%%%%%
\section{Labels of puckering patterns}
\label{app:pucker-labels-details}

In this appendix, we give detailed ways to label and enumerate
possible configurations of a puckering unit 
(the short labels were explained in Sec.~\ref{sec:pucker-labels}).

To exhaustively specify the puckering configuration around 
the center, we list the mirror-layer Al atoms, 
giving the angular placement $l$ of each
[meaning angle $(2\pi/5)l$], 
and its puckering sense ($+$ if found in mirror layer 0, or $-$ 
if in mirror layer 2):  
The zero angle is defined as the shared \Dec~ edge 
(for type $A$ puckering unit)
or the ray through the \Dec~ center (for any type $B$ puckering unit).
An integer-plus-half angle is a ``merged'' channel site, as
discussed in Sec.~\ref{sec:pucker-units}.
%%%%%%%%%%%%
\remark{Of course, the actual angles can take other angles too: 
recall these atoms sit in the ring-shaped trough of the Al 
potential function in each mirror layer.}
%%%%%%%%%%%%
The constraints on these sequences are (i)
two atoms in the same layer must differ in angle
by at least 1.5 steps; (ii) for every $l=0,...,4$ at 
least one atom must either have that $l$, or else have $l\pm 0.5$.
%%%%%%%%%%%%%%%%%
\remark{One can add, as many as possible are integer angles.}

The short name of a $B$ pattern can have a parity superscript
``$+$'' or ``$-$'' added, 
meaning that the atoms at (or near) angles 1 and 4
(on the \Dec~ edges) have the same or opposite puckering sense, 
respectively: this indicates how the puckering sense propagates
around the \Dec.
As we noted in Sec.~\ref{sec:pucker-labels}, the $-$ parity
is commonest, suggesting $\Jpuck{\alpha\beta}$
is ``antiferromagnetic'' for {\it second} nearest neighbor channels
around a puckering unit center. 
A straightforward explanation would be that the ideal sites 
of mirror-layer Al atoms, two angle steps apart,
are separated by $2\sin (\pi/5) \tau^{-1} a_0 \approx$ 2.88\AA{}, 
which is close enough that the Al-Al repulsion is significant.
\CLH{To Nan, MM: is there a better explanation?}
In an ``$A$'' pattern, the parity is undefined, since it 
would be $+$ on one of the \Decs~ and is $-$ on the other one. 
(That follows since the mirror-layer Al atoms at angles 2 and 3,
over the unshared \Dec~ edges meeting at the vertex,  
always have opposite puckering senses.)

Some common arrangements are listed in Table~\ref{tab:pucker-unit-long}.
The two standard kinds of ``crooked cross'' appear here as ``$p4^+$''
(one arm aligned radially and the other tangentially), or
``$p4^-$'' (arms about 45$^\circ$ from the radial axis).
%%%%%%%%%%%%%%%%
\remark{A way to describe the $p4^-$ kind of crooked-cross is that 
the Al mirror atoms nominally angles 1 and 4
(along the \Dec~ perimeter) get displaced a bit sideways, 
so the arms of the cross are $45^\circ$ from that ray.  
So, actually $B_1(p4^-)$ looks a bit like $(0.5+, 2-, 3+, 4.5-)$.}
%%%%%%%%%%%%%%%%
(The $p4$ pattern in the ``$A$'' environment is a ``crooked cross''  
which, when labeled as if on an unshared vertex, would be $p4^+$ 
from one cluster's viewpoint and $p4^-$ from the other's.)
The common $A(p5)$ pattern could be described as 
another way to resolve the frustration of the puckering
sense.  As in Subsec.~\ref{sec:pucker-units}), 
imagine we alternate puckering senses $\puck_\alpha$ 
around the five channels, necessarily with one adjacent 
pair having the same sign.  Instead of merging  these
atoms, keep all five and accomodate  the steric constraint
by displacing one of them by half an angle step.

\remark{In the overpacked, unpuckered $m=6$ patterns, the Al atoms are
always the same places in both mirror layers.  This could
mislead one into thinking this is necessarily true when one
overpacks a channel.  However, I think the real story is
that certain channels have a Ni neighbor and are much less
prone to puckering.  So, when one packs additional Al, 
these are the first place it will go (there is less 
puckering energy to lose). But the reason the atoms go
into the same places is the single-Al energy potential
due to the TM's.}

\remark{The $B_0(p5^+)$, a flattened cross, is 
less common.  Notice there are two different patterns called
$p5^+$ here.  This $B_0(p5^+)$ also appears as a variant
of the $A(p5)$ environment, but due to the different convention for angle 0,
transcribed as $A(0+,1-,2+,2.5-,4.5-)$.  From the viewpoint of the
other \Dec, it would have been a kind of $B(p5^-)$, transcribed as
$B(0.5+,1-,2+,3-,3.5+)$.}

%%%%%%%%%%%%%%%%
\remark{Of the cases of $B_1(u6)$ in the W-cell, about half are 
a semi-puckered variant $(0+,1-, 2.5\pm, 4\pm)$.
(In our original terminology, one of the four atoms from ring 3 
has moved into ring 2.5 of the \Dec.)}
%%%%%%%%%%%%%%%%

%%%%%%%%%%%%%%%%%%
\begin{table}
\begin{tabular}{|lll|}
\hline
Location & Short name & Long name \\
\hline
\quad $A$   & \quad  $p5$   & $(0+,1-,2+,3-,4.5-)$ \\
\quad $A$   & \quad  $u6$   & $(0\pm, 1.5\pm, 3.5\pm)$ \\
\quad $B_0$ & \quad $p4^-$  & $(1+, 2-, 3+, 5-)$  \\
\quad $B_0$ & \quad $p5^+$   & $(0+,1-,1.5+,2.5-, 4-)$  \\
\quad $B_1$ & \quad $p4^+$  & $(0+, 1-, 2.5 +, 4-)$  \\
%% \quad $B_1$ \quad & $p5^-$   & $(0+,1+,2.5\pm, 4-)$  \\
\quad $B_1$ & \quad $p5^+$   & $(0+,1-,2.5\pm, 4-)$  \\
\quad $B_1$ & \quad  $u6$   & $(1\pm, 2.5\pm, 4\pm)$ \\
\hline
\end{tabular}
\caption{ \footnotesize
 Common puckering patterns.}
\label{tab:pucker-unit-long}
\end{table}

\section{20\AA{} decagon models?}
\label{app:20A}

In this appendix, we compare our results to three recent experiment-based
structure models, all based on some sort of 20\AA{} diameter decagon
(the first two being similar to the Burkov cluster of 
Subsec.~\ref{sec:ideal-Burkov}).
We then report on a trial simulation of our own, using the methods
of Sec.~\ref{sec:fixed-site} but with an enriched site list for 
the lattice gas, from which a 20\AA{} decagon packing emerges that
is competitive in energy with the 13\AA{} model developed in this paper.

\LATER{YES, ut somewhere.
[From MM's note in Intro from 9/9/05 edit].
HREM images of the ``superstructure type I'' modification,
of composition roughly $d$(Al$_{72.5}$Co$_{18}$Ni$_{9.5}$,
show 20\AA{} clusters forming a 
nearly perfect rectangle-triangle tiling (with 20\AA{} edge length),
probably decorated by two kinds of cluster alternating on
even and odd vertices.}

\subsection{Structure models based on PD4 approximant}
\label{app:20A-PD4}

\remark{Let's not forget the near-impossibility of distinguishing Co/Ni 
by X-rays when comparing to our model.}

After most of our work was completed, a tentative structure solution 
appeared for the approximant PD4 of the Co-rich phase~\cite{oley06}
with nominal composition
Al$_{72.5}$Co$_{18}$Ni$_{9.5}$.
\LATER{Is that the composition as given in Grushko, Holland-Moritz,
and Wittmann, J. All. Compd. 280, 215 (1998)?}
%%%%%%%%%
The $c$ projection (see their Fig.~3) clearly shows an 
arrangement of decagonal clusters of diameter $20$\AA{}
(larger than the \Decagon~ by exactly the factor $\tau$).
This 20\AA{} cluster appears practically the same as
the Burkov cluster which -- as we explained in 
Subsec.~\ref{sec:ideal-Burkov} -- appears in our structure model.
Namely,  this decagon has rings 1,2, and 3 like our \Dec~
of Sec.~\ref{sec:clusters}; 
their ring 3 Al show deviations into ring 2.5, similar to
what occurs in our structures.  In half of their clusters,
ring 1 is missing a couple of Al atoms; these have more
irregularity of their ring 2.5/3 Al atoms.

The big difference is that, in PD4, the 20\AA{} decagons do not
overlap; instead, they decorate a Fat rhombus with edge 10.4\AA{}.
Thus, the cluster-cluster linkages in this model are of length
$\tau^3 a_0=10.4$\AA{}, for 20\AA{} decagons sharing an edge, or
$1.176 \tau^3 a_0=12.2$\AA{}, for 20\AA{} decagons related by
the short diagonal of a Fat rhombus.
%%%%%%%%%%%%%%%%%%%%
\CLH{to MM:  Can you be sure what is the tile geometry these clusters 
are decorating, if we had a decagonal model based on PD4?}
%%%%%%%%%%%%%%%%%%%%
It may also be noted that the cluster orientation pattern
in PD4 is neither ``ferromagnetic'' nor ``antiferromagnetic'';
this would suggest that (in terms of Sec.~\ref{sec:decLRO}
the 12.2\AA{} linkage induces an ``antiferromagnetic'' 
interaction.
%%%%%%%%%%%%%%%%%%
\remark{That means frustration, since the network in 
that structure is an anisotropic triangular lattice
in which the other bonds are  all the same kind.}

\CLH{to MM: I didn't quite understand your comment in
email of 1/19/06; can you write it in the fashion 
which would insert in the paper?
``Another observation:
A common description for these classes of structures remains binary-PT
with 10.4A edge length.
To go from one structure to another, we just exchanger role of 
$L \leftrightarrow S+$
(or $S-$) type vertices.
(For W cell, structure is invariant under such exchange).
As a consequence, Al6Co5 motifs are now 
$\sqrt{3-\tau} \quad \tau \times$ 10.4\AA{} = 19.8\AA{} apart.''}

Although the ``\Star'' motifs no longer sit at vertices of this
19.7\AA{} edge network, that atom cluster is still in evidence.
The difference is that in the PD4 structure, 
every \Dec~ is encircled by ten such \Stars, 
whereas this number was smaller in our model of Fig.~\ref{fig:RulesWideal}.
The PD4 atomic structure,  like the rigid-site-list models of 
Sec.~\ref{sec:ideal-deco},
can be decomposed into a DHBS (Decagon-Hexagon-Boat-Star)
tiling with edge $a_0=2.45$\AA{}.

In fact, a decagonal model based on PD4 is a concrete example of the
structure models intermediate between the ``basic Ni'' decoration
of Ref.~\onlinecite{alnico01} and the decoration of our 
Sec.~\ref{sec:ideal-deco}.  Such a hybrid model would have 8\AA{} decagons,
which are absent in the former decoration, but have a larger proportion of 
2.45\AA{} HBS tiles than in the latter decoration. 
(In particular, every edge of the 20\AA{} decagon is the centerline of 
a 2.45\AA{} Hexagon, decorated typically by a Ni-Ni pair.)
We expected such an intermediate model to be favored at an intermediate 
composition such as Al$_{70}$Co$_{15}$Ni$_{15}$, but the energy 
differences are quite small, so it may well be competitive at
the compositions the present paper focuses on.  

\subsection{Structure model from
    Al$_{71}$Ni$_{22}$Co$_7$ approximant}
\label{app:20A-bNiapp}

Another decagonal approximant was discovered with composition 
Al$_{71}$Ni$_{22}$Co$_7$, close to the ``basic Ni'' phase, 
and  a structure model developed based on
electron diffraction and Z-contrast imaging~\cite{abe06-ICQ9}.
This model consists of edge-sharing 20\AA{} decagons of the
kind we have been describing.
They have Al$_6$TM$_5$ cores, {\it alternating} in orientation
(the cluster network happens to be bipartite).
There are no ring 2.5 Al; about half the edges of the 
\Decagons~ (contained in the 20\AA{} one) have two
ring 3 Al, the other half of the edges have only
one ring 3 Al.  The edges of the 20\AA{} decagons, without
exception, have two TM atoms.

A good interpretation of the cluster network in Ref.~\onlinecite{abe06-ICQ9}
is that the 20\AA{} (edge 6.4\AA{}) decagons occupy the Large sites of 
a Binary tiling with edge 16.8\AA{}, while pentagonal bipyramids (\PBs)
occupy the Small sites.
The vertices of the 20\AA{} decagons, as in the PD4 model, are occupied
by motifs like the \Star. Such \Stars~ also occur between pairs of
adjacent \PBs. There, they define additional tile
vertices which divide the area between the 20\AA{} decagons 
into Hexagons, Boats, and Stars with edge 6.5\AA{}.
%%%%%%%%%%%%%%%%%%%%%%%
\remark{So we now have a 6.5\AA{} DHBS tiling to add to the 
2.45\AA{} and 4.0\AA{} DHBS tilings!   But it is not useful to
compare these HBS tiles to those of the basic Ni decoration,
as these ones are always centered by the PB-type cluster.}
%%%%%%%%%%%%%%%%%%%%%%%

\CLH{To MM: in email 1/19/06, you wrote
"the traces of the Ni-rich 6.5A-HBS decorations are more pronounced."
I have referred, instead, to Ni-rich 2.45-HBS.  Which is more proper?
Would you like to contribute a couple sentences here?}

\subsection{Structure model based on W(AlCoNi) approximant}
\label{sec:deloudi-W5D}

     Deloudi et al~\cite{del06} have presented a structure model
(formulated  in a 5D-cut framework) for the same Co-rich 
composition we address here.  They formulate a 20\AA{} pentagonal
cluster, which is built around the W-phased pentagon cluster;
these can overlap in various ways.  
The Al/TM assignments in the model are based entirely
on those reported in Sugiyama's $W$(AlCoNi) refinement~\cite{Su02}.

Examination of their
Fig. 1 reveals the relation between their cluster and our
\Decagon~ motif. Take each vertex of their 20\AA{}~ decagon that is
{\it not} a vertex of the inscribed large pentagon, and draw
an arc around it through the adjacent decagon vertices 
(which {\it are} vertices of the inscribed pentagon); this 
arc encloses 3/10 of a \Decagon.  Their model, being based
on a stacking period $2c\approx 8$\AA{}, accomodates not
only the pentagonal bipyramid (\PB), but also
the puckering of the Al atoms we called ring 2.5/ring 3 
(or better, channel atoms) along the arcs just mentioned.

In Fig.~3(a) of Ref.~\onlinecite{del06} it can be
seen that \Decagons, although not recognized at all in
their formulation, are naturally generated in the interstices
between the large pentagons. The overall pattern
could be decribed as a packing of edge-sharing 32\AA{} 
super-decagons (inflated by one factor of $\tau$), with 
\PBs~ placed on the super-decagon centers and odd 
vertices, and \Decagons~ on the super-decagon even vertices.
(It is not surprising to find descriptions on different 
length scales: their model, being essentially a decoration of 
the quasiperiodic Penrose tiling, acquires its inflation symmetry.)

We question whether the details of this model are good energetically.
Our study suggests that good models are built by combinations of
the 8\AA{} decagon (the central part of the \Decagon), of 2.45\AA{}
edge HBS tiles, plus the \PB.  To the extent that this
model appears to have incomplete fragments of the 8\AA{} decagon,
we suspect it will have an increased energy.  It is conceivable
that only Al atoms are wrong, and the TM atoms are correctly
placed, which would still give good agreement with electron
micrographs of all kinds.

The \Decagons~show up prominently in
the experimental HAADF-STEM image of Ref.~\onlinecite{del06}
(their Fig.~3(b)) as white pentagons,
all oriented the same way. These are not as frequent as
they would be in the decoration of our Fig.~\ref{fig:RulesWideal}, 
suggesting that our model is not correct for that composition.

\subsection{Preliminary simulations with 20\AA{} clusters}
\label{app:20A-MC}

%%%%%%%%%%%%%%
\begin{figure}
\includegraphics[width=2.35in,angle=0]{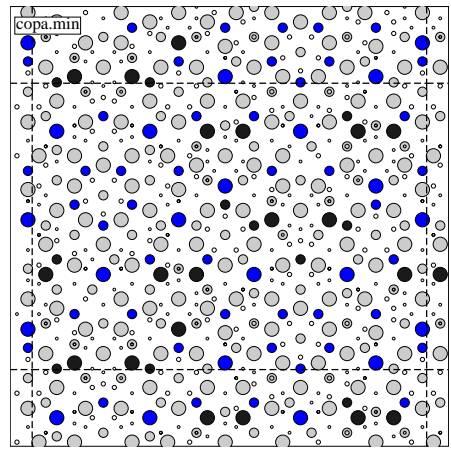}
\caption{Fixed-site simulation in which 20\AA{} decagons emerge.}
\label{fig:20A}
\end{figure}

Fig.~\ref{fig:20A} shows an exploratory fixed-site lattice-gas 
simulation on a 4\AA{} tiling in the 32$\times$23 cell, similar 
to Sec.~\ref{sec:MC-4A}, but with a richer set of candidate sites 
decorating the tiles, and a much longer annealing  time.  
Two 20\AA{} decagons are seen in this configuration, in place of
the four 13\AA{} decagons that typically emerged in our main
simulations.  The atomic structure is quite similar to those
described above;  a minor difference is that 
every edge of the \Decagon~ has exactly one ring 3 Al, and their
placement alternates perfectly. (That alternation was necessarily
disrupted in our model of Sec.~\ref{sec:ideal-deco}, wherever
the \Dec's shared edges.)  Also, the edges of the 20\AA{} decagon
are not only Ni-Ni, but often Al-Co or Ni-Co in Fig.~\ref{fig:20A}.
It would take much more work to settle the optimum decoration
of these sites, and the optimum atom content for a decoration
based on such configurations.

\remark{Following data in MM email 1/19/06.}
The 20\AA{} based configuration of Fig.~\ref{fig:20A} is lower
in energy by 1.8 meV/atom as compared to the best result of
the 13\AA{} sort produced by our 4\AA{} simulations, if
ideal-site configurations are compared.  However, after
the RMR protocol in an 8\AA{} cell (see Sec.~\ref{sec:relax}), 
the 20\AA{} type structure was slightly {\it higher} in energy,
by 0.6 meV/atom, than the 13\AA{} type structure.
(The latter probably contains more Al channels, and thus
offers more opportunity to reduce energy by puckering.)

\CLH{to MM: would you agree with speculation in the preceding sentence?}

\remark{If we apply RMR in a 4A stacked cell -- which is not
very physical -- then MM (1/19/06) tells the 13A structure is
better by 3.1 meV/atom.}

\subsection{Ways our approach can mislead}

{\sl FROM email: mvic-colong-disc.out0622)}

\CLH{to MM: I asked your opinion in email 6/22/05, but you did not respond,
I think}
\MM{NEW Please see what I said below. Regarding ``misleadings'', further 
below.}
Overall, it is our impression that -- despite the deficiencies
(known and still unknown) in our potentials, the biggest problem
for achieving a correct structure is the search pathway to find it.
On the one hand, we think our method has been remarkably
successful at predicting characteristic features such as the
appearance of the \Decagon in Co-rich compositions.
\CLH{AND, I THINK, THEY WOULD PREDICT THE PUCKERING PATTERN VERY WELL
IN THE W-PHASE.}
\MM{NEW I think so too. Overall, we didnt really get to the stage
of refining structure with respect to optimal channeling/puckering.
I have done already somewhere a calculation (may be need to repeat),
in which I compared energy of relaxed W structure, with the result
of our RMR procedure applied to the same composition/density/cell.
The result was W structure was significantly better 
(was it like 20meV/atom?) -- clearly showing we didnt exploit
all the predictive power of the pair potentials.
}

On the other hand, we have recounted four ways in which our procedure
misled us: (i) the fixed site list (ii) the 8 \AA{} periodicity (iii)
the possibility of a larger cluster than \Dec (iv) the cluster
orientation order

\CLH{I need to think of a SPECIFIC EXAMPLE for (i) and (ii)}
\MM{NEW I dont quite understand here. I would rather say we were
not mislead at all, quite surprisingly!
(i) fixed site list  brought us very close to W-phase structure (well,
say by another small step, we already new a lot from the primary 2.45A 
general simulation);
(ii) 8A periodicity: yes, clearly the 8A periodicity doesnt show up
from ideal sitelist simulation without puckering. And I do think it
is hardly possible/effective to include puckering in the initial
exploratory stages. But RMR has lead us to all the ideas about
puckering, and now I would say we are ready to set up another
discrete MC stage of exploration that would incorporate puckering, 
once we understand what's going on.
(iii) yes, the small-size restriction is an issue. We are now looking
for objects as big as 20A in diameter, so one would say our cells
are just around the very limit when they {\em perhaps} could form.
The difficulty is that we dont even know whether we could expect
this object emerging from 4A simulation. There are indication about
opposite: 5-channel objects that form around Co in the 3-ring are
very favourable, yet we dont know a good way of maximizing their
number, due to the way the \Dec clusters pack.
(i) orientations: I dont quite see how we were mislead here?
}

A separate note is that our calculations are good only to predict
stability between similar decagonal approximant structures.  Many
deficiencies of our potentials -- the dependence on electron density
(and hence on composition, in principle); inaccuracies in the
nearest-neighbor TM-TM and TM-Al potential wells; cutoffs at some
interaction radius; omission of three- or four-body terms; and poor
handling of vacancies -- will tend to cancel, in such a comparison.
But any phase may, of course, be preempted by a coexistence of two
dissimilar phases that happens to have a slightly lower energy; our
pair potentials are more likely to give a wrong answer in this
situation.  To construct a global phase diagram, it is necessary to
follow up the kind of search described in this paper, by ab-initio
total energy calculations, with an attempt to imagine {\it all}
possible competing phases and include them in this database.  
\CLH{to MM: can you cite YOUR BEST examples of such a project?}  
\MM{NEW You could cite Hennig ea TiCrSi work, then d-BMgRu and d-CaCd 
by MM+Widom.}  
Since it would be prohibitive to try out a large set of
candidate structures with the ab-initio codes, the present sort of
study is a prerequisite to the phase-diagram studies.


\begin{thebibliography}{99}
\bibitem{FN-bracketdensity}
In the initial stage Monte Carlo simulations,
a better procedure would have been to use 
(at least) two densities, bracketing the 
physical one, as elaborated in Sec.~\ref{sec:disc-discrete}.
\bibitem{FN-highT} In some runs, samples were only cooled to $\beta =
 4$, corresponding to $T=$3000 K (which would be well above the
 melting point, but the constraint of a discrete site list makes a
 large difference) But low energy structures obtained at these high
 temperatures were quite similar in configuration to those obtained
 from low temperatures, in that the decagonal rings were pretty well
 bound together: the structures that are favorable $\beta = 4$ are, in
 general, the same as those favorable at $\beta = 20$.
\bibitem{FN-sitelist} 
The augmented site list used in Sec.~\ref{app:20A-MC}
does produce exactly the same site list, no matter
how the 4\AA{} rhombi are packed between 13\AA{}
star decagons.
\CLH{CLHNEW: added this footnote}
\bibitem{FN-4Aanneal} 
The temperature is lower because these
simulations have fewer degrees of freedom than the $2.45$\AA{}
simulations.
\bibitem{FN-AlpotwithAl}
More generally, a potential map may defined whenever a securely
occupied subset of sites can be identified. Its efficacy depends on
how nicely the problem of placing the remaining atoms factorizes
into independent problems.  In the present case, we might (after
filling the deep wells with Al) define a second potential map,
which would be more realistic since the undetermined Al atoms
obviously do interact with the well-determined ones.  However, the
images of such a potential function (which is dominated by
forbidden zones where the atoms have been filled in) are harder to
relate to the atomic structures, and we did not use them in the
present work.
\bibitem 
{FN-anisowell} Observe how the ring-1 Al well extends in the $xy$
plane outward from the cluster center 
(in the \Star as well as \Dec).
helping explain shifts in the position under relaxation
(subsec.~\ref{sec:relax-bilayer}), 
and predicting an anisotropic Debye-Waller factor in
diffraction experiments owing to fluctuations in that well.
\bibitem{FN-rotateDec}
If we simply take the idealized model
of Fig.~\ref{fig:RulesWideal}, and rotate every alternate \Dec~ core
by the 10-fold screw axis without any other modification,
the energy increases by by $\sim$1 eV per cluster.
This confirms how the ring 3 and \Star~ atoms adapt to various 
\Dec~orientations so as to make them nearly degenerate.
     %%%%%%%%%%%
\remark{Previous note from MM 9/05. He added another
confirmation of the same thing: 
``Another interesting observation: with a site-list
that only contained fully occupied deterministic model positions, plus
alternative Al-cluster-center positions, I needed to go to 8000K (!)
to end up with the favourable orientations.''}
\bibitem{FN-seed} In the simulations reported here, runs were performed
 as a series of repeated cooling cycles. Each succeeding cycle started
 from the last configuration of the previous cycle, at a very high
 temperature, so that we didn't need to re-initialize random seed. Eventual
 variant runs were performed with different run parameters, or started
 from different seed configuration.
\bibitem{FN-U0value} $U_0$ has $\pm 1$ eV variations between channels
that do not affect this discussion.
\bibitem{FN-omit2ndAl} Apart from the hard core, $V_{\rm AlAl}(r)$ is
so weak that we are fully justified in omitting $V(z_{m+2}-z_m)$ from
Eq.~\ref{eq:Etot}.
\bibitem{FN-solve-Echannel} This solution took many steps: to manifest
the symmetres was helpful to work using the basis vectors
$(-1,0,1)/\sqrt {2} $ and $(1,2,1)/\sqrt{6}$ for the subspace of
allowed $(u_{-1},u_0, u_{1})$
\end{thebibliography}
\end{document}